\documentclass[11pt]{article}
\usepackage{graphicx,dblfloatfix}
\usepackage{amssymb,amsmath,amsfonts,eurosym,geometry,ulem,color,setspace,sectsty,comment,footmisc,caption,pdflscape,array}
\usepackage{mathrsfs}
\usepackage{tabu}
\usepackage{epstopdf}
\usepackage{float}
\usepackage{array}
\usepackage{rotating}
\usepackage[justification=justified]{caption}
\usepackage{caption}
\usepackage{subcaption}
\newenvironment{proof}[1][Proof]{\noindent\textbf{#1.} }{\ \rule{0.5em}{0.5em}}
\usepackage[colorlinks,citecolor=blue,urlcolor=blue]{hyperref}
\allowdisplaybreaks

\newcolumntype{L}[1]{>{\raggedright\let\newline\\arraybackslash\hspace{0pt}}m{#1}}
\newcolumntype{C}[1]{>{\centering\let\newline\\arraybackslash\hspace{0pt}}m{#1}}
\newcolumntype{R}[1]{>{\raggedleft\let\newline\\arraybackslash\hspace{0pt}}m{#1}}
\UseRawInputEncoding
\newcolumntype{P}[1]{>{\centering\arraybackslash}p{#1}}
\newcolumntype{M}[1]{>{\centering\arraybackslash}m{#1}}
\usepackage{tikz}

\usepackage{amssymb}

\usepackage{pifont}
\newcommand{\cmark}{\ding{51}}%
\newcommand{\xmark}{\ding{55}}%
\usepackage[natbibapa]{apacite}
   \usepackage{amsmath}
    
\usepackage[section]{algorithm}
\usepackage{algorithmic}
    \usepackage{amssymb, amsfonts}
    \usepackage{adjustbox}
    \usepackage[]{mathtools}
\usepackage{verbatim}

        \newtagform{eqbrackets}[]{(}{)}
        \usetagform{eqbrackets}
\usepackage{babel,blindtext}

    \usepackage{etoolbox}
    \patchcmd{\subequations}{\alph{equation}}{.\arabic{equation}}{}{}
    \AtBeginEnvironment{alignat}{\addtocounter{equation}{0}}
\newtheorem{thm}{Theorem}

\newtheorem{Assump}{Assumption}

\begin{document}
\date{\today}
\onehalfspacing
\title{Preference Estimation in Deferred Acceptance with Partial School Rankings\medskip\\ \normalsize{(JOB MARKET PAPER)}}

\author{
Shanjukta Nath\thanks{email: nath@umd.edu. I am grateful to Profs Guido Kuersteiner and Sergio Urzua for their invaluable suggestions and insights. I also thank Profs Nolan Pope, Ethan Kaplan, Chenyu Yang and seminar participants at the Applied Microeconomics brown-bag seminar at the University of Maryland, College Park, for their comments and suggestions. I thank the Ministry of Education of Chile for providing access to anonymized data. All views and opinions expressed are mine and not of the Ministry of Education.}
\\
Department of Economics, University of Maryland, College Park }
\date{\parbox{\linewidth}{\centering%
  \today\endgraf\bigskip }}
\maketitle
\begin{center}

\end{center}

\begin{abstract}
\singlespacing
\footnotesize

The Deferred Acceptance algorithm is a popular school allocation mechanism thanks to its strategy proofness. However, with application costs, strategy proofness fails, leading to an identification problem. In this paper, I address this identification problem by developing a new Threshold Rank setting that models the entire rank order list as a one-step utility maximization problem. I apply this framework to study student assignments in Chile. There are three critical contributions of the paper. I develop a recursive algorithm to compute the likelihood of my one-step decision model. Partial identification is addressed by incorporating the outside value and the expected probability of admission into a linear cost framework. The empirical application reveals that although school proximity is a vital variable in school choice, student ability is critical for ranking high academic score schools. The results suggest that policy interventions such as tutoring aimed at improving student ability can help increase the representation of low-income low-ability students in better quality schools in Chile. 

\end{abstract}

\bigskip 

\bigskip 


\newpage

\section{Introduction}
Policymakers are using centralized allocation algorithms to match students to schools all across the globe. These algorithms require parents to submit a ranking over schools, which is a critical input to the matching process.\footnote{Variants of centralized student assignment algorithms have been used in New York City, Boston, Chicago, Mecklenburg County in the United States \citep{abdulkadirouglu2017welfare,abdulkadirouglu2005boston,pathak2013school,hastings2005parental},
Finland \citep{salonen2014matching}, Hungary \citep{biro2012university}, Ghana \citep{ajayi2013school}, Tunisia \citep{luflade2017value} and several other countries.} Among such algorithms, the Deferred Acceptance (DA) algorithm is extensively used as it is strategy-proof \citep{abdulkadirouglu2003school}. Parents are expected to report school rankings according to their true underlying preferences and not manipulate their behavior to achieve favorable allocations.\footnote{Another common assignment algorithm is the Boston mechanism \citep{ergin2006games,pathak2008leveling}. However, research indicates that it gives sophisticated parents an advantage as there are gains associated with strategic behavior.} 

However, strategy proofness of DA fails if there are positive application costs \citep{fack2019beyond}. Empirical studies studying parental rankings in DA find that parents often report a partial list. The partial list can result from hard limits on the length of the list\citep{haeringer2009constrained,luflade2017value}. But, it is observed that parents do not reveal the complete ranking over schools, even without any hard limits. Multiple factors can contribute to such behavior. Often parents have access to a guaranteed school, and they do not want to list schools below this outside option. Parents are also aware of the oversubscribed schools and would like to skip reporting such schools if the likelihood of admission is very low and does not compensate for the application cost. Partial ranking due to these factors can lead to non-strategy proof equilibria and pose an issue with identification. 

This paper addresses identification under partial student ROL in DA. I incorporate the value of the outside option and the probability of admission in an additive cost framework in the school choice model to identify the determinants of true preferences over schools \citep{causalinference}. The school choice framework developed here differentiates between ranked and non-ranked alternatives using the rank cut-off. Ranked schools are assumed to provide higher utility (accounting for the probability adjusted cost) than the cut-off, and non-ranked do not offer as much utility as the cut-off. Notably, the cut-off is modeled as a choice variable and is endogenous. I contribute to the methodology of rank-ordered choices by modeling the student's decision to rank schools as a single step process. In my model, the random error component of latent utilities is known to the parents and is drawn once and assumed to be fixed for the entire school choice process. The one-step process is computationally intractable as the likelihood involves an m-dimensional definite integral. For instance, if a parent ranks $m$ schools, the likelihood involves $O(2^{m})$ terms, which becomes computationally prohibitive if $m$ is large. Therefore, I develop a novel recursive algorithm to compute the likelihood efficiently. The new model predicts the length of student ROL along with its determinants by allowing for variation in rank cut-offs. This is applied to Chilean DA for high school assignments. I find that student ability plays a critical role in ranking behavior in Chile that may dominate the costs imposed due to low household income or travel distance to schools.

There are several unique features of the Chilean system that leads to partial ROL and variations in the rank cut-off. Parents submit partial ROLs even though there are no monetary application fees or limits on length of ROL. About 60\% of parents rank at most three schools for high school admission. Although there are no direct monetary costs, there might be implicit costs. Such implicit costs are likely to be a function of the level of sophistication of the parents. 

The outside option is an essential component of the cut-off at which the parents decide to stop ranking other schools. In Chile's case, the outside option depends on two features for middle to high school transition. If the student's pre-DA school offers the grade at which the student seeks admission, the student is guaranteed admission at pre-DA school when the algorithm fails to allocate the student at any of the listed schools in ROL. On the contrary, if the pre-DA school does not offer a higher grade, which is the case for many middle schools, the student gets allocated to the nearest public school with a vacancy. Moreover, there is a high negative correlation between vacancy and school academic quality. Students enrolled in high score schools pre-DA are less likely to participate in DA. This correlation makes it critical to account for indicators of the probability of admission explicitly in the school choice model. 

In this paper's school choice estimator, identification can be achieved for Chile with full support. If there is a positive density of all types of students around different school types, this variation in student location of a similar type can be critical for identification. In other words, let's assume one can use student income and the ability to define student type. School types are defined by academic quality as a simplifying assumption. If similar typed low income and low ability students are located around high and low academic quality schools, then this distance variation based on location can be used for identification. The student location provides variation in the outside option. If there is variation in outside options for similar type students, this can be used to explain the length of ROL for these students. Empirical analysis reveals a notable variation in the placement of similar type students around different schools and the outside value due to this geographic location. 

The parametric framework allows me to account for the outside option value and the additive costs adjusted by the assignment probability. There are likely systematic differences in costs based on unobserved characteristics. I account for unobserved heterogeneity using the correlated random effect.  Correlated random effects are a commonly used technique to account for unobserved heterogeneity, particularly in situations with a small number of observations per agent, making fixed effect estimation inconsistent. I use the EM algorithm to account for the unobserved component in the underlying preferences. 

A crucial source of variation for identification in the Chilean context is the distance to school as it modifies the outside option. Consequently, it is critical to computing this variable precisely. Most of the current work in related literature uses crude proxies of this distance measure either due to the precise student address's unavailability or due to the computational costs of calculating such travel distances for large administrative data-sets \citep{laverde2020unequal,burgess2015parents,burgess2010school}. The Chilean government has provided precise student and school geo-coordinates to the researchers to analyze the new system. I take advantage of this information and employ the Open Source Routing Machine (OSRM) API to compute travel distances for each student school combination for the set of all schools in a student's school market to use as an input into the rank order model.\footnote{These computations were done on secure servers in collaboration with the Ministry of Education in Chile. The resulting data-sets are owned by the Ministry of Education, Chile.}

A decentralized system preceded the centralized system with vouchers for students belonging to poor income families. However, school segregation in Chile based on socioeconomic status has increased in the last decade.\footnote{The voucher system led to a proliferation of voucher schools in Chile funded partially by government vouchers but managed by private entities}. This is due to the disproportionate flight of students belonging to middle-income households from the public to voucher schools. Consequently, the Chilean government's key motivation behind introducing centralized student assignments is to reduce the existing levels of school segregation. However, \cite{KNU2020} does not find any evidence suggesting an unambiguous decline in school segregation due to the government's new policy. This result is crucial and necessitates the study of parental ranking preferences.

My results illustrate that travel distance is a critical element in the decision-making process. Nevertheless, parents also care about the extent of the match between student ability and school academic rigor. The impact of distance on the ROL varies substantially by student ability and income. I observe two critical results. Higher-income households might have an advantage in overcoming the travel cost to good quality schools. However, student ability proxied by pre-centralization test scores is the most crucial determinant of listing the best quality schools in the students' school market. Once I condition the marginal effects on student income, the results showcase that higher ability students do end up applying to good quality schools irrespective of income levels. In other words, the ability can compensate for income levels and induce parents to incur the additional travel cost as parents care about the ability match between the student and school. 

The above result is critical as it suggests that reducing travel costs alone might not improve students' representation from low socioeconomic status in higher-quality schools. Policies geared towards improving student ability, especially for students belonging to low-income households, can go a long way in improving their representation in high-quality schools. Optimal reallocation can go a long way in improving student outcomes. It can help reduce absenteeism, drop out rates \citep{hanushek2008students} and improve academic performance \citep{kirabo2010students,hastings2008information,glewwe1994student}.

This paper is organized as follows. In section 2, I discuss the theoretical framework for the school choice model. I present the estimation strategy and the recursive algorithm in section 3. Section 4 discusses Chile's institutional setting, critical features of centralized allocation, followed by a detailed description of the analysis's data-sets. I also provide some reduced-form evidence on the impact of school quality on student attendance and the extent of school diversity before and after the reform. I apply the new estimator to student data in Chile and present the results in section 5. I discuss alternative simulations and make policy recommendations in section 6. Lastly, I conclude in section 7.

\section{Theoretical Framework}
I follow the school choice model provided in \cite{fack2019beyond} and \cite{causalinference}. I assume that every student $i\in\{1,2,..,n\}$ chooses from a set $\mathcal{J}=\{1,2,...,J\}$ schools.\footnote{$\mathcal{J}$ corresponds to the set of the total number of participating schools in DA, and this is likely different from the total available schools in the schooling market. For instance, only public and voucher schools participated in DA and not the non-voucher schools in the Chilean schooling market.} The three components of DA comprises of student preferences, priority indices for each student school pair and the set of vacancies $q=\{q_{1},...,q_{J}\}$ at the DA participating schools. Given these three components, the student submits a ROL $L_{i}=\{l_{2,i},l_{1,i}....,l_{K_{i},i}\}\in \mathcal{L}$, which are manifested ranks over latent utilities. Here, $l_{1,i}$ is the top ranked school, $l_{2,i}$ is the second choice and so on and so forth till the student ranks the least preferred school $l_{K_{i},i}$. I allow the cut-off where the student decides to stop ranking schools to vary across individuals and is denoted by $K_{i}$. 

I use a linear index for the student utilities from schools. This linear index is the sum of a component $V_{ij}$ that is observed to the researcher and a random component $\epsilon_{ij}$ not observed by the researcher. Consequently, the set of student utilities is described as $u_{i}=\{u_{1,i},....,u_{J,i}\}$, where $u_{ij}=V_{ij}+\epsilon_{ij}$. 

$V_{ij}$ comprises of three types of explanatory variables i) covariates varying for each student school pair, ii) student specific characteristics and iii) school specific characteristics. Finally, $V_{ij}$ also comprises of any type of advantage that student $i$ enjoys for school $j$. Such advantages for school $j$ can be a function of factors such as sibling enrolled in school $j$, parents employed in school $j$, former students or for students with special needs. Conditional on these factors, DA uses a random lottery to generate priority indices for each student school pair used for tie breaking in each iteration in DA. In other words, students with an advantage at school $j$ will be assigned a higher lottery number relative to a student without an advantage all else equal. 

The popularity of DA is associated with its property of being strategy-proof. This mainly depends on parents revealing a complete ordering over schools and no costs associated with adding additional schools to ROL. However, the ROL observed by the researcher in real applications is a partial list instead of a complete ordering, and second, there are often some positive costs of the application. 

First, the partial list in DA can be attributed to the guaranteed seat or outside option. Since students have a positive probability of getting assigned to one of the listed schools, the ranked schools must offer at least as much utility as the guaranteed school. For instance, in the Chilean DA, the value of the outside option is determined by one of the following two components. Student $i$, while making the transition from middle to high school through DA, has a guaranteed seat in the middle school if it offers the high school grade. On the contrary, if the middle school does not offer high school grades, the student gets automatically allocated to the nearest public school with vacancies.\footnote{In Chile, this cut-off for the nearest public school with vacancies is 17 km from the student's place of residence.} This generates the outside option $u_{i,0}$ for student $i$. The outside option is observed to the researcher in the Chilean DA. Consequently, all schools in $\mathcal{J}$ that do not offer at least $u_{i,0}$ will not be ranked in ROL ($u_{i,j}-u_{i,0}<0$).

Second, DA is often associated with some costs of adding schools to ROL. The costs have multiple interpretations in student assignment. In some DA applications such as Hungary, there is an application fee associated with additional schools. On the contrary, \cite{fack2019beyond} interestingly suggests that there could be an implicit cost of application even without an explicit fee. Often this could correspond to the mental cost of obtaining information about several aspects of the school and the effort of listing additional schools on ROL. 

Variation in the value of an outside option can often result in partial lists. However, the critical question is even when I account for $u_{i,0}$ in the school choice model, do I observe a partial list based on true ranking for the schools strictly preferred over the outside option. This might not hold if there are costs associated with listing additional schools or parents are trying to exclude schools that are impossible to get into due to low vacancies. Proposition 1 in \cite{fack2019beyond} illustrates that the truth-telling property is no longer the equilibrium strategy for all under a positive application cost. Moreover, the authors highlight that the cost's magnitude need not be considerable to deviate from truth-telling. Even if this cost component is minuscule, the marginal benefit of adding school can be meager if the probability of admission to the additional school is close to 0 or there is a high chance of admission to a higher ranked school. All these factors poses difficulty for identification of student preferences in DA. 

\cite{causalinference} extends the results in \cite{fack2019beyond} and illustrates that observed ROL's are a subset of actual preferences by parameterizing the reasons for leaving out certain schools. This parameterization can be used for identification. We show that under no costs or a linear cost specification where the probability of admission enters the utility function in a additive manner one can identify the parameters of the student preferences over schools.

In Chilean DA, cost comprises of the implicit mental cost of application as there is no application fee. Such cost are less likely to be school specific and hence I assume a linear cost function. Using the linear index parameterization for utilities and proposition 2 in \cite{causalinference}, the utilities from a ranked alternative has to compensate for the value of outside option and the cost component adjusted by the probability of admission. In other words, 
\[
   u_{i,j}-\frac{c_{i}}{p_{j}^{s}}= 
\begin{cases}
    X_{ij}\beta+\epsilon_{ij}-\frac{c_{i}}{p_{j}^{s}}> u_{i,0},& \text{if}\hspace{2mm} j \hspace{2mm}\text{is ranked}\\
    X_{ij}\beta+\epsilon_{ij}-\frac{c_{i}}{p_{j}^{s}}\leq u_{i,0},              & \text{otherwise}
\end{cases}
\]
Further, I impose the following assumptions. 
\begin{Assump}
\label{assump1}
The school choice decision is made in a single step where student $i$ draws from the error component $\epsilon_{ij}$ from an i.i.d Type 1 extreme value distribution.
\end{Assump}

According to Assumption \ref{assump1} the unobserved component of the utility is known to the student and I assume it to be fixed for the school choice decision process. This deviates from the sequential choice process used in the urn model where the agent is assumed to draw the unobserved component repeatedly in each step. 

\begin{Assump}
The individual specific component $c_{i}$ is modeled as a correlated random effect as follows
\begin{equation*}
c_{i}=Z_{i}\gamma+w_{i}
\end{equation*}
where $Z_{i}$ is a student specific observed characteristic such as the outside option value and $w_{i}$ is drawn from $N(0,\sigma^{2})$. 
\end{Assump}
Identification in data using the above set up can be achieved with the assumption of full support. Full support implies that there should exist variations in the placement of all types of students across different school types. This geographic variation will generate variation in the outside value or the quality of the guaranteed school, which will help to pin down the parameters of parental preferences in the presence of partial lists. I explain the argument using a simple example. To keep the illustration simple, I assume two students type low income and high income. I also assume two school types of high and low ability.  

The student and school types are part of the observed covariates vector $X_{ij}$. Besides, this vector contains information for travel distance for every school student pair. The placement of every student type around all types of schools will make the outside option school for each type of student accessible for students of other types. This variation in the outside option for similar types of students such a low income due to the geospatial location will reveal the preference ordering of low-income students over all types of schools. 
\section{Estimation}
The school choice model outlined in section 2 provides a framework to identify the determinants of true parental preferences with the partial ROL. In this section I lay out the steps used for estimation. I compute the marginal distribution that student $i$ ranked $K_{i}$ schools by integrating over the distribution of the unobserved individual correlated effect $c$. 
\begin{align*}
f_{l}(L_{i})=\int g(L_{i}|c_{i};\beta)f_{w}(c;\gamma,\sigma^{2})dc
\end{align*}
Since individual $i$ ranks $|L_{i}|=K_{i}$ schools out of a choice set of $\mathcal{J}$ schools, $g(L_{i}|c_{i};\beta)$ is given as follows as the individual draws the random component $\epsilon_{ij}$ once and then for the researcher it is fixed for the school choice process. This deviates significantly from the distribution on errors in a sequential decision making process where the random components are drawn in every draw from the urn \citep{glazerman2017market}.\footnote{I drop the $i$ subscript for the following derivation to keep the notation simple.} 
\small
\begin{align*}
g(L|\alpha;\beta)&=P(u_{1}-\frac{c_{i}}{p_{j}^{s}}>u_{2}-\frac{c_{i}}{p_{j}^{s}}>....>u_{K}-\frac{c_{i}}{p_{j}^{s}}>u_{0},u_{K+1}-\frac{c_{i}}{p_{j}^{s}}<u_{0},..,u_{J}-\frac{c_{i}}{p_{j}^{s}}<u_{0})\\
&=\bigg[\int_{-W_{1}}^{\infty}\int_{-W_{2}}^{W_{1}+\epsilon_{1}-W_{2}}......\int_{-W_{K}}^{W_{K-1}+\epsilon_{K-1}-W_{K}}f(\epsilon_{K})d\epsilon_{K}.....f(\epsilon_{1})d\epsilon_{1}\bigg]\prod_{l=K+1}^{J}\int_{-\infty}^{-W_{l}}f(\epsilon_{l})d\epsilon_{l} 
\end{align*}
\normalsize
The derivation from the second to the third equality holds as the events $u_{j}$ for $j\in\{K+1,....,J\}$ are independent conditional on the covariates and $c$.  Additionally, the movement from step three to four works as the event $u_{1}>u_{2}>....>u_{K}>-c$ is independent of $u_{K+1}<-c,u_{K+2}<-c,..,u_{J}<-c$. 
 
\textbf{Recursive Algorithm}: With the single step decision, the likelihood becomes hugely complicated and computationally expensive as now every integral in the above density has two bounds. For instance, the above problem with $J$ alternatives has $2^{J}$ terms. Due to this complexity, the literature on ROL models has been agnostic on the rank cut-off heterogeneity. I solve this computational challenge by developing a recursive algorithm for the Threshold Rank Order estimator. 

\begin{thm}\label{theorem1}
If there are a total of K ranked alternatives from a set of J schools such that K+1,..,J are non-ranked, then the likelihood of the observed ranks conditioned on covariates and the latent variable is given as
\footnotesize
\begin{align*}
P(u_{1}&-\frac{c_{i}}{p_{j}^{s}}>u_{2}-\frac{c_{i}}{p_{j}^{s}}>....>u_{K}-\frac{c_{i}}{p_{j}^{s}}>u_{0},u_{K+1}-\frac{c_{i}}{p_{j}^{s}}<u_{0},..,u_{J}-\frac{c_{i}}{p_{j}^{s}}<u_{0})\\
&=\bigg[\kappa_{1}\kappa_{2}...\kappa_{K-2}\kappa_{K-1}(1-F(-W_{K})...F(-W_{1}))-\bigg\{\kappa_{2}..\kappa_{K-1}F(-W_{K})F(-W_{K-1})..F(-W_{2})\boldsymbol{I(1)}\bigg\}\\
&-......-
\bigg\{\kappa_{K-1}F(-W_{K})F(-W_{K-1})\boldsymbol{I(K-2)}\bigg\}-\bigg\{F(-W_{K})\boldsymbol{I(K-1)}\bigg\}\bigg]\prod_{l=K+1}^{J}F(-W_{l})
\end{align*}
where $\boldsymbol{I(k)}=\int_{-W_{1}}^{\infty}\int_{-W_{2}}^{W_{1}+u_{1}-W_{2}}......\int_{-W_{k}}^{W_{k-1}+u_{k-1}-W_{k}}f(u_{k})du_{k}.....f(u_{1})du_{1}, \kappa_{n-j}=\frac{1}{\sum_{j=n-j}^{n}e^{-(W_{n-j}-W_{j})}}$ and \\
$F(-W_{j})=e^{-e^{W_{j}}}$.
\end{thm}
The proof of theorem 1 is provided in Appendix A.

The urn model is the most popular model used to analyze ranked data in the literature. According to Plackett and Luce, the ranking process for J items can be thought of as an aggregation of $J$ independent iterations. In the first iteration, the individual chooses the most preferred item from the set of $J$ items. In the next iteration, individual chooses the best from the left over items after the top choice is removed. This process continues till all the items are ranked. The likelihood of the event $u_{1}>.....>u_{J}$ under a logit specification is 
\begin{align*}
P(U_{1}>.....>U_{J})&=\frac{e^{W_{1}}}{\sum_{j=1}^{J}e^{W_{j}}}\times \frac{e^{W_{2}}}{\sum_{j=2}^{J}e^{W_{j}}}\times.....\times \frac{e^{W_{J-1}}}{\sum_{j=J-1}^{J}e^{W_{j}}}\\
&=\kappa_{1}\kappa_{2}...\kappa_{n-2}\kappa_{n-1}
\end{align*}
where $\kappa_{n-j}=\frac{1}{\sum_{j=n-j}^{n}e^{-(W_{n-j}-W_{j})}}$. I use the same notation as the threshold rank order model for comparison (see Theorem \ref{theorem1}). The Plackett-Luce (urn model) multiple stage process for ranking data did not explicitly discuss impartial rankings. Some discussion on ties and partial rankings is done in \cite{allison1994logit,skrondal2003multilevel,guiver2009bayesian}. For instance, \cite{allison1994logit} suggested that one can assume an underlying rank over non-ranked alternatives, which is not observed by the researcher. However, one can account for all permutations of possible rank orderings over the non-ranked alternatives in the likelihood. Suppose, the individual ranks two alternatives out of 4 and last two alternatives are non-ranked. I follow the model in  \cite{allison1994logit} and  modify the likelihood of the urn model as 
\begin{align*}
P(u_{1}>.....>u_{4})&=\frac{e^{W_{1}}}{\sum_{j=1}^{4}e^{W_{j}}}\times \frac{e^{W_{2}}}{\sum_{j=2}^{4}e^{W_{j}}}\times \bigg[\frac{e^{W_{3}}}{e^{W_{3}}+e^{W_{4}}}+\frac{e^{W_{4}}}{e^{W_{3}}+e^{W_{4}}}\bigg]\\
&=\frac{e^{W_{1}}}{\sum_{j=1}^{4}e^{W_{j}}}\times \frac{e^{W_{2}}}{\sum_{j=2}^{4}e^{W_{j}}}
\end{align*}
 The above likelihood has several differences with the assumptions used for the school choice model in this paper. First, this method does not differentiate between ranked and non-ranked alternatives in terms of the underlying utility. Second, splitting the decision process into multiple stages assumes that the individual cares about the most preferred alternative for that stage and this decision is completely independent of the decision process during other stages. On the contrary, I work with the assumption that the individual ranks all the ranked schools in the same step. 
 
Next, I discuss the computation of the parameters for the manifested variable ranks and those determining $c$ with unobserved individual heterogeneity. I use the Monte Carlo EM algorithm to estimate the parameter (\cite{dempster1977maximum} and \cite{sammel1997latent}). The parameter space for the manifested ranks consist of $\beta$ (manifested ranks) and individual specific effect consist of $\gamma,\sigma^{2}$. If I were to observe the $c_{i}$, the log-likelihood for the complete data is 
\begin{align}
log f_{rc}(L_{i},c_{i})=\sum_{i=1}^{N}\{log[g(L_{i}|c_{i};\beta)]+log[f_{w}(c_{i};\gamma,\sigma^{2})]\}, \label{eqem1}
\end{align} 
However, the latent cost is unobserved and the computation of the observed likelihood ($f_{l}(L_{i})=\int g(L_{i}|c_{i};\beta)\\f_{w}(c_{i};\gamma,\sigma^{2})dc
$) of listed ROL requires to integrate over the unobserved cost. EM algorithm provides an iterative solution to maximum likelihood estimation. 
\begin{algorithm}[h]
  \caption*{\textbf{Algorithm 1} Estimate school choice model}
  \begin{algorithmic}
  \STATE {\bfseries Input:} ROL $L$, Explanatory variables $X,Z$
  \STATE {\bfseries Result:} Parameter estimates for manifested ranks $\beta$, latent variable parameter estimates $\gamma,\sigma$
  \end{algorithmic}
  \begin{algorithmic}[1]
  \STATE Start with initial guess of parameters $\{\beta^{0},\gamma^{0},\sigma^{0}\}$
  \STATE Draw $T$ Monte Carlo samples from the prior distribution of latent variable $c$, which is $f_{w}(c_{i};\gamma,\sigma^2)$.
  \STATE Using the recursive algorithm of theorem 1 compute $g(R|c;\beta)$ and calculate expectation of the score function
  \STATE Solve for $\gamma,\sigma^2$ using equation (2) and (3). Update $\gamma,\sigma^2$.
  \STATE Use the $f_{w}(c_{i};\gamma,\sigma^2)$ and $g(L|c;\beta)$ obtained in step 2 and 3 respectively to formulate the $Q$ as derived in equation 4.
  \STATE Maximize and update $\beta$.
  \STATE Repeat steps 1 to 6 till all the parameters converge.
  \STATE \textbf{return} $\hat{\beta},\hat{\gamma},\hat{\sigma}$
  \end{algorithmic}
\end{algorithm}
In the E step, the algorithm computes the expectation of the observed likelihood using the initial guess of the parameters, and then the initial guess is updated in the M step by optimization of the expectation obtained in the E step.\footnote{There exists an extensive work on the use of the EM algorithm in the context of latent class structure \cite{croon1993latent}, \cite{francis2010modeling} and \cite{marden2014analyzing}\footnote{See chapter-10 in \cite{marden2014analyzing} for a detailed discussion on the application of EM algorithm to obtain MLE in the context of ranked data.}. However, for my setting, $c$ is a continuous latent variable, and I need an added approximation in the E step to compute the integral (\cite{booth1999maximizing,ibrahim1999monte}). Such approximation is not required with finite latent classes as the integral becomes a weighted sum of the conditional likelihood over the posterior distribution of discrete latent classes.}

Equation \ref{eqem1} is critical as it informs that the parameter space can be split into two parts. The first part can be used to estimate $\beta$, and the second component can be used to obtain estimates of $\gamma,\sigma^{2}$ corresponding to the correlated latent variable $c$. Since $c$ is unobserved, so instead of the above likelihood, I use the EM algorithm and compute the expectation of the gradient of the log-likelihood function (score) corresponding to $\gamma$ and $\sigma^{2}$. I follow the methodology in \cite{sammel1997latent} very closely for the EM algorithm. \footnote{The methodology in \cite{sammel1997latent} has been modified for my problem of rank-ordered ROL as \cite{sammel1997latent} developed the EM method for multinomial choice. Second, the distributional assumptions on error term is different in \cite{sammel1997latent}. Moreover, this paper did not solve for the variance of the distribution of the unobserved cost. I modify their proof used for the mean of the latent variable to additionally solve for the variance.} The posterior distribution $h(c_{i}|L_{i})$ is used to form the expectation of the score function.\bigskip\\
\textbf{Step 1, solving for $\hat{\gamma},\hat{\sigma}$}: The expectation of the score function w.r.t $\gamma$ is given below. I can equate $E_{c}S_{i}(\gamma)=0$ and solve for $\hat{\gamma}$. The proof for the mean is provided in \cite{sammel1997latent}, section 3.3, page 671. 
\begin{equation}
  \begin{aligned}
\hat{\gamma}&=\sum_{i=1}^{N}(z_{i}'z_{i})^{-1}(\sum_{i=1}^{N}z_{i}'\bigg[\frac{\sum_{t=1}^{T} c_{i}g(L_{i}|c;\beta)}{\sum_{t=1}^{T} g(L_{i}|c;\beta)}\bigg] )\\
&\hat{\sigma}^{2}=\frac{1}{n}\sum_{i=1}^{N}\bigg[\frac{\sum_{t=1}^{T}  (c_{i}-z_{i}\gamma)^{2}f(L_{i}|c;\beta)}{\sum_{t=1}^{T} f(L_{i}|c;\beta)}\bigg] 
 \end{aligned}
 \end{equation}
The proof for $\sigma^{2}$ is provided in Appendix A as \cite{sammel1997latent} does not solve for the variance.\bigskip\\
\textbf{Step 2, solving for $\hat{\beta}$}: In the E step, I compute the expectation of the likelihood w.r.t the distribution of the latent variable $c_{i}$ conditional on the observed data $L_{i}$. This is based on the same steps as \cite{sammel1997latent} on how Monte Carlo samples can be used for approximation of the integral. The steps for computation are given below. These are comparable with equation six and the next (unnumbered) equation on page 671 in \cite{sammel1997latent} modified for the likelihood in this paper. 
\begin{align*}
 E_{h(c|L)}g(L_{i},c)&=\int f_{lc}(L_{i},c)h(c|L;\gamma,\sigma^2)dc\\
%&=\int f_{lc}(L_{i},c) \bigg[\frac{g(L_{i}|c;\beta)f_{w}(c)}{\int g(L_{i}|c;\beta)f_{w}(c)dc}\bigg] dc\\
&=\frac{\int f_{lc}(L_{i},c)g(L_{i}|c;\beta)f_{w}(c)dc}{\int g(L_{i}|c;\beta)f_{w}(c)dc}
\end{align*}
I use Monte Carlo approximation at this step. I draw a sample of $c$ generated from the distribution $f(c)$.\footnote{$f(c)$ follows a normal distribution as $w\sim N(0,\sigma)$, but the means are adjusted by $Z\gamma$. } Using this sample the Monte Carlo approximation of the integrals are in equation 4. In the M step, I maximize $Q(\beta,\gamma,\sigma^{2}|\beta^{(0)},\gamma^{(0)},\sigma^{2}(0))$ and update the estimate to $\beta^{(1)},\gamma^{(1)},{\sigma^{2}}^{(1)}$. The optimization over $\beta$ and $\gamma,\sigma^{2}$ takes place in two steps within the maximization step. I keep iterating between the E and M step till the estimates converge. The steps of the algorithm needed to compute the estimates are provided as Algorithm 1. 
\begin{equation}
  \begin{aligned}
E_{h(c|L)}g(L_{i},c)&= \frac{\sum_{t=1}^{T}f_{rc}(L_{i},c_{t})f(L_{i}|c_{t})}{\sum_{t=1}^{T}g(L_{i}|c_{t};\beta)} \hspace{2mm}\text{,}\hspace{2mm}
Q(\beta,\gamma,\sigma^{2}|\beta^{(0)},\gamma^{(0)})&=\frac{\sum_{t=1}^{T}{g(L_{i}|c_{t};\beta)}^{2}f_{w}(c)}{\sum_{t=1}^{T}g(L_{i}|c_{t};\beta)}
 \end{aligned}
\end{equation}

\textbf{Monte Carlo Simulations}: I discuss the performance of the estimator for a model with multiple covariates (four covariates) and no latent individual heterogeneity. First, I generate the observed covariates $X_{1},X_{2}$ as random draws from $U[-3,1]$, $X_{3}$ is drawn from $U[-1,3]$ and lastly, $X_{4}$ from $U[-1,2]$.  The unobserved component $u$ follows a Type 1 extreme value distribution. Using the utility model, I generate the matrix of observed ranks. For an individual, I observe a partial ordering of ranks over schools from which the individual obtains a positive net benefit. An individual does not rank a school if the net benefit from that school is zero. 
\begin{table}[h]
\centering
 \fontsize{10}{10}\selectfont
\captionsetup{width=13cm}
\caption{Simulation: Multiple covariates and without latent variable}
\label{simtable1}
\begin{tabular}{p{1.5 cm}p{1.5 cm}p{1.5 cm}p{1.5 cm}p{1.5 cm}p{1.5 cm}p{1.5 cm}}\hline\hline
N & s & $\beta$ & MSE & MAE & Median bias & Mean bias \\\hline
100 & 15 & 0.5 & 0.0030 & 0.0410 & 0.0071 & 0.0061 \\
1000 & 15 & 0.5 & 0.0003 & 0.0129 & 0.0005 & -0.0004 \\
2000 & 15 & 0.5 & 0.0001 & 0.0085 & 0.0029 & 0.0011 \\\\
100 & 15 & 0.8 & 0.0041 & 0.0502 & 0.0003 & 0.0017 \\
1000 & 15 & 0.8 & 0.0004 & 0.0160 & 0.0025 & 0.0031 \\
2000 & 15 & 0.8 & 0.0002 & 0.0121 & 0.0013 & -0.0002 \\\\
100 & 15 & -0.9 & 0.0049 & 0.0569 & 0.0054 & 0.0051 \\
1000 & 15 & -0.9 & 0.0004 & 0.0168 & -0.0033 & 0.0000 \\
2000 & 15 & -0.9 & 0.0002 & 0.0109 & -0.0005 & -0.0021 \\\\
100 & 15 & -0.8 & 0.0059 & 0.0611 & -0.0123 & -0.0058 \\
1000 & 15 & -0.8 & 0.0007 & 0.0217 & -0.0061 & -0.0068 \\
2000 & 15 & -0.8 & 0.0004 & 0.0152 & -0.0002 & -0.0007\\
\hline\hline
\multicolumn{7}{c}{ \begin{minipage}{13.5 cm}{\footnotesize{Notes: This table illustrates the error distribution for the parameter estimates using the recursive model. The measures of error distribution are shown for four parameters. The number of schools, $s$, has been kept constant at 15, and $N$ increases from 100 to 1000 to 2000.}}
\end{minipage}} \\
  \end{tabular}
\end{table}
The observed data for the econometrician consists of $L,X1,X2,X3,X4$. The likelihood for the observed ranks can be obtained using the recursive solution provided in theorem 1. The data is generated to match closely with the actual ranking data that I examine in section 4. Figure \ref{simulationnolatent} shows the frequency distribution of ranks for one random simulation in this analysis (see (e) in figure \ref{simulationnolatent}). The figure shows that the frequency distribution tapers before the maximum ranked schools which is 15 in this case. Table \ref{simtable1} provides a summary of the simulations. The parameters to be estimated are $[0.5,0.8,-0.9,-0.8]$. The sample size $n=[100,1000,2000]$ and the number of simulations are 100. I use multiple starting values for optimization where starting values are random draws from $U[-1,1]$. The mean squared error as illustrated in column 4 in table \ref{simtable1} drops as I increase the sample size from 100 to 2000. Moreover, I also find better concentration of the estimates as the sample size increases, shown in panel (A) to (D) in figure \ref{simulationnolatent}. 

Here, I discuss the performance of estimator for the parameters $\theta=\{\beta_{1},\beta_{2},\beta_{3},\beta_{4},\gamma\}$. I generate the student ranks over $J$ in the following way. First, I generate the observed covariates $X_{ij}$ and $Z_{i}$. $X_{ij}'s$ are generated from uniform distribution $U[-3,3]$ and $Z_{i}$ is drawn from a standard Gaussian distribution.  Since the correlated effect $c_{i}$ varies at the level of individual, the covariate $Z_{i}$ is average characteristic over schools for an individual. The unobserved component $u_{ij}$ follows a Type 1 extreme value distribution and $w_{i}\sim N(0,0.25)$. The latent variable $c$ is generated as a linear function of observed component $Z\gamma$ and the unobserved random component $w$. Using the utility model as described in section 2.1, I obtain the matrix of observed ranks $L$.
\begin{table}[h]
\centering
 \fontsize{10}{10}\selectfont
\captionsetup{width=13cm}
\caption{Simulation: Multiple covariates with latent variable}
\label{simtable22}
\begin{tabular}{p{1.5 cm}p{1.5 cm}p{1.5 cm}p{1.5 cm}p{1.5 cm}p{1.5 cm}p{1.5 cm}}\hline\hline
N & s & $\beta$ & MSE & MAE & Median bias & Mean bias \\\hline
100 & 10 & 0.8 & 0.0062 & 0.0643 & 0.0014 & 0.0005\\
1000 & 10 & 0.8 & 0.0009 & 0.0237 & -0.0138 & -0.0118 \\\\
100 & 10 & 0.4 & 0.0052 & 0.0530 & 0.0091 & 0.0066 \\
1000 & 10 & 0.4 & 0.0003 & 0.0140 & 0.0018 & 0.0017 \\\\
100 & 10 & -0.8 & 0.0067 & 0.0651 & 0.0242 & 0.0143 \\
1000 & 10 & -0.8 & 0.0009 & 0.0233 & 0.0101 & 0.0131 \\\\
100 & 10 & -0.9 & 0.0079 & 0.0668 & -0.0016 & -0.0114 \\
1000 & 10 & -0.9 & 0.0006 & 0.0201 & 0.0070 & 0.0070 \\\\
100 & 10 & 0.5 & 0.0127 & 0.0869 & -0.0076 & -0.0139 \\
1000 & 10 & 0.5 & 0.0029 & 0.0331 & -0.0113 & -0.0115\\
\hline\hline
\multicolumn{7}{c}{ \begin{minipage}{13.5 cm}{\footnotesize{Notes: This table illustrates the error distribution for the parameter estimates using the recursive model. The measures of error distribution are shown for four parameters. The number of schools, $s$, has been kept constant at 10, and $N$ increases from 100 to 1000.}}
\end{minipage}} \\
  \end{tabular}
\end{table}
The observed data comprises of $R,X,Z$. I use the recursive solution derived above for density of ranks conditioned on the latent variable $c$. Since $c$ is the latent component in the utility, I use EM algorithm for estimation as described in section 2. For the posterior distribution of the latent variable  $h(c|L;\gamma,\sigma)=\frac{g(L_{i}|c;\beta)f_{w}(c;\gamma,\sigma^2)}{\int g(L_{i}|c;\beta)f_{w}(c;\gamma,\sigma^2)dc}$, I need to integrate over the latent variable $c$. I approximate for the integral using monte carlo simulations. I draw $T=500$ random samples of $w$ from $N(0,0.25)$ and obtain the latent variable $c=Z\gamma+w$. The denominator $\int g(L_{i}|c;\beta)f_{w}(c;\gamma,\sigma^2)dc=\frac{1}{T}\sum_{t=1}^{T}g(L|c;\beta)$. Similarly, I use the average over Monte Carlo samples for the numerator in $Q$. The initial values are random draws from a $U[-1,1]$.\\
The maximization step is divided into two components. First, I solve for the parameters determining the distribution of latent cost. Second, I use MLE estimator for $\beta$ and solve for it. Based on parameter estimates obtained in this step, I update it for the next iteration. I keep on iterating alternatively between the E and M step till the estimates converge. \\
Table \ref{simtable22} illustrates the simulation results for the parameters of this model. For all the parameters, I observe a decline in mean squared error as I am increasing the sample size from 100 to 1000. I also illustrate the concentration of the parameters as I increase the sample size in figure \ref{simulationlatent}. The estimated values for the parameters get closer to the population value as the number of individuals increase in the simulations. 

\textbf{Comparison with the urn model}: Here, I compare the performance of the new preference model estimator with the existing estimators used in the literature to study rank data. I use the urn model for comparison. Figure \ref{dataurn} shows that the threshold rank order model estimates are closer to the actual parameter than the estimates obtained using the urn estimator for each of the four parameters in the model.\footnote{Observed covariates $X_{1},X_{2}$ are random draws from $U[-3,1]$, $X_{3}$ is drawn from $U[-1,3]$ and lastly, $X_{4}$ from $U[-1,2]$.} Table \ref{simtableuurnrec} provides a detailed comparison of the mean squared error as I increase the sample size for the two models. The mean squared error for every parameter and $N$ is lower for the recursive estimator as compared to the urn model. 
\begin{table}[H]
\centering
 \fontsize{10}{10}\selectfont
\captionsetup{width=13cm}
\caption{Simulation: Comparison of urn model and recursive estimator}
\label{simtableuurnrec}
\begin{tabular}{p{1.5 cm}p{1.5 cm}p{1.5 cm}p{2.5 cm}p{2.5 cm}}\hline\hline
N & s & $\beta$ & MSE (recursive) & MSE (urn)   \\\hline
100 & 15 & 0.5 & 0.0033 & 0.0052   \\
500 & 15 & 0.5 & 0.0005 & 0.0009  \\
1000 & 15 & 0.5 & 0.0002 & 0.0005   \\
 &  &  &  &    \\
100 & 15 & 0.8 & 0.0043 & 0.0053  \\
500 & 15 & 0.8 & 0.0007 & 0.0010   \\
1000 & 15 & 0.8 & 0.0003 & 0.0006  \\
 &  &  &  &    \\
100 & 15 & -0.9 & 0.0053 & 0.0064   \\
500 & 15 & -0.9 & 0.0008 & 0.0011   \\
1000 & 15 & -0.9 & 0.0004 & 0.0006   \\
 &  &  &  &   \\
100 & 15 & -0.8 & 0.0061 & 0.0078   \\
500 & 15 & -0.8 & 0.0014 & 0.0017   \\
1000 & 15 & -0.8 & 0.0006 & 0.0009  \\\hline\hline
\multicolumn{5}{c}{ \begin{minipage}{11.5 cm}{\footnotesize{Notes: This table illustrates the error distribution comparison for the underlying agent utility parameters between the recursive and the urn estimator. The comparison has been made keeping $s$ fixed at 15, and $N$ varies from 100 to 500 to 1000.}}
\end{minipage}} \\
  \end{tabular}
\end{table}

A critical feature of the recursive estimator is that it can calculate the probability of a school getting listed. I compute these probabilities for each student school pair in the simulated data using $\hat{\beta}$. I compare these predicted probabilities with the actual probability using the population parameters for every student school pair. 
\begin{figure}[h] 
  \caption{Error distribution}
  \centering
    \label{lengthrol} 
  \begin{minipage}[b]{0.45\linewidth}
    \centering
    \includegraphics[width=\linewidth]{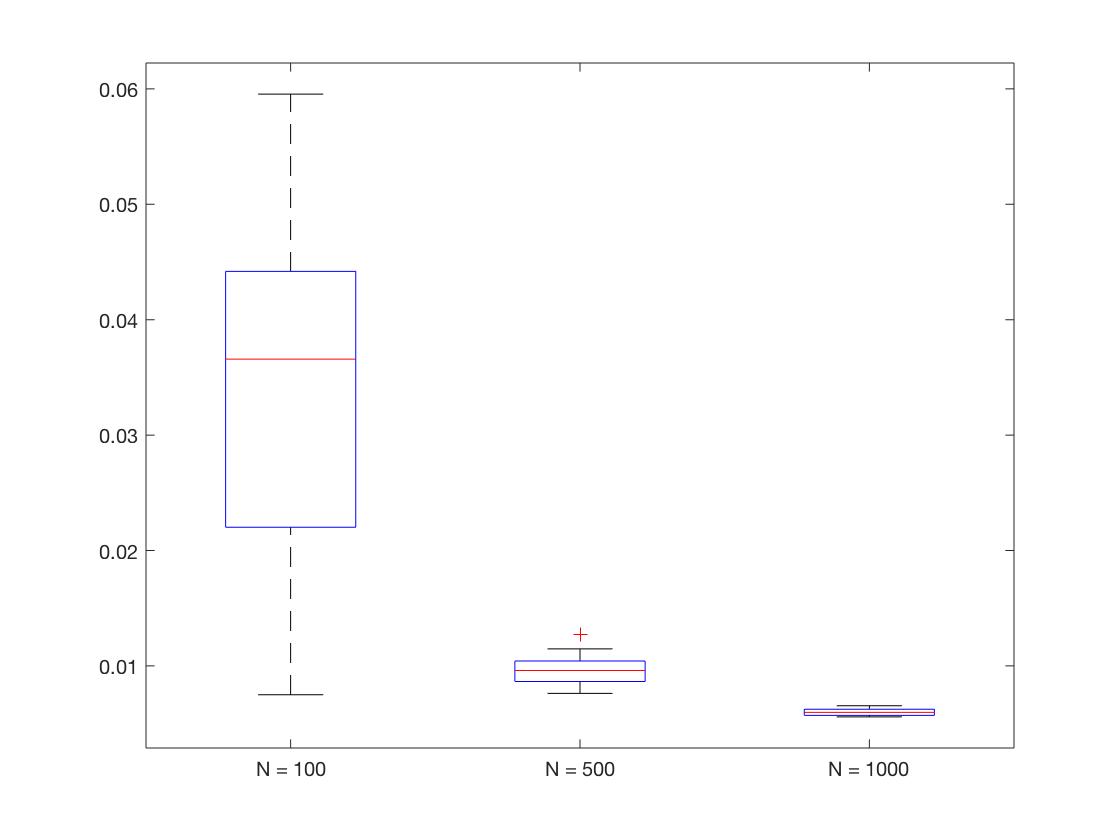} 
        \caption*{A. School popularity}
    \vspace{0ex}
  \end{minipage}%%
  \begin{minipage}[b]{0.45\linewidth}
    \centering
    \includegraphics[width=\linewidth]{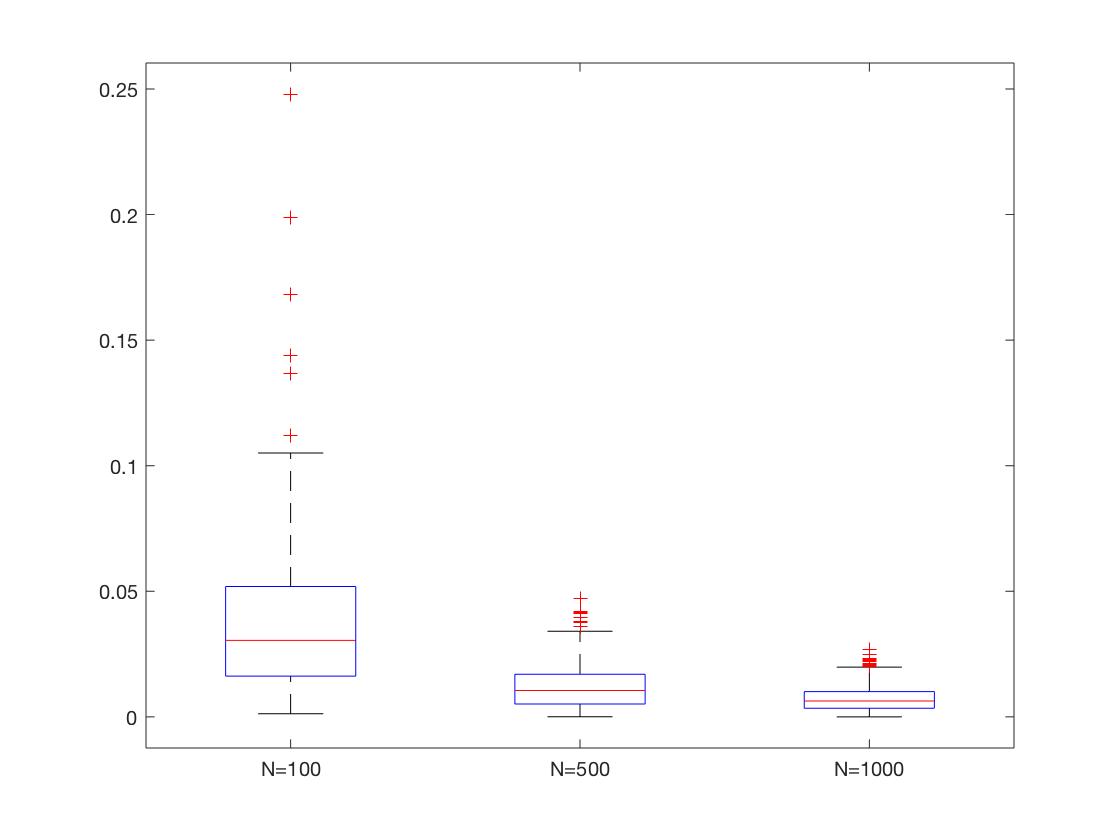} 
    \caption*{B. Expected ROL}
    \vspace{0ex}
  \end{minipage} 
  \begin{minipage}{17.5 cm}{\footnotesize{Notes: Panel (A) depicts the distribution of error in the total predicted probability of ranking schools in the simulated data by students. Panel (B) shows the error in the length of expected ROL. The error distribution diminishes as sample size increases from 100, 500 to 1000.}}
\end{minipage} 
\end{figure}
An essential limitation of the urn model gets highlighted in this comparison. The urn estimator does not differentiate between ranked and non-ranked alternatives. Every student ranks every school. Consequently, any exercise that intends to compute a school's popularity using the sum of predicted probabilities across students will be possible using the recursive estimator but will always put a probability 1 under the urn model. Moreover, using the recursive model I can predict the expected ROL for each student. Such an exercise is not possible using the urn estimator as there is no cut-off to distinguish between the ranked and non-ranked alternatives. 

I plot the error distribution between the actual and predicted total probability of ranking each school in panel (A) in figure \ref{lengthrol}. This can be interpreted as a measure of expected school popularity. The size of this error consistently shrinks as I increase the sample size. Panel (B) shows the error between the true and predicted expected ROL.

\section{Application: Chile's Schooling System}\label{sec:back}
\subsection{Background}
Chile has undertaken several significant education reforms. The first round of reforms happened in the early 1980s, followed by another reform in 1993. The last two rounds of reform happened in 2008 and 2015, respectively. Back in the 1980s, the government decided on the decentralization of primary and secondary education in Chile. Consequently, they transferred the public school system from the jurisdiction of the central government to local municipalities (school districts).\footnote{School admissions in the municipalities (school districts) in Chile works differently than the United States as students are allowed to apply to schools outside their municipality of residence.} This transfer was complemented with the introduction of a school voucher system. 

Due to this reform public, private voucher, and private non-voucher schools were created. The government financed and administered public schools. On the contrary, private voucher schools were managed privately but received government vouchers, and lastly, private non-voucher schools had no financial or administrative intervention from the government. 

The second round of reforms happened in 1993, where the private voucher schools were allowed to charge partial tuition from students in return for some reduction in government vouchers. Following this, in 2008, the government introduced additional vouchers to schools if they enrolled students from poor socio-economic backgrounds (also known as priority students).

Although the government intended to make the education system more inclusive through these reforms, education research in Chile has documented that school segregation has been on the rise in the last couple of decades \citep{valenzuela2014socioeconomic}. \cite{hsieh2006effects} has shown that the voucher system resulted in the disproportionate flight of students belonging to the middle class to the private sector from public schools (10-15 year old for years 1982-1996).  
\begin{figure}[H] 
  \caption{Time-line of Schooling Reforms in Chile}
  \centering
  \begin{minipage}[b]{0.7\linewidth}
    \centering
    \includegraphics[width=\linewidth]{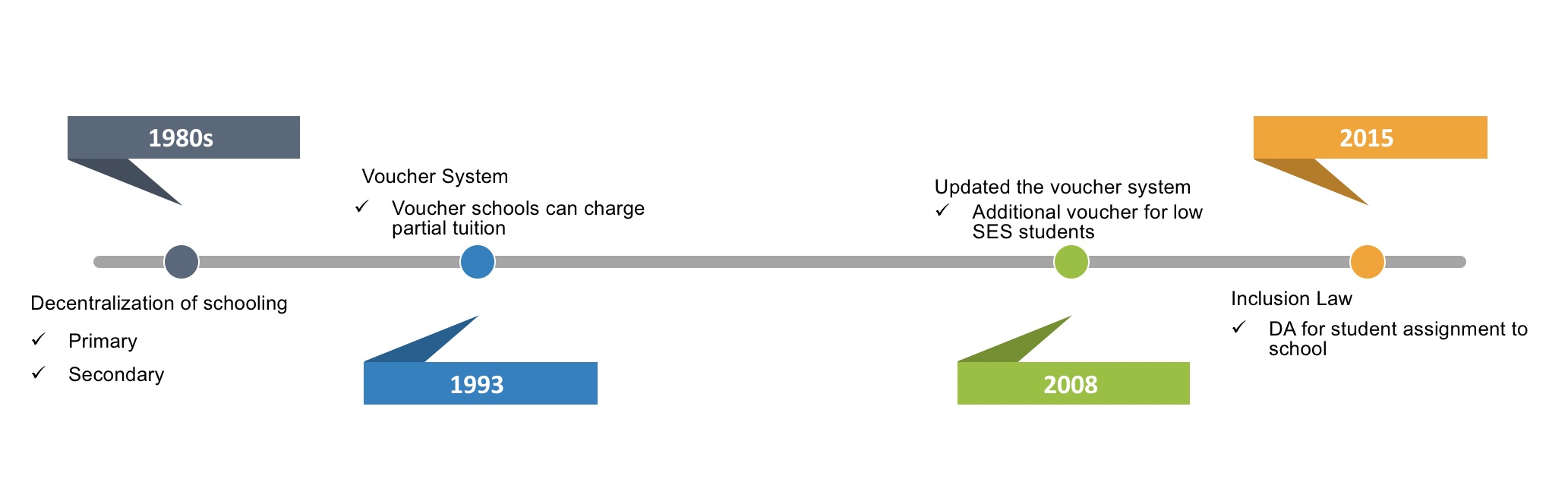} 
    \vspace{0ex}
  \end{minipage} 
  \begin{minipage}{15.5 cm}{\footnotesize{Notes: This time-line displays the key reforms that have happened in Chilean schooling sector. The latest reform studied as part of this paper is a key component of the Inclusion Law introduced in the Congress in 2015.}}
\end{minipage} 
\end{figure}
The Congress in Chile introduced the Inclusion law in 2015 to address the ongoing concerns over school segregation and to promote inclusiveness in the schooling system. Under this law, the centralized system of school assignment was launched. The government transferred all the schools from the jurisdiction of the municipalities to that of the central government. Parents were required to apply for school admission through a common web portal, and admissions were no longer decentralized. 

Chile has adopted the deferred acceptance (DA) algorithm \citep{gale1962college,abdulkadirouglu2003school} for student assignment. The first region that underwent the reallocation under DA in 2016 is Magallanes. In 2017, the new system was implemented in four other regions, namely Tarapaca, Coquimbo, O'Higgins, and Los Lagos. In 2018, the new system was extended to the rest of the country except Metropolitana. Finally, in 2019, the Ministry extended it to Metropolitana, and therefore, students all over the country could participate in the new system.

\subsection{Data}
I compile data-sets from multiple sources for the empirical application. There are three critical inputs to the algorithm-student ranks over schools, student priorities, and school vacancy. All these variables are obtained from the DA files. Students who want to change school ($>$pre-K) or enter the public schooling system (pre-K) can list all their preferred schools (ranks) in the common application. If the student decides to continue in the existing school, there is no requirement to participate in the centralized system. The vacancy at each grade for a school is the difference between the capacity for that grade and the number of students who get promoted to that grade and decide to continue in the same school.  In other words, the vacancy at ninth grade is computed as the difference between ninth grade capacity and the number of previous year's eighth-graders who get promoted and decide not to change school. Lastly, the special priorities include applicants who belong to a lower socioeconomic status based on priority index, sibling studying in the same school, school officials' children, and previous alumni of the school\footnote{The last special priority for previous alumni of the school excludes students who were expelled from the school.}.\footnote{In the existing literature on school choices in Chile such as  \citealp{gallego2008determinants,chumacero2011would}, the researcher can mostly observe only the final school choices, and the student ranks over schools are unobserved. Here, however, I observe the ranking of schools in addition to the final allocation. In this regard, my work relates to \cite{hastings2005parental,hastings2007no}, \cite{ajayi2013school} and       \cite{fack2019beyond} where the researchers observe parents' preferences. Estimating a parametric model on the preferences provides useful information on demand-side heterogeneity on school ranks, which is not possible to capture if researchers can merely observe the final allocation.}

Although I observe DA for all grades, the highest participation is for two grades-pre-K and ninth grade. The primary reason for the highest student application in these two grades is that they are the entry points for the primary and secondary school in Chile, respectively.\footnote{It is mandatory for parents who are seeking admission for their children in public/voucher schools in pre-K to participate in DA. Students who intend to switch schools between primary and secondary must participate in DA unless they are seeking admission in private non-voucher schools.} I focus on the ninth grade cohort for my analysis due to the availability of background variables. I obtain the student background characteristics by matching unique student identifiers in DA files with the standardized test scores in Chile, also known as SIMCE. Such files can be obtained for students already in the education system (ninth-grade) and not for students entering primary education through DA (pre-K). Additionally, the Chilean education system has witnessed more pronounced SES segregation in high schools than primary schools \citep{valenzuela2014socioeconomic,torche2005privatization}. This makes it compelling to study the schooling choices in the transition from middle school to high school. 

I illustrate the participation for ninth-grade admission by region in Table \ref{datatable1}. The number of regions vary by year, as DA was sequentially implemented. The first three columns summarize the number of high schools that participated in the new system. It is critical to note that only public and private voucher schools participated in DA. Private non-voucher schools did not participate in DA.
Moreover, columns 2 and 3 display substantial variation in the distribution of participating schools by type. This difference emanates from variation in local schooling structure across regions in Chile. There are regions such as Tarapac\'a and Coquimbo, which had a much higher supply of private voucher schools relative to other regions such as Los Lagos and Ays\'en, which had an almost balanced availability of both public and private voucher schools. Overall, the fraction of private voucher schools is higher among the participating schools indicating a higher presence of such schools in most of the regions. Earlier work in Chile has shown that the former government schooling reforms aimed at the school voucher led to the proliferation of private voucher schools. Such heterogeneity in school supply can have important implications for the parental decisions on school ranks. The type of school is strongly correlated with school fees. School fee is likely a critical component in the school choice decision, particularly for low income households. Table \ref{datafee} illustrates that less than 1\% of public schools were charging any fee during the DA implementation. On the contrary, around 50\% private voucher schools were charging fee. Columns 5 and 6 in Table \ref{datatable1} displays the number and the percent of total ninth graders who participated in DA by region. The participation of students vary between [32\%,67\%] with an average around 50\%. 
\footnote{All the descriptive statistics corresponds to the information in the regular allocation files.}

I distinguish between the ranked and non-ranked alternatives in my school choice framework. The Chilean data display specific characteristics that make this distinction critical. Some of those features are; i) substantial variation in the number of ranked alternatives and cut-off varies across individuals, ii) significant fraction lists three or fewer schools in ROL, iii) a sizable number of students do not end up in their top choice, iv) high disparity in the academic quality of guaranteed school and v) strong negative association between vacancies and school academic quality. 

\begin{figure}[H] 
  \caption{Distribution of ranks: outcome variable for choice model}
    \label{rankdist} 
  \begin{minipage}[b]{0.5\linewidth}
    \includegraphics[width=1.0\linewidth]{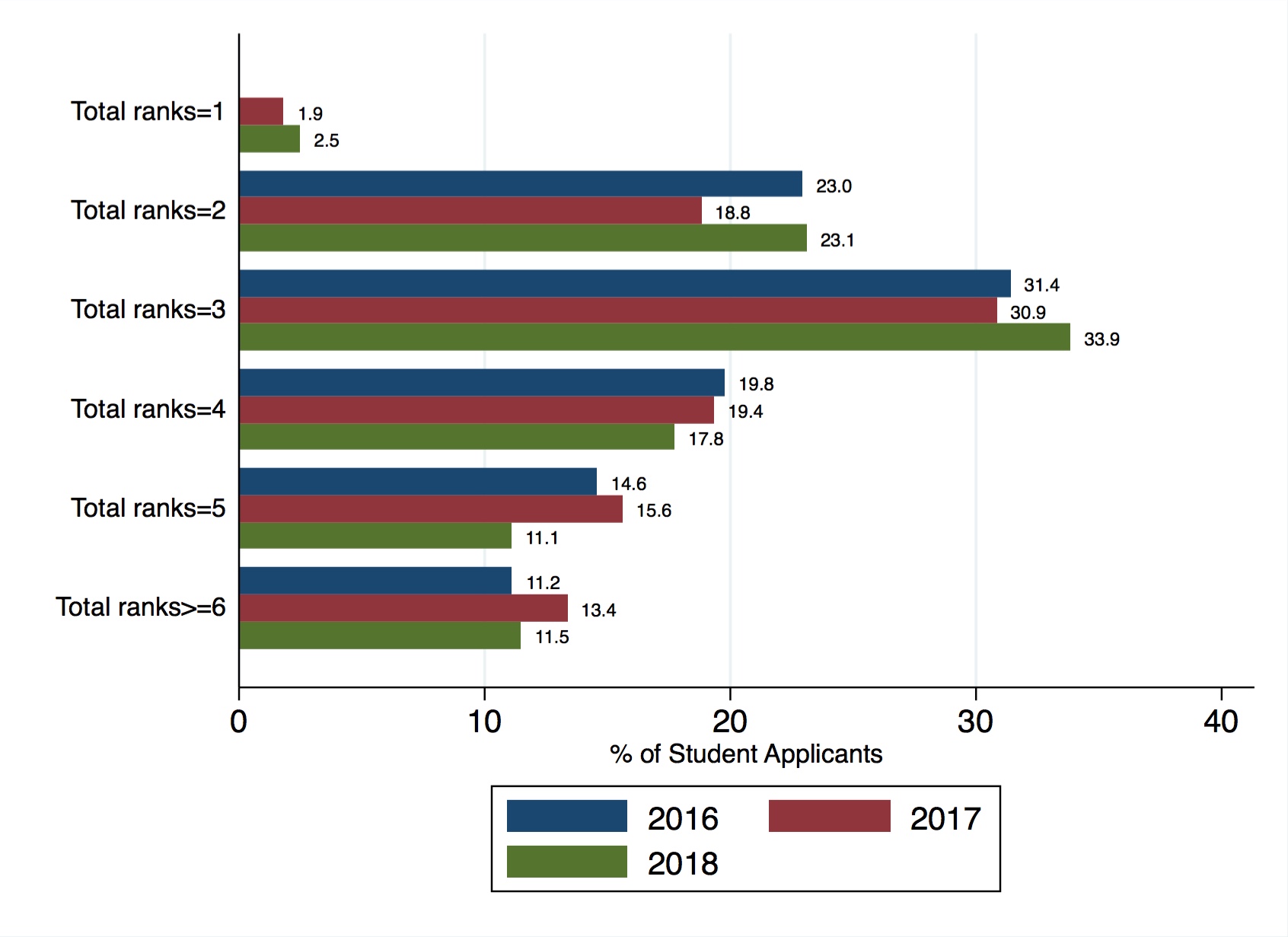}
    \caption*{A. Rank dist. for ninth-graders} 
  \end{minipage}%%
  \begin{minipage}[b]{0.5\linewidth}
    \includegraphics[width=1.0\linewidth]{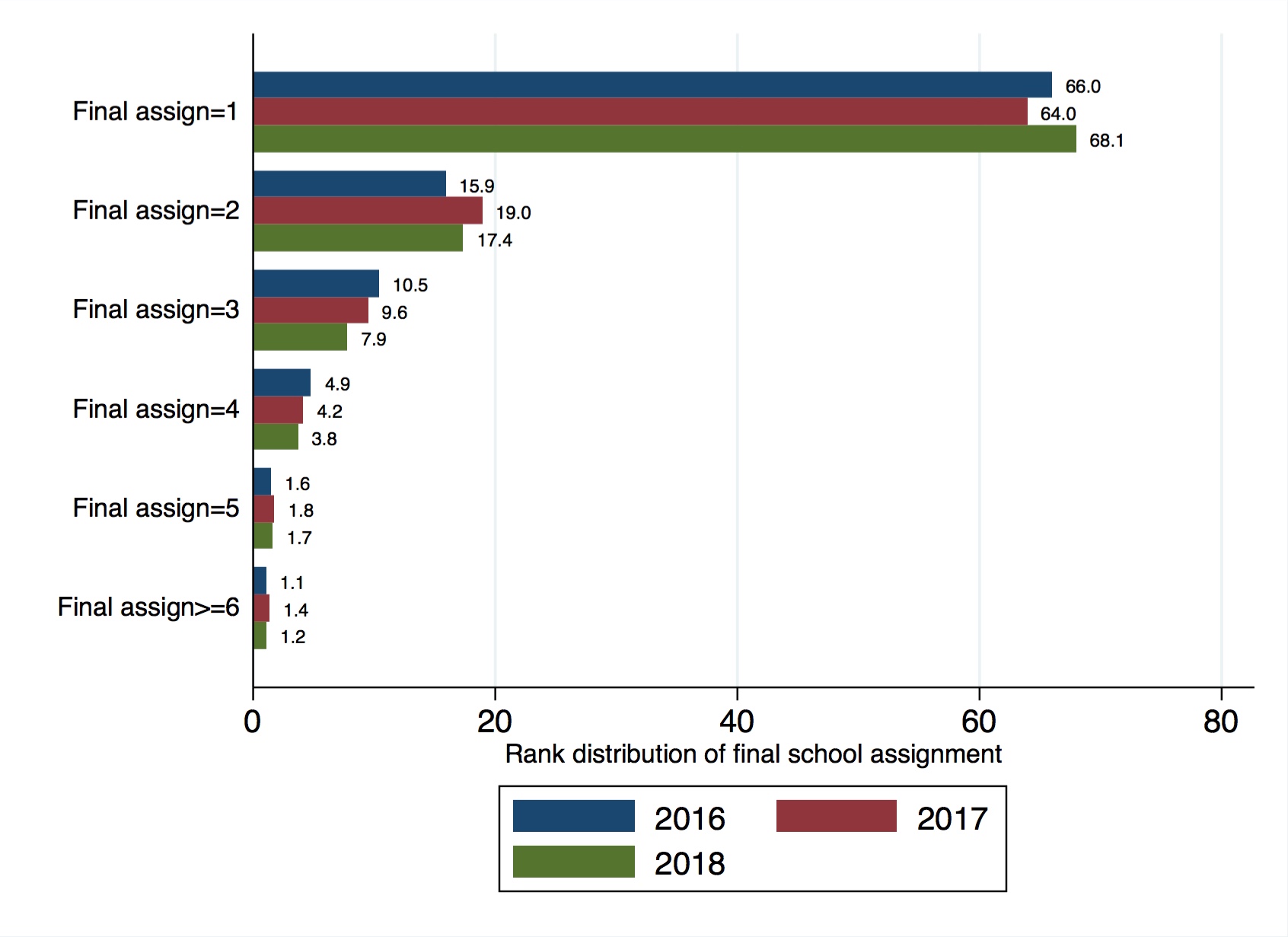} 
        \caption*{B. Rank of final assignment} 
  \end{minipage} 
    \begin{minipage}{17.5 cm}{\footnotesize{Notes: The figure in panel (A) illustrates that there is significant variation in reported number of ranked schools in student applications. This figure uses the data for ninth-graders for 2016, 2017 and 2018. In panel (B), I illustrate the association between final assignment and the rank for the assigned school in student application. The sample for this analysis consist of participants who were allocated to one of the ranked schools through the algorithm. For students for whom the algorithm failed to assign a school in ROL, they were assigned the closest school based on their place of residence. Consequently, the sample for this analysis is slightly lower than the actual number of participants.}}
\end{minipage} 
\end{figure}

I display the distribution of total ranks for ninth-graders in panel (A) in figure \ref{rankdist}. At least 54\% of families listed three or fewer ranks in their application in 2016. The corresponding figures for 2017 and 2018 are 52\% and 60\%, respectively. This suggests that the cut-off where families stop listing schools can be extremely critical in determining their final school assignment through the centralized algorithm. Moreover, I illustrate in panel (B) in figure \ref{rankdist}, the fraction of students who got allocated to their top choice. About two-thirds of students get allocated to their top choice. Nonetheless, one-third get allocated to their second, third, or latter choices. The key takeaway is that there is a possibility that students end up in lower-ranked schools on their list. This compels the need to study factors that determine both the ranking order as well as the cut-off of ranks.

\begin{figure}[H] 
  \caption{Variation in outside value/guaranteed school quality by student income}
    \label{OV_income} 
  \begin{minipage}[b]{0.5\linewidth}
    \includegraphics[width=1.0\linewidth]{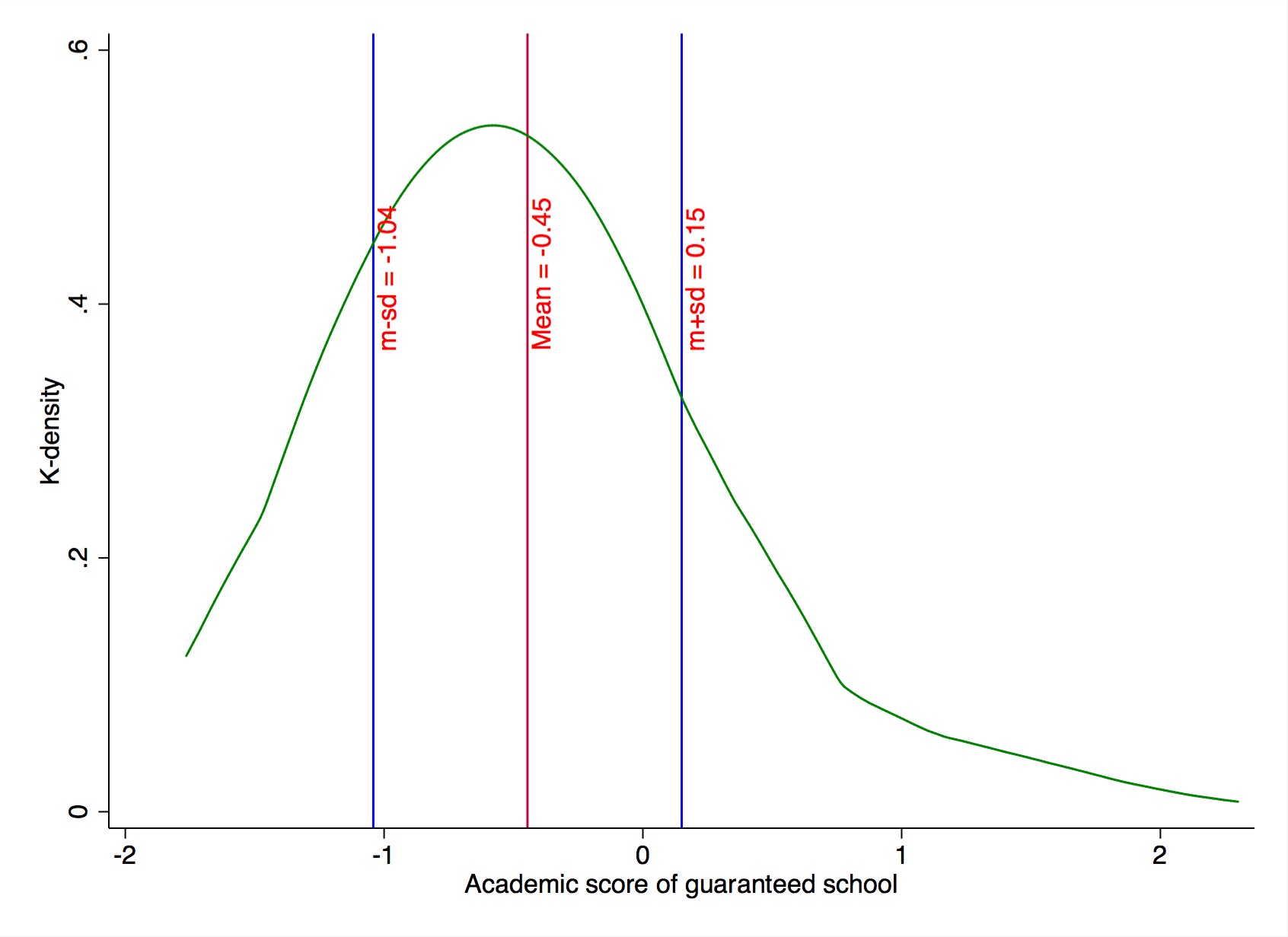}
    \caption*{A. High income} 
  \end{minipage}%%
  \begin{minipage}[b]{0.5\linewidth}
    %\centering
    \includegraphics[width=1.0\linewidth]{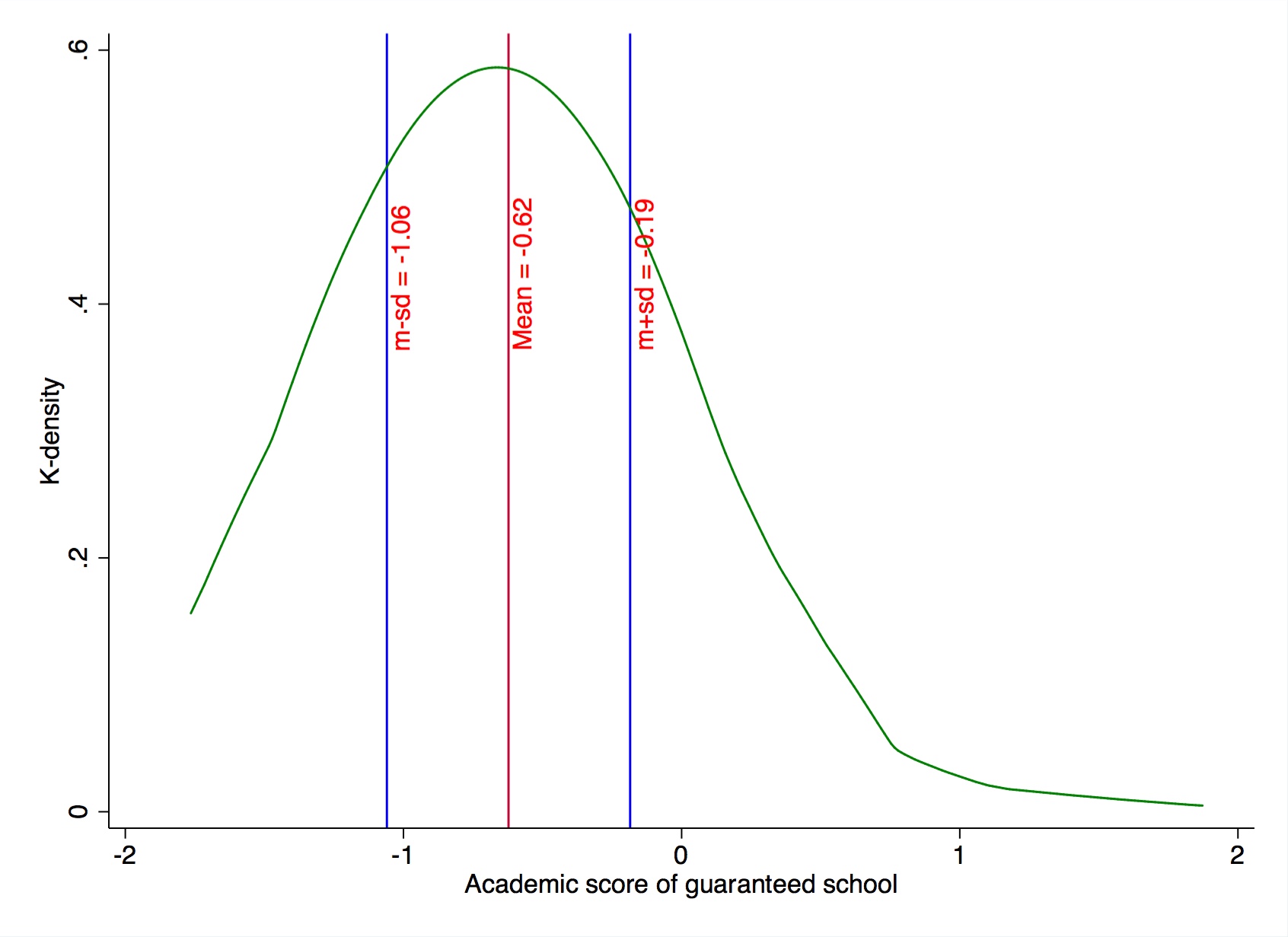} 
        \caption*{B. Low income} 
  \end{minipage} 

   \begin{minipage}{17.5 cm}{\footnotesize{Notes: These graphs display the kernel density plots for academic quality of the guaranteed school. The academic quality has been obtained as an average of math and language SIMCE scores measured before DA. These test scores have been adjusted by the mean and standard deviation. The resulting test score distribution has $\mu=0$ and $\sigma=1$.}}
\end{minipage} 
\end{figure}

Student cut-offs in ROL are likely to vary by the quality of the guaranteed school. There is a positive probability associated with the DA algorithm in Chile of not allocating a student to any of the schools on ROL. The Ministry of Education in Chile provides detailed guidance on allocation in this scenario. There are two possibilities: first, if the student's old school, the school in which the student is enrolled before DA reallocation, offers the grade to which the student seeks admission, then the student is guaranteed a seat there DA fails to allocate. Second, if the prior school does not offer the grade, the student is guaranteed admission to the nearest public school with vacancies. This rule creates a significant variation in the value of the outside option for the participating student. I plot the distribution of pre-DA test scores for the guaranteed school in figure \ref{OV_income}. I observe substantial differences in outside value across students. Moreover, on an average high income students (Panel (A)) have a higher outside value than the low income students (Panel (B)). 

The second important component that can result in parents revealing a partial set of rankings is their expectation on the probability of acceptance at different schools. It is often seen that high academic quality schools are oversubscribed, and this behavior impacts the likelihood of admission. In DA, the likelihood of admission is never zero as admission at every iteration of the algorithm is tentative, and ties are resolved by lotteries. However, parents, even under DA, can modify their behavior if they expect the chances of admission to a high-quality school are low due to fewer vacancies. 

\begin{figure}[H] 
    \centering
  \caption{Vacancies and pre DA test score}
    \label{vacancyscores} 
      \begin{minipage}[b]{0.45\linewidth}
    \includegraphics[width=\linewidth]{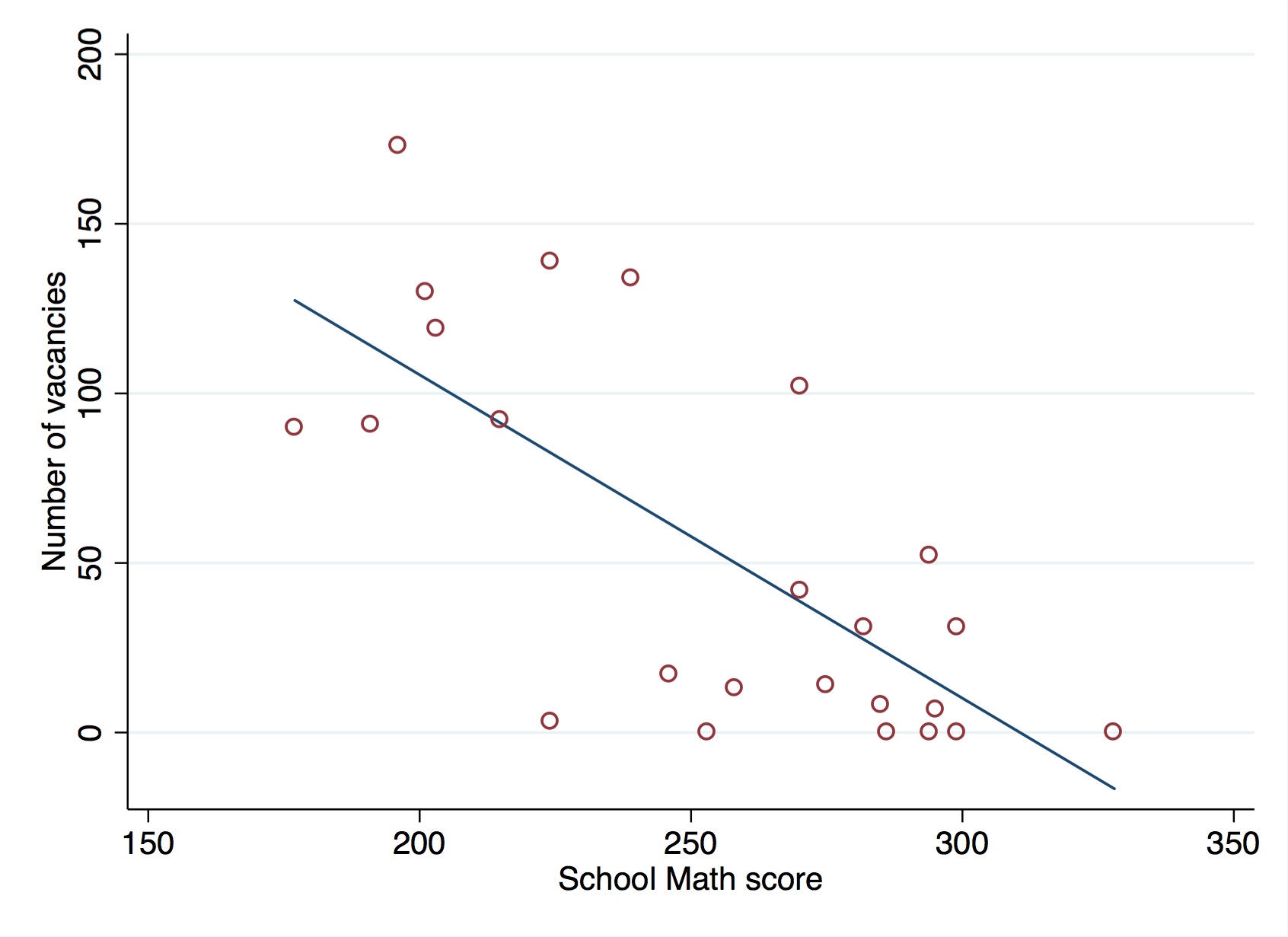} 
       \caption*{A. 2016} 
              \end{minipage} 
  \begin{minipage}[b]{0.45\linewidth}
    \includegraphics[width=\linewidth]{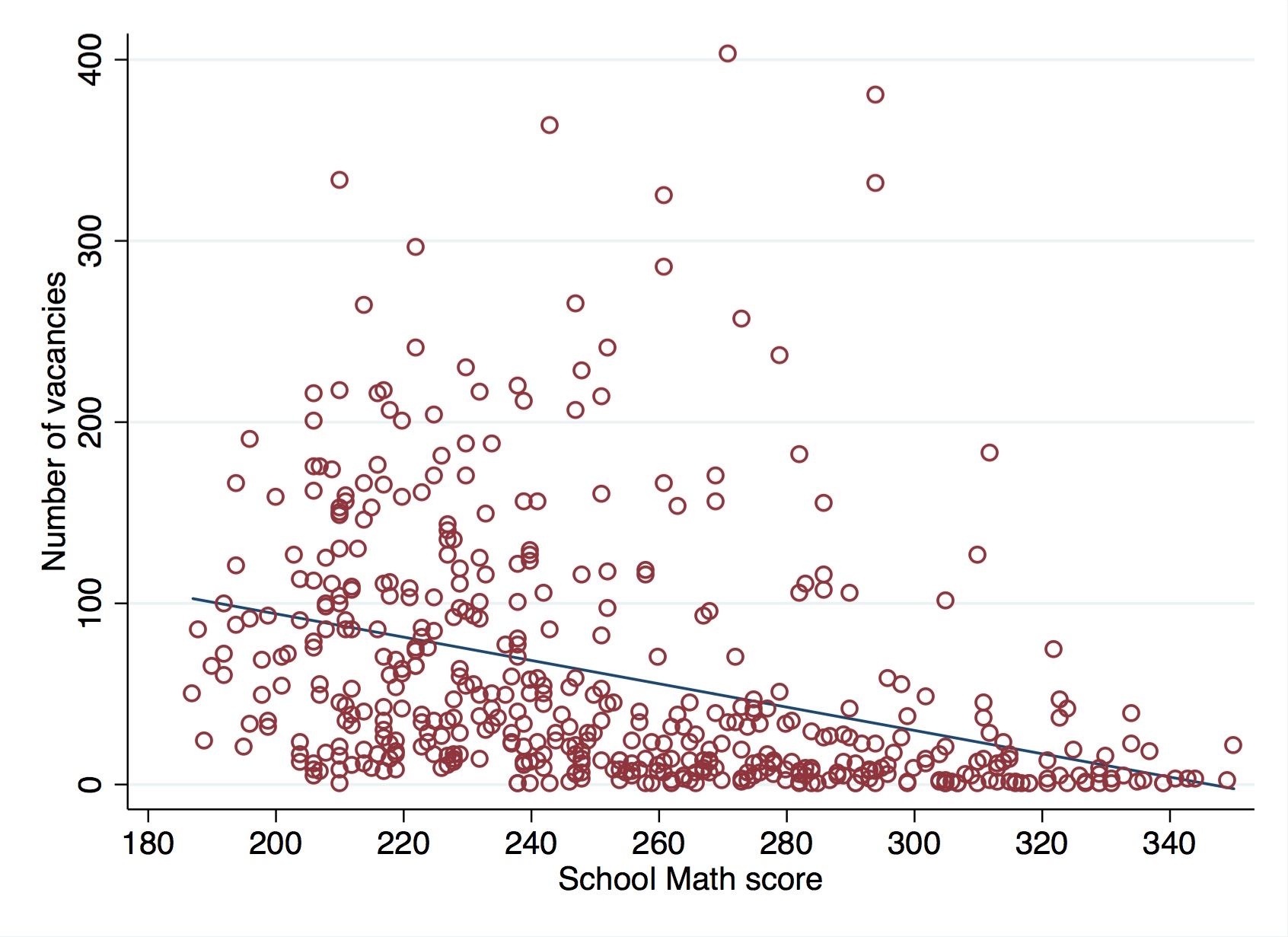} 
       \caption*{B. 2017} 
              \end{minipage} 
               \begin{minipage}{17.5 cm}{\footnotesize{Notes: These graphs show the relationship between school quality and vacancies in DA. I study the transition from middle to high schools and some students can choose to continue in their pre-DA school if it offers high school grades. The expected probability of admission takes advantage of this relationship in the empirical specification.}}
\end{minipage} 
  \end{figure}

Since I focus on the transition between the middle and high school, the vacancies are directly a function of the fraction of eighth-graders participating in DA to switch schools. Figure \ref{vacancyscores} illustrates that there is a strong negative correlation between vacancies and school academic quality. This is likely as students enrolled in good academic quality schools are less likely to switch schools in ninth-grade. Consequently, high academic quality schools post fewer openings for ninth-grade admissions. 
  
  I provide suggestive reduced-form evidence on the relationship between the length of ROL reported in DA and the value of the outside option and school vacancies. For this exercise, I use the eighth-grade cohort in 2017 who participated in DA for ninth grade admissions. I construct the measure of outside options using the pre-DA school test scores. Next, I account whether parents consider the expected probability of admission in their ranking process by computing the average vacancies in schools to which the student did not apply but was part of the student's choice set. Additionally, I account for various background characteristics of the student. Table \ref{ROLdet} illustrates the results of this association. Once I condition on student characteristics, the value of the outside option is negatively associated with the length of ROL (Model (2)), which is in line with expectation. But I do not observe a significant association between length of ROL and vacancies at non-listed schools. This might be because the expected likelihood of admission might be less relevant for the Chilean parents on average. Nevertheless, I do account for the likelihood of admission in my school choice model as there might be parents at the margin accounting for such probabilities. 

  \begin{table}[h]
  \centering
   \fontsize{10}{10}\selectfont
  \caption{Length of ROL, outside value and vacancies}
  \label{ROLdet}
  \begin{tabular}{p{8 cm}cc} \hline\hline
 & ROL length & ROL length \\
VARIABLES & (1) & (2) \\ \hline
 &  &  \\
Vacancy to capacity in non-applied schools & 0.416 & 0.612 \\
 & [1.014] & [1.015] \\
Outside option value & 0.001 & -0.046* \\
 & [0.021] & [0.023] \\
Student pre-DA score &  & 0.055** \\
 &  & [0.022] \\
Mother's education &  & 0.034*** \\
 &  & [0.008] \\
Income &  & 0.070*** \\
 &  & [0.007] \\
Total school availability & 0.011*** & 0.011*** \\
 & [0.003] & [0.003] \\
Constant & 3.059*** & 2.352*** \\
 & [0.578] & [0.593] \\
 &  &  \\
Observations & 15,125 & 10,558 \\
 R-squared & 0.030 & 0.050 \\ \hline\hline
 \multicolumn{3}{c}{ \begin{minipage}{15.5 cm}{\footnotesize{Notes: Robust standard errors in brackets. *** p$<$0.01, ** p$<$0.05, * p$<$0.1. This analysis uses the data on eighth grade cohort that participated in DA in 2017 in five regions-Tarapaca, Coquimbo, O'Higgins, Los Lagos and Magallanes.}}
\end{minipage}} \\
\end{tabular} 
  \end{table}
 
To determine the factors explaining the student rank order list (ROL), I closely follow the literature on school choice. I incorporate determinants for both benefits and costs associated with an application to a school. On the benefits side, parents care about the academic quality of the school. \cite{black1999better,hastings2008information,reback2008demand,hanushek2007charter,hastings2009heterogeneous} show that school average test scores are important determinants of school choice. Moreover, it is often the case that value attached to academic quality varies by parental income. Parents with higher incomes tend to put a higher value on school test scores than parents with lower levels of income \citep{burgess2015parents}. Since I focus on ninth-graders in this analysis, I use the tenth-grade average (school) math and language SIMCE test scores as primary measures of school quality (see Table \ref{dataschooltenth} for details).\footnote{Unique school identifier in the DA files can be matched with SIMCE files to obtain the school academic quality variable.} I observe significant variation in school test score distribution across all three years displayed in Table \ref{dataschooltenth}. I use the SIMCE data from 2015, 2016, and 2017 as the parents need to observe the tenth-grade test scores when making school choice decisions in 2016, 2017, and 2018 respectively, and the SIMCE tenth grade scores for the same year will not be reported at the time of application. 

School fees can be a major barrier to private school enrollment \citep{alderman2001school,glick2006demand}. As discussed in section \ref{sec:back} school fee structure is closely associated with the school type in Chile. Table \ref{datafee} shows the fee structure for high schools in Chile. As of 2018, secondary education in most of the public schools was free. On the contrary, 47\% private voucher schools charge an add on fee to parents. This is a critical difference between the two types of schools that participated in DA (public and private voucher). 

School choice literature such as \cite{hastings2008information}, \cite{gallego2010school} illustrate school proximity is a key determinant of parental preferences. For computing the travel time to school, I use precise student residential and school addresses, provided by the Ministry of Education. I use open street maps API to calculate commuting time to schools. Travel time or distance using actual road network provides a much more accurate measure than any geodetic or straight-line measures used in related literature for similar analysis \citep{frenette2006too,chumacero2011would, laverde2020unequal}. 

\begin{figure}[H] 
  \caption{Travel time and route to ranked school}
    \label{datafig2} 
  \begin{minipage}[b]{0.3\linewidth}
    %\centering
    \includegraphics[width=0.95\linewidth]{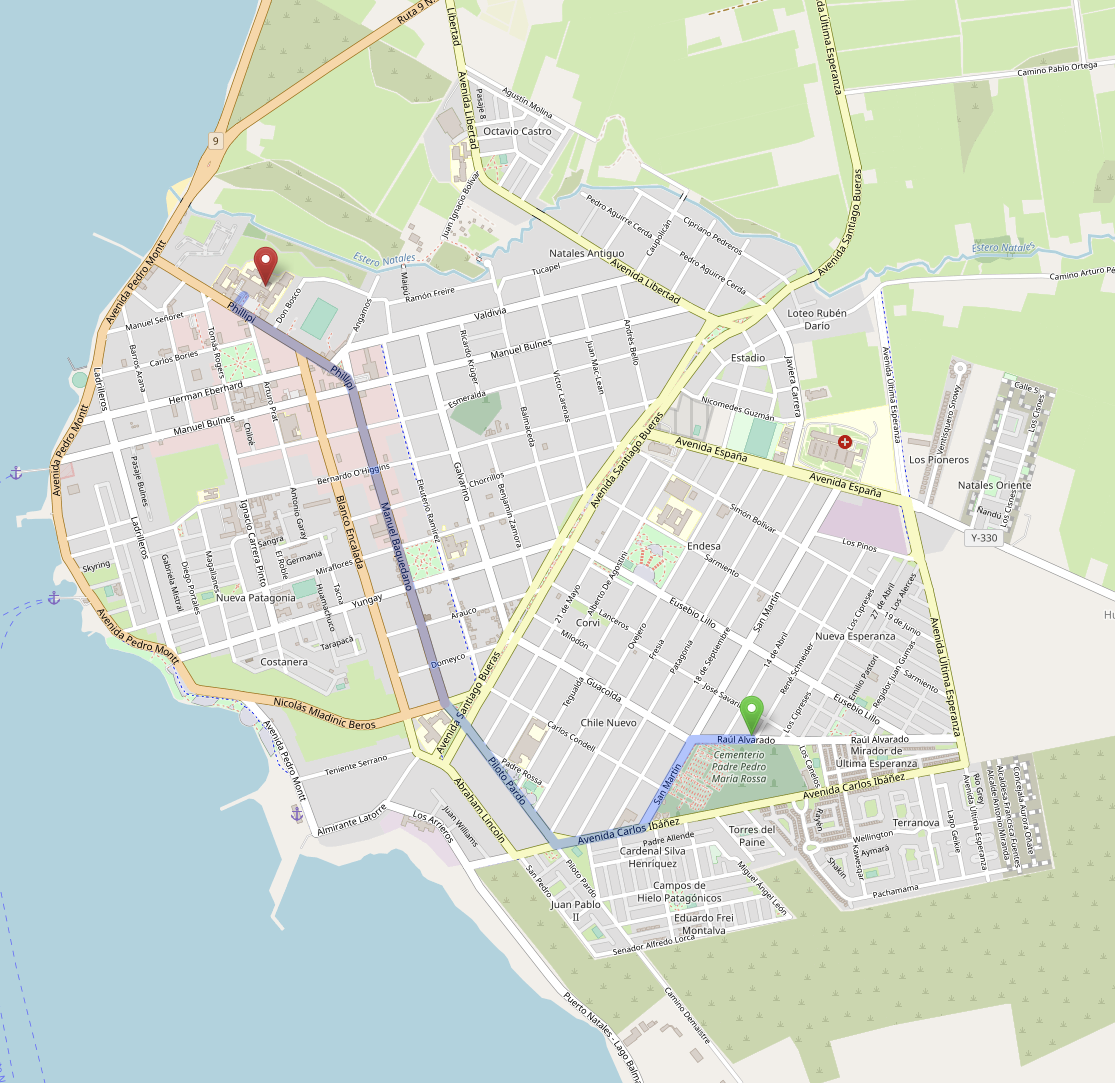}
    \caption*{A. Student A} 
  \end{minipage}%%
  \begin{minipage}[b]{0.3\linewidth}
    %\centering
    \includegraphics[width=0.89\linewidth]{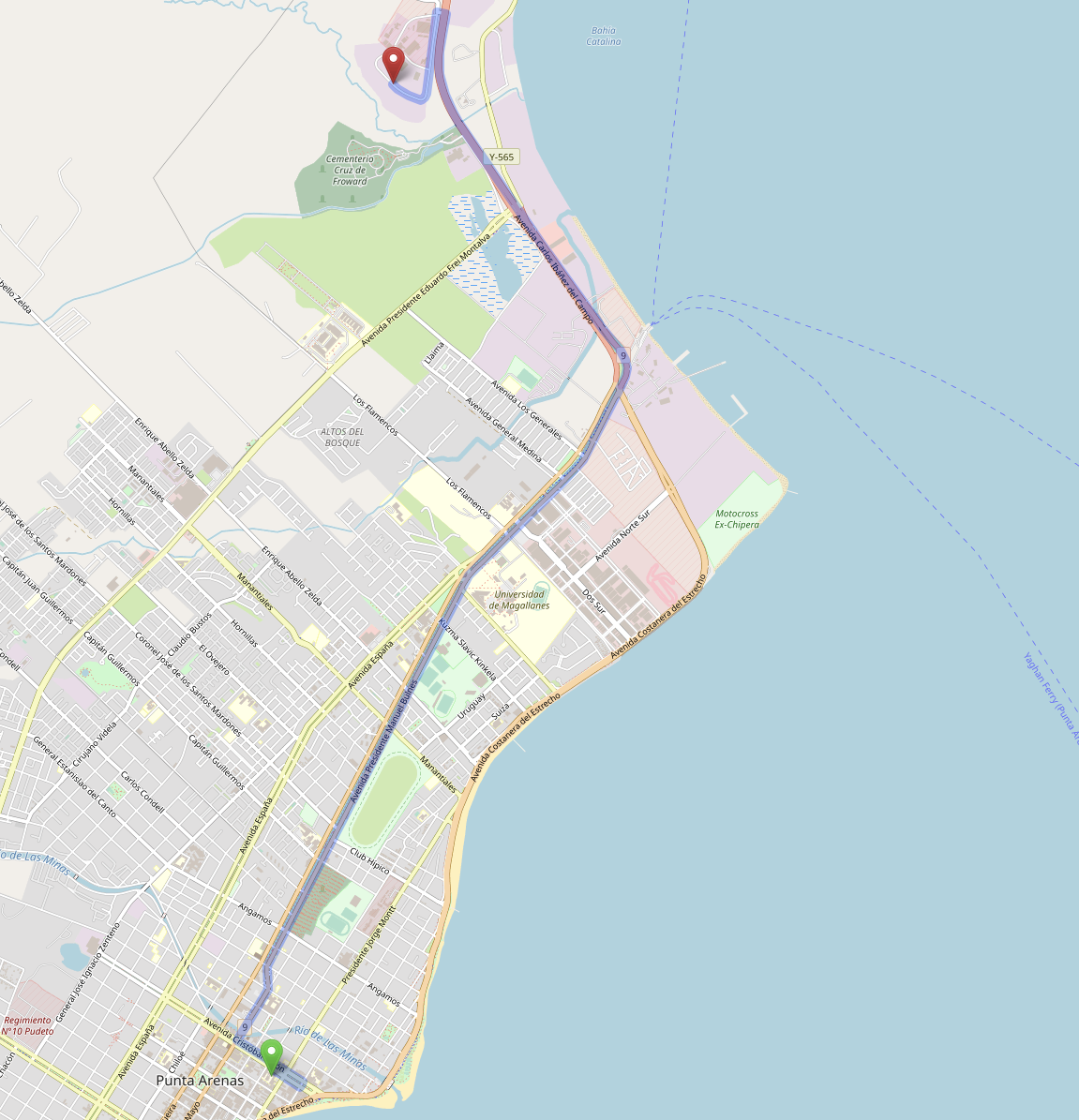} 
       \caption*{B. Student B} 
  \end{minipage} 
  \begin{minipage}[b]{0.3\linewidth}
    %\centering
      \includegraphics[width=1.25\linewidth]{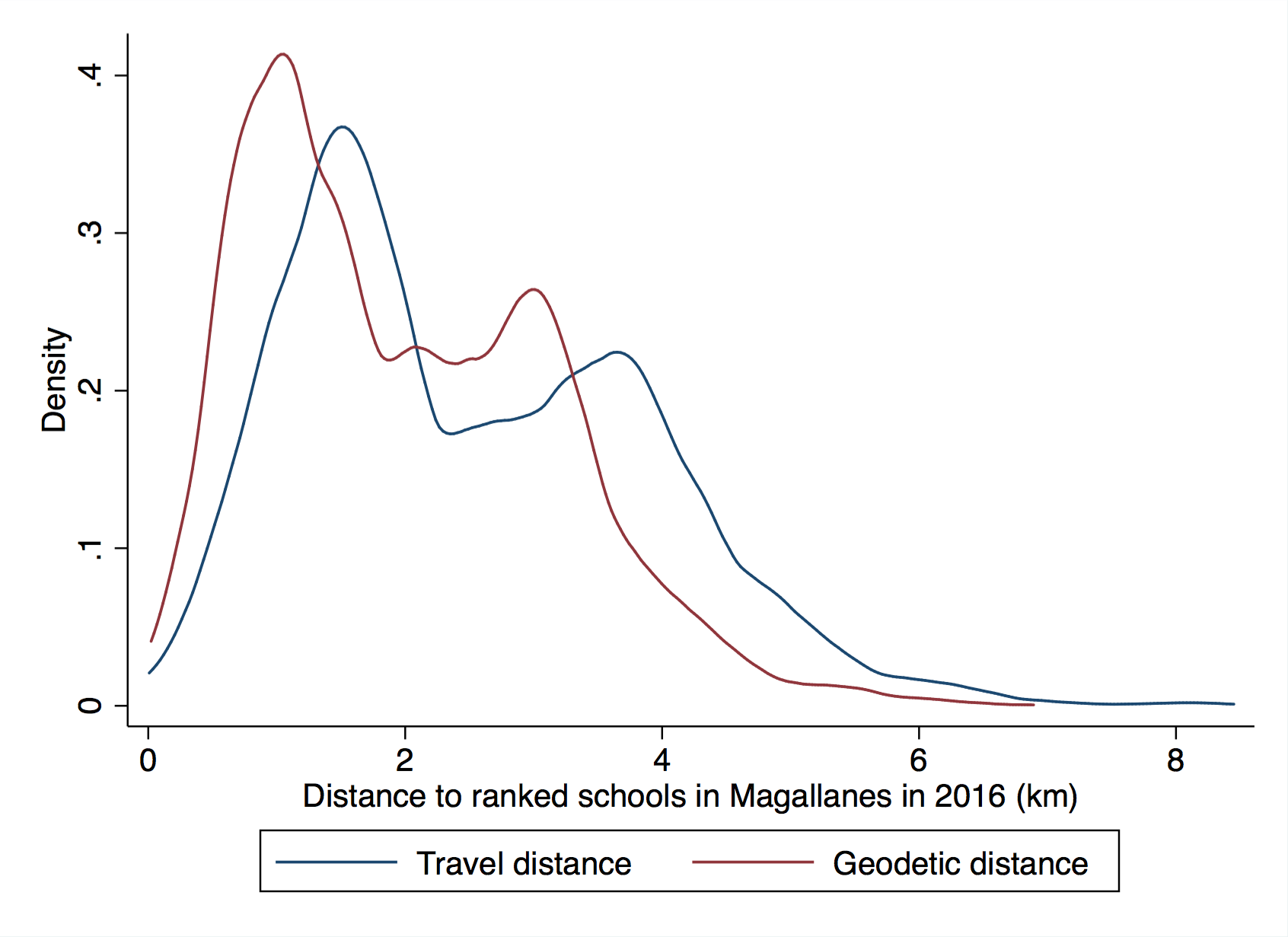}
       \caption*{C. Travel vs. geodetic} 
  \end{minipage} 
  \begin{minipage}{17.5 cm}{\footnotesize{Notes: The figure in panel (A) and (B) illustrate the travel distance by car computed using the actual road network between the student's residence and school. Panel (C) depicts the kernel density plots of travel and geodetic distance to the schools listed in ROL. }}
\end{minipage} 
  \end{figure}

Figure \ref{datafig2} displays the commuting route and travel distance using a car for two ninth graders who applied in DA in Magallanes in 2016. The travel distance for Student A shown in panel (A) is 2.4 km to the preferred school. The geodetic distance computes this as 1.95 km. Similarly, travel distance for student 2 is 8.3 km (Panel (B)). However, the corresponding geodetic measure is 6.3 km. Next, I also illustrated the kernel density plots for all the student school rankings of ninth graders in Magallanes in 2016 (Panel (C)). I observe significant variation in the two densities. Moreover, geodetic distances consistently underestimate the distance to school. 
Beyond the above variables, there are several other determinants of applying to a school. \cite{fack2019beyond} and \cite{hastings2005parental} suggest that parents care about the socio-economic make up of a school. Particularly, parents seek schools that have students coming from a similar socio-economic background. 
Besides, parents might also favor schools where the academic standards match the student's academic ability \citep{light2000determinants,fuller1982new}. I do control for ability and SES match in my school choice model.

Lastly, parents' decision to stop ranking schools critically hinges on three critical variables in the Chilean context. First, if students do not get allocated to any of the schools listed in their ROL, they are guaranteed a seat in their old school, conditional it offers ninth-grade. Else, they are allocated to the nearest public school with a vacancy. Second, there could be heterogeneity in the psychological cost of listing additional schools, which can be strongly correlated with the extent of parent sophistication. Lastly, parents might modify their behavior based on the expected likelihood of admission, and school academic quality variables can account for these differences. I account for such components in my school choice model. 

\begin{figure}[H] 
  \caption{Geographic variation in placement of school types and student types}
    \label{exampleidentification} 
  \begin{minipage}[b]{0.5\linewidth}
    %\centering
    \includegraphics[width=1.0\linewidth]{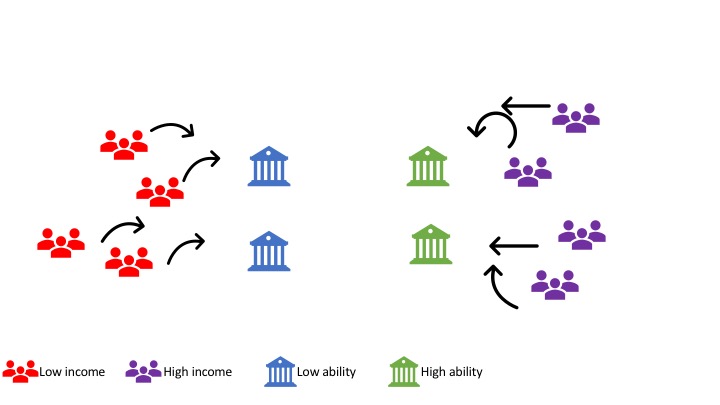}
    \caption*{A. Identification fails} 
  \end{minipage}%%
  \begin{minipage}[b]{0.5\linewidth}
    %\centering
    \includegraphics[width=1.0\linewidth]{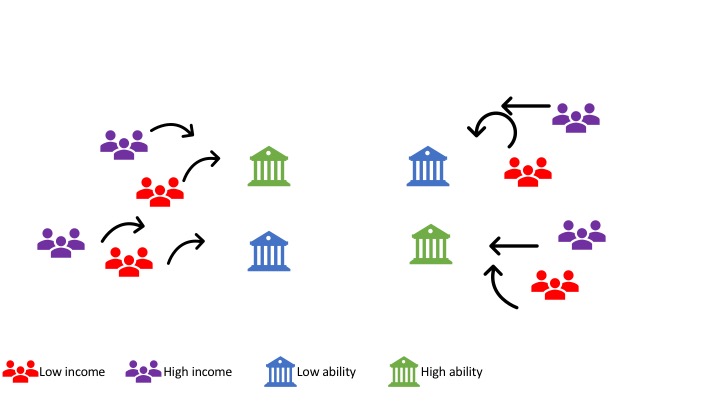} 
        \caption*{B. Identification possible} 
  \end{minipage} 
   \begin{minipage}{17.5 cm}{\footnotesize{Notes: Panel (A) and (B) illustrate the variation required in the data to satisfy the assumption of full support. Students of the same type should be placed geographically around every school type for the required variation in the value of guaranteed school (outside value).}}
\end{minipage} 
\end{figure}

The identification argument for partial lists illustrated in \cite{causalinference} requires two critical components. First, empirically one needs to illustrate the existence of full support. In other words, there should exist variations in the placement of all types of students across different school types. This geographic variation will generate variation in the outside value or the quality of the guaranteed school, which will help to pin down the parameters of parental preferences in the presence of partial lists. I explain the argument using a simple example. To keep the illustration simple, I assume two students types low income and high income. I also assume two school types high and low ability. Figure \ref{exampleidentification} shows the type of variation necessary for identification in case of partial lists. 
Assume that parents care about school quality and distance while listing schools. In panel A, I observe that low-income students are all geographically clustered around low ability schools, and consequently, these parents list only the low-income schools. On the contrary, since the high-income students are clustered around the high ability school, only the high ability school features on their ROL. One of the primary reasons driving this partial listing for the low-income group could be that the outside option (closest school) is higher than the utility provided by listing a faraway high ability school. For the high-income students, there is less reason to list a low ability school further away from residence. In this context, it is hard to pin down determinants of parental ranks for the complete set of schools, so identification fails. 

In panel B, I reshuffle this set-up, and now there is variation in the placement of different student types around high and low ability schools. Under this set-up, even with partial ranking, I observe the lists of low-income parents for both high and low-income schools due to differences in placement and variation in the quality of the guaranteed school. Using this variation and the partial ROL, I can back out the complete ordering of parental preference parameters in a school choice set-up.

\begin{figure}[H]
\centering{}
\caption{Variation in spatial location of school type and student type}
\label{region5identify}
\begin{subfigure}{0.25 \textwidth}
\includegraphics[width=\linewidth]{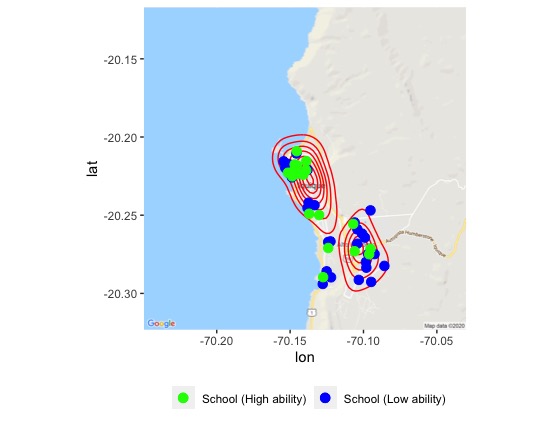}
\end{subfigure}%
\begin{subfigure}{0.25 \textwidth}
\includegraphics[width=\linewidth]{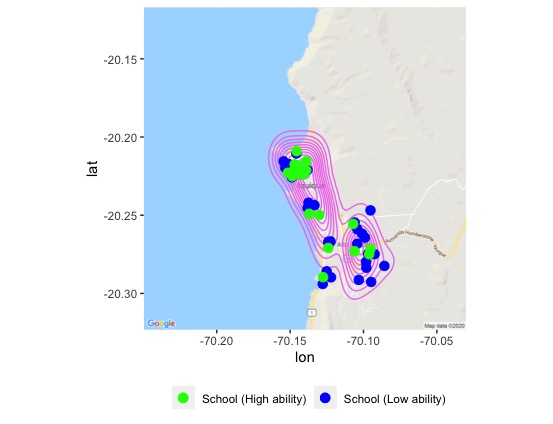}
\end{subfigure}%
\begin{subfigure}{0.25 \textwidth}
\includegraphics[width=\linewidth]{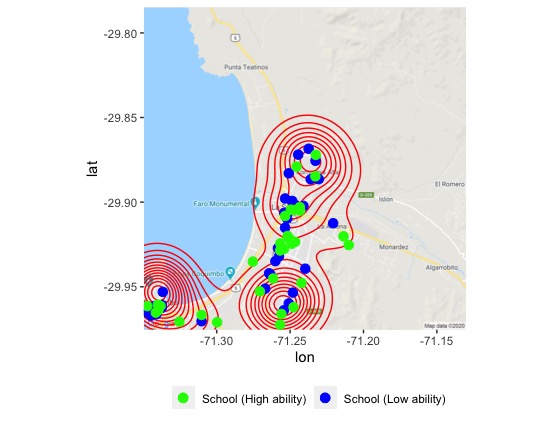}
\end{subfigure}%
\begin{subfigure}{0.25 \textwidth}
\includegraphics[width=\linewidth]{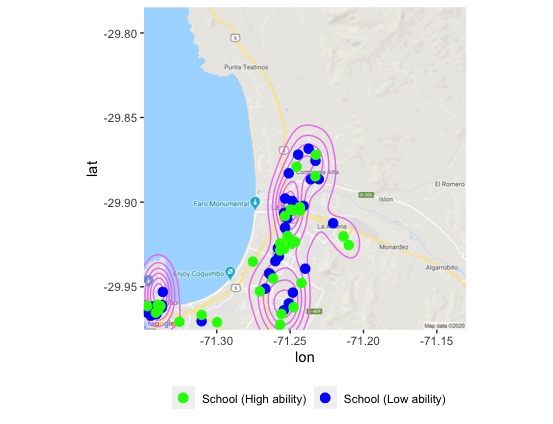}
\end{subfigure}
\begin{center}
 A. Tarapaca \hspace{150pt} B. Coquimbo 
\end{center}
\begin{subfigure}{0.25 \textwidth}
\includegraphics[width=\linewidth]{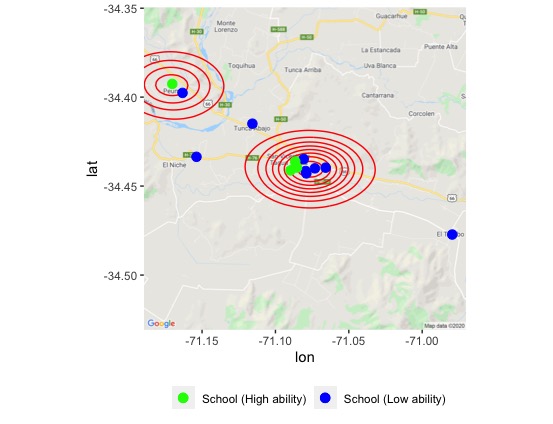}
\end{subfigure}%
\begin{subfigure}{0.25 \textwidth}
\includegraphics[width=\linewidth]{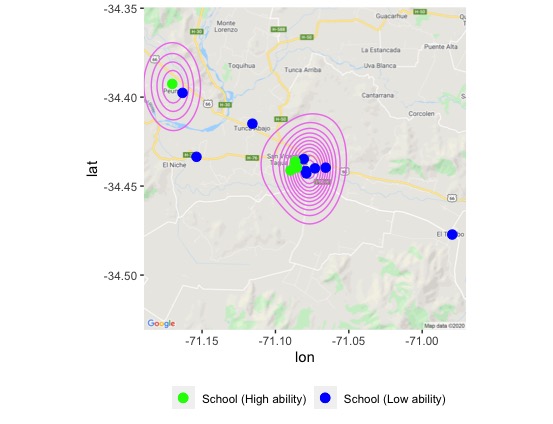}
\end{subfigure}%
\begin{subfigure}{0.25\textwidth}
\includegraphics[width=\linewidth]{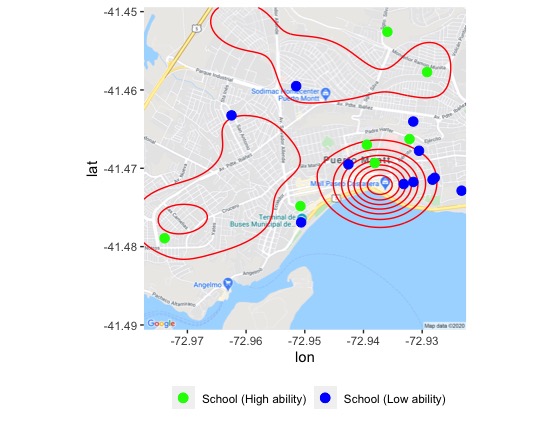}
\end{subfigure}%
\begin{subfigure}{0.25 \textwidth}
\includegraphics[width=\linewidth]{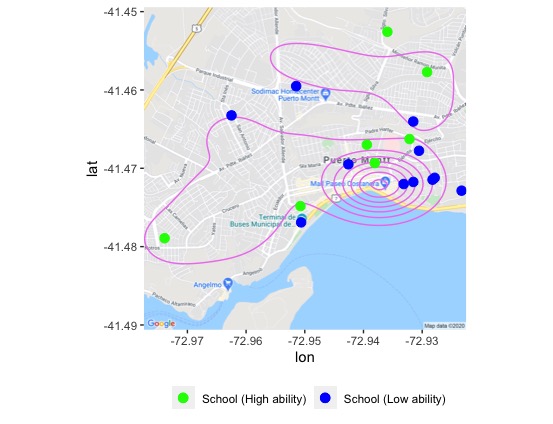}
\end{subfigure}
\begin{center}
 C. O'Higgins \hspace{150pt} D. Los Lagos
\end{center}
\begin{subfigure}{0.25 \textwidth}
\includegraphics[width=\linewidth]{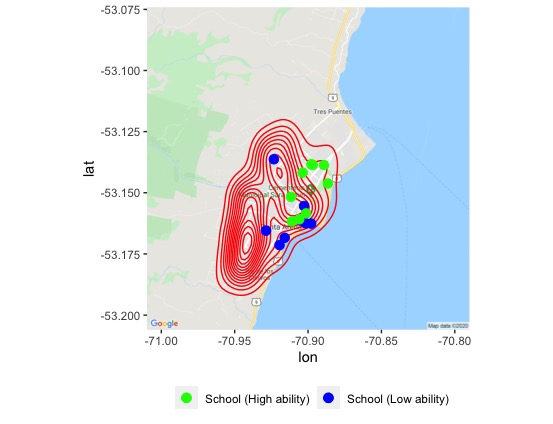}
\end{subfigure}%
\begin{subfigure}{0.25 \textwidth}
\includegraphics[width=\linewidth]{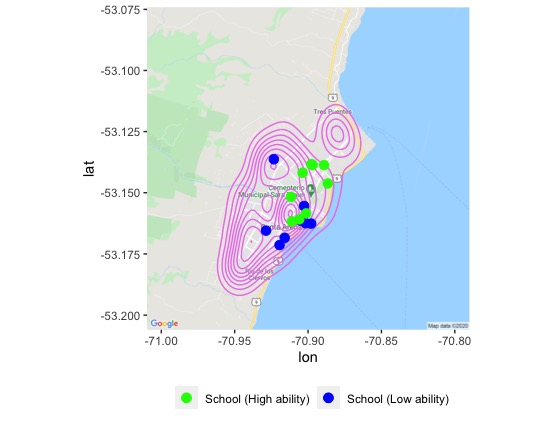}
\end{subfigure}
\begin{center}
\hspace{50pt} E. Magallanes
\end{center}
  \begin{minipage}{17.5 cm}{\footnotesize{Notes: These graphs display the spatial density plots for high and low income students around high ability and low ability schools respectively. The plots with red density contours correspond to the low income students and the plots with violet density contours correspond to high income students.}}
\end{minipage} 
\end{figure}

It is important to identify the source of the above variation in data. For this, I show the geospatial makeup of student and school types in the regions that participated in DA in 2017. Figure \ref{region5identify} provides suggestive evidence on full support. I observe a significant mix of low and high-income students around every school type in each of the regions. This variation is critical for identification in DA with partial lists. The dimension of student and school types are kept at two for the simplicity of illustration. However,  I expand the dimension of student types and condition on both income and ability later in the paper. 

\subsection{Reduced form evidence}
The government's key motivation behind the introduction of centralized assignment in Chile was to reduce the existing segregation levels based on socio-economic status (SES). The voucher system introduced in the 1990s and later modified in 2008 had led to an out-migration of middle-income category students to voucher schools. This resulted in overcrowding of low-income students in the free public schools that raised segregation in schools. 

School segregation poses additional constraints if low SES students are under-represented in high quality schools. In figure \ref{seg20152016}, I divide the schools into two types, high and low academic quality and display the composition of student type across these schools. I do this analysis for the ninth grade cohort just before and after the introduction of DA for student assignment. 

I divide the total enrolment in high and low quality schools by student income.\footnote{The percentage of low and high income students should add to 100 for each school type in figure \ref{seg20152016}.} Figure \ref{seg20152016} displays that low income students constitute a much lower fraction of total enrollment in high-quality schools. Such under-representation by school quality can have ramifications on student academic performance and the frequency of absenteeism suspension and drop out rates \citep{figlio2016school,hanushek2008students}. In particular, lower school absenteeism is a precursor to achieving better student outcomes in the short and long term\citep{bergman2019leveraging,liu2019short,jackson2018test,gottfried2017students}. 
Moreover, it is often students at the margin who show up at the left tail of the attendance distribution. Given the existing segregation levels in the Chilean context, it might be relevant from the policy perspective to understand how school quality impacts attendance for those at the lower end of the distribution. 

I display the differences in absenteeism by schools in the Chilean context. Panel A in Figure \ref{attendanceschooltype} shows the distribution of attendance for the ninth graders in 2018. I observe overall the distribution for low income students is shifted marginally to the left of the distribution for high income students. However, for both groups there are students at the left tail of the distribution. I intend to capture to what extent school quality impacts attendance. Panel B displays the coefficients on school dummies for the following reduced for model
\begin{align*}
\text{Attendance}_{is}=\sum_{s=1}^{M}\text{School dummy}_{s}+X_{i}\beta+\epsilon_{is}
\end{align*}
The dependent variable in the above specification is the student $i's$ annual attendance rate in school $s$ in 2018. I regress this on school dummies and a set of observed student-level characteristics. This suggests that there are differences in attendance rates by schools in Chile. Lastly, in panel C, I repeat the above analysis but using the two school types, high and low academic quality. The results of this estimation indicate there is a positive correlation between attendance and school quality.  
\section{Main Findings}
I provide the main findings on the determinants of the underlying parental preferences in Chilean DA for the year 2016 and 2017. Chilean DA was implemented first in Magallanes in 2016 but it was expanded to Tarapaca, Coquimbo, O'Higgins and Los Lagos in 2017. I use province as the definition of schooling market in my analysis. Preliminary examination for the 2017 participants suggests that 91.6\% of the students apply to schools within their province of residence. The choice set for student $i$ residing in province $p$ consists of the schools present in $p^{th}$ provinces.\footnote{The students who applied for ninth-grade admission in 2017 switched to new schools in 2018. Similarly, the students who applied for admission in 2016, they were admitted to the new schools in 2017.}

\subsection{Results: 2016}

In this section, I use the threshold rank model to estimate the determinants of student ROL in 2016. In 2016, the government introduced DA in Magallanes. This analysis focuses on the ninth-grade applications. Panel (A) in Table \ref{magallanes1} illustrates the key characteristics of students who participated in the new system in 2016. I observe that the average number of listed schools in ROL is 4.1 with a standard deviation of 1.6. Additionally, I observe significant heterogeneity in their academic ability and background characteristics. 
\begin{table}[H]
\centering
\captionsetup{width=15 cm}
\caption{School choice for ninth graders in Magallanes in 2016 (Parameter estimates)}
\label{resultstable1}
\begin{tabular}{p{7 cm}P{3.0 cm}P{3.0 cm}}\hline   \hline
Outcome variable:& \multicolumn{2}{c}{Rank ordered list}  \\ 
Variables& (1)  & (2)  \\ \hline
Travel distance (km)&-0.691*&-0.493*\\
&[-1.030,-0.250]&[-1.116,-0.045]\\
Fee dummy&-0.124*&	-0.516*\\
&[-0.945,-0.036]&[-0.803,-0.127]\\
Student income&0.395&0.487\\
&[-0.047,0.789]&[-0.152,0.812]\\
Student score&0.358&0.335\\
&[-0.030,1.030]&[-0.038,0.757]\\
School score&0.589*&0.167\\
&[0.002,0.671]&[-0.087,0.650]\\
School score $\times$ Student score&&0.418*\\
&&[0.001,0.679]\\
Travel distance $\times$Student Income&&0.043\\
&&[-0.380,0.836]\\
Constant&0.498&-0.021	\\
&[-0.435,0.597]&[-0.462	,0.610]\\\\
Unobserved cost & \xmark &\xmark\\
J(Schools)&17&17\\
N(Students)&499&499\\\hline\hline
\multicolumn{3}{c}{ \begin{minipage}{15 cm}{\footnotesize{Notes: The 90\% bootstrap confidence intervals are provided in the square brackets. The sample for these specifications include the eighth grade students who applied for ninth-grade admission in 2016.}}
\end{minipage}} \\
  \end{tabular}
\end{table}
In panel (B) in Table \ref{magallanes1}, I display the summary statistics for the top-choice school for the participants. As discussed in related work on school choice, parents do seem to have a strong preference for proximity. The average commuting distance by car to the top choice school is around $\sim 2.4$ km, with a standard deviation of 1.5 km. As for the academic quality, the average math and language test score for the top choice school is marginally below the average for all public and voucher schools in Magallanes in 2016 (252 vs. 254: math and 242 vs. 243: language). Moreover, this average is significantly below the school with the highest academic record in Magallanes (328: math and 299: language). This suggests that not all parents are necessarily aiming for the best academic school in their application. 

Moreover, Figure \ref{dataspatialall} illustrates the spatial location of students and schools used for this analysis. Panel (A) depicts actual school assignment for ninth-grade admission for low-ability students in 2017. Panel (B) replicates the same information for high ability students. A comparison of the two graphs suggests that the enrollment of low ability students is lower in some of the high test score schools, particularly those schools in the northeast direction. I explain the determinants for such systemic variations in the actual assignment using the threshold rank order model.

I begin with the Threshold Rank Order Model with travel distance to school by car, student income, student test score, and the school average test score. The results for this model are displayed in column 1 of Table \ref{resultstable1}. Since it is a non-linear model, I need to evaluate the marginal effects for the impact of each of the covariates.

\begin{figure}[H] 
  \caption{Likelihood of applying to the average school}
  \centering
    \label{margin1} 
  \begin{minipage}[b]{0.5\linewidth}
    \centering
    \includegraphics[width=\linewidth]{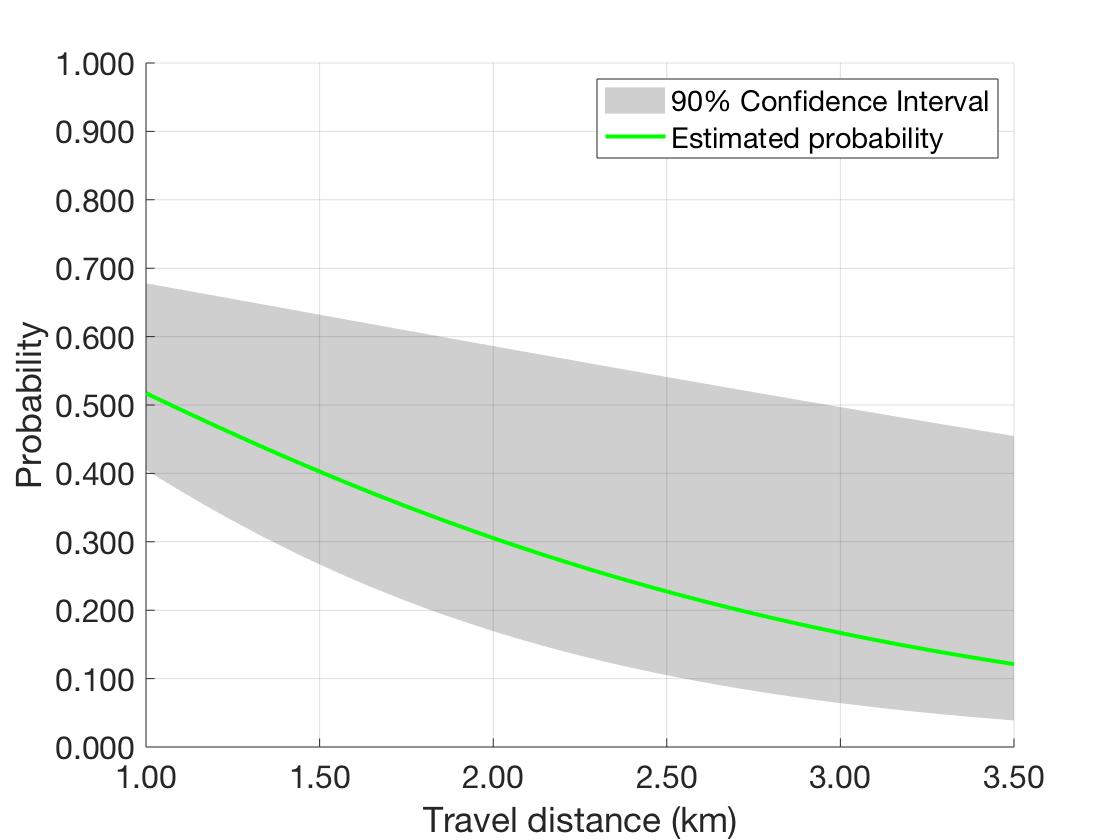}
    \caption*{A. Travel distance by car} 
  \end{minipage}%%
   \begin{minipage}[b]{0.5\linewidth}
    \centering
    \includegraphics[width=\linewidth]{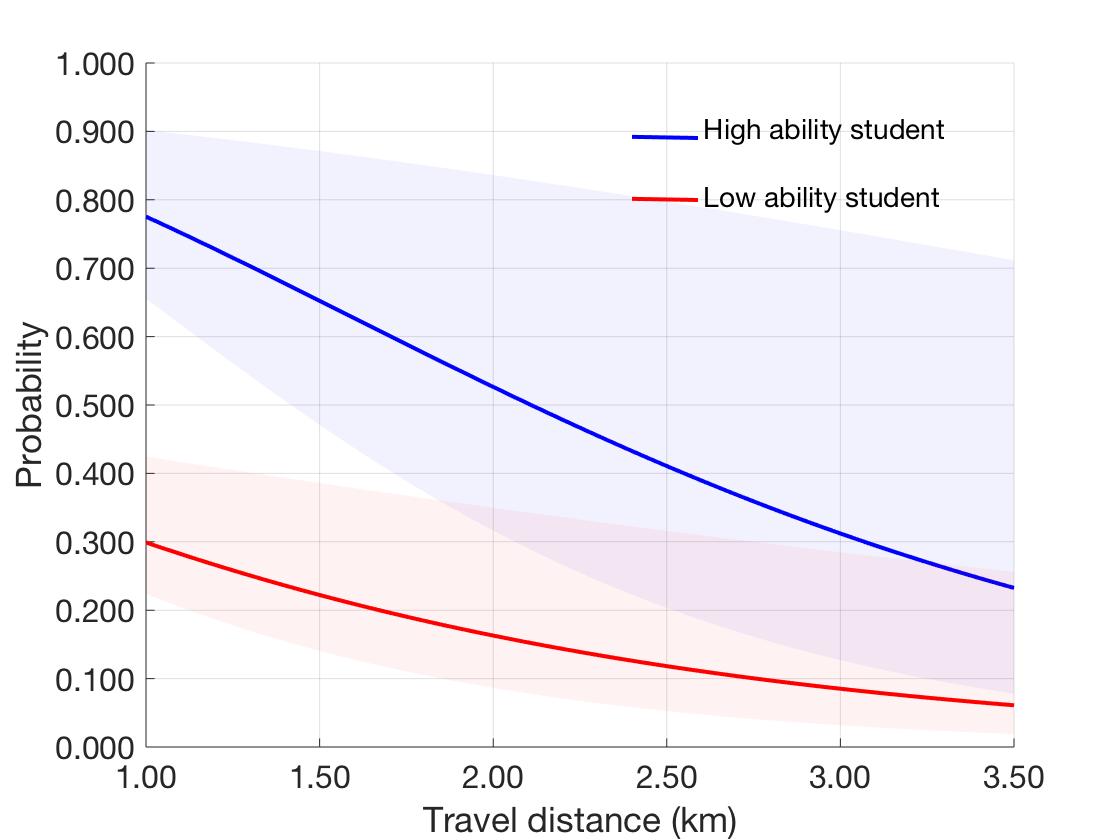}
    \caption*{B. School ability and distance} 
  \end{minipage}%%
  \end{figure}

First, I compute the sensitivity to distance (model (1) in Table \ref{resultstable1}). The other covariates, such as school test scores, student test scores, and student income for this analysis, have been fixed at their average values. The school fee dummy is set at one suggesting that this school charges a fee. I compute the confidence intervals using the bootstrap samples used in Table \ref{resultstable1}. I observe a consistent drop in the likelihood of applying to this average school as the student gets further away. Second, I explore differences in average sensitivity to distance for high and low ability students. Panel (B) in Figure \ref{margin1} suggests although high ability student has a higher probability of applying to the average school, the sensitivity to distance might vary by student ability. 
I expand the set of covariates to incorporate additional functional forms for travel distance and interactions between student test scores and school test scores (see Model (2) in \ref{resultstable1}). I also do a model selection test. The AIC for model (1) is 16653 with smaller set of covariates as compared to 14618 for the model with additional covariates for additional functional forms for distance and the match between student and school ability. It suggests that the model fit is higher for the specification with additional covariates relative to model 1. 
\begin{figure}[H] 
  \caption{Likelihood of applying to the top school by student ability}
    \label{dataspatial1} 
  \begin{minipage}[b]{0.23\linewidth}
    \centering
    \includegraphics[width=\linewidth]{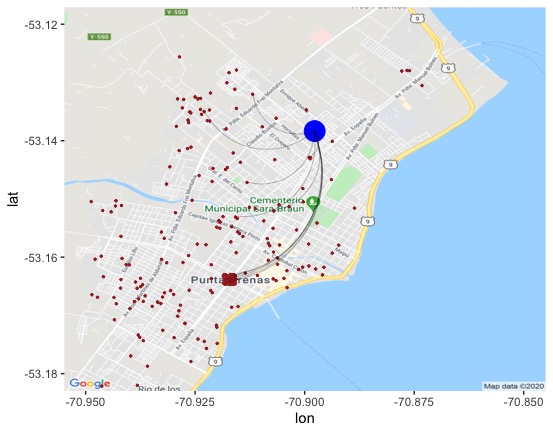}
    \caption*{A. Low ability} 
  \end{minipage}%%
  \begin{minipage}[b]{0.23\linewidth}
    \centering
    \includegraphics[width=\linewidth]{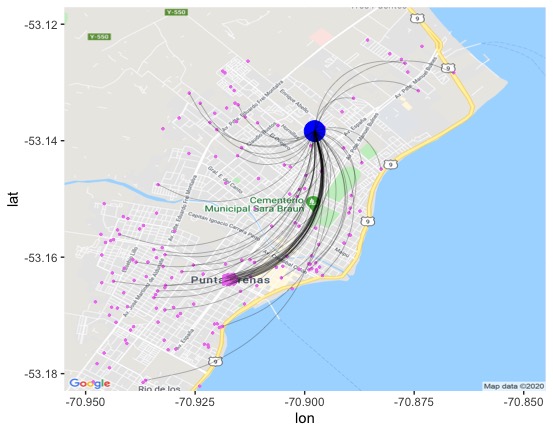} 
       \caption*{B. High ability} 
  \end{minipage} 
   \begin{minipage}[b]{0.23\linewidth}
    \centering
    \includegraphics[width=\linewidth]{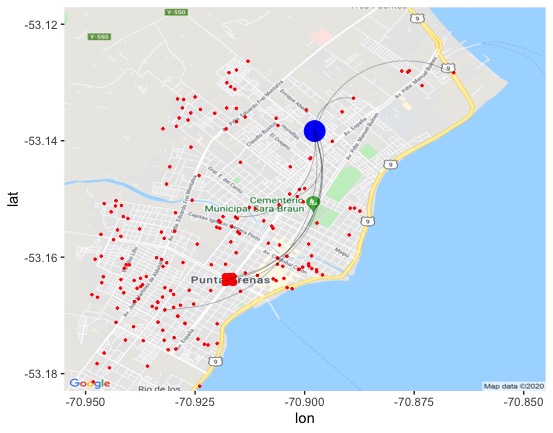}
    \caption*{C. Low income} 
  \end{minipage}%%
  \begin{minipage}[b]{0.23\linewidth}
    \centering
    \includegraphics[width=\linewidth]{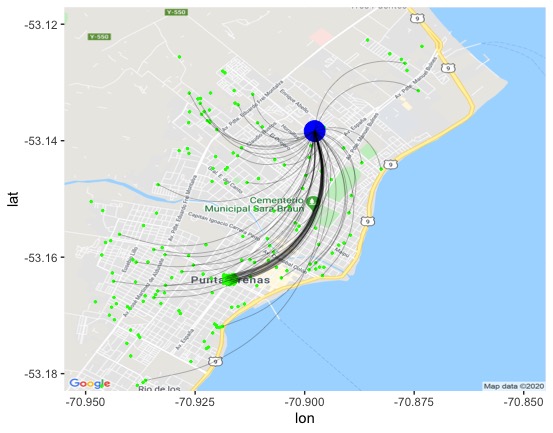} 
       \caption*{D. High income} 
  \end{minipage} 
  \end{figure}
I examine whether parents prefer schools that are a better match for student ability. Such assessments are possible in model 2, where I allow for interaction between student test scores and school test scores. I display the students who have a high probability ($\geq$0.5) of applying to the top school by ability (Figure \ref{dataspatial1}). If the probability of application is high, it is indicated using the black curve. The absence of a curve joining the student and school suggests that the student had a low likelihood of applying to the top school by the model predictions. Panel (A) and (B) in Figure \ref{dataspatial1} show the presence of a higher density of curves for high ability as compared to low ability students. This illustrates that parents do have a preference to apply to schools closer to the student's ability.
 
I observe in all columns of Table \ref{resultstable1} that distance is consistently negative. This supports the hypothesis that parents have a preference for school proximity. However, there might exist heterogeneity in this preference across income groups. Such differences will be captured by the interactions between travel distance and student income in the rank order model. 

Panel (C) in Figure \ref{dataspatial1} shows the probability of applying to the best school. I use a curve to join the student location and school location if the probability of applying to the school is larger than 0.5. The density of such curves is much higher for high income students as compared to low income students. 

\begin{figure}[H] 
  \caption{Marginal effect of travel distance on ranking the median school}
    \label{distancelatent} 
    \centering
    \includegraphics[width=0.3\linewidth]{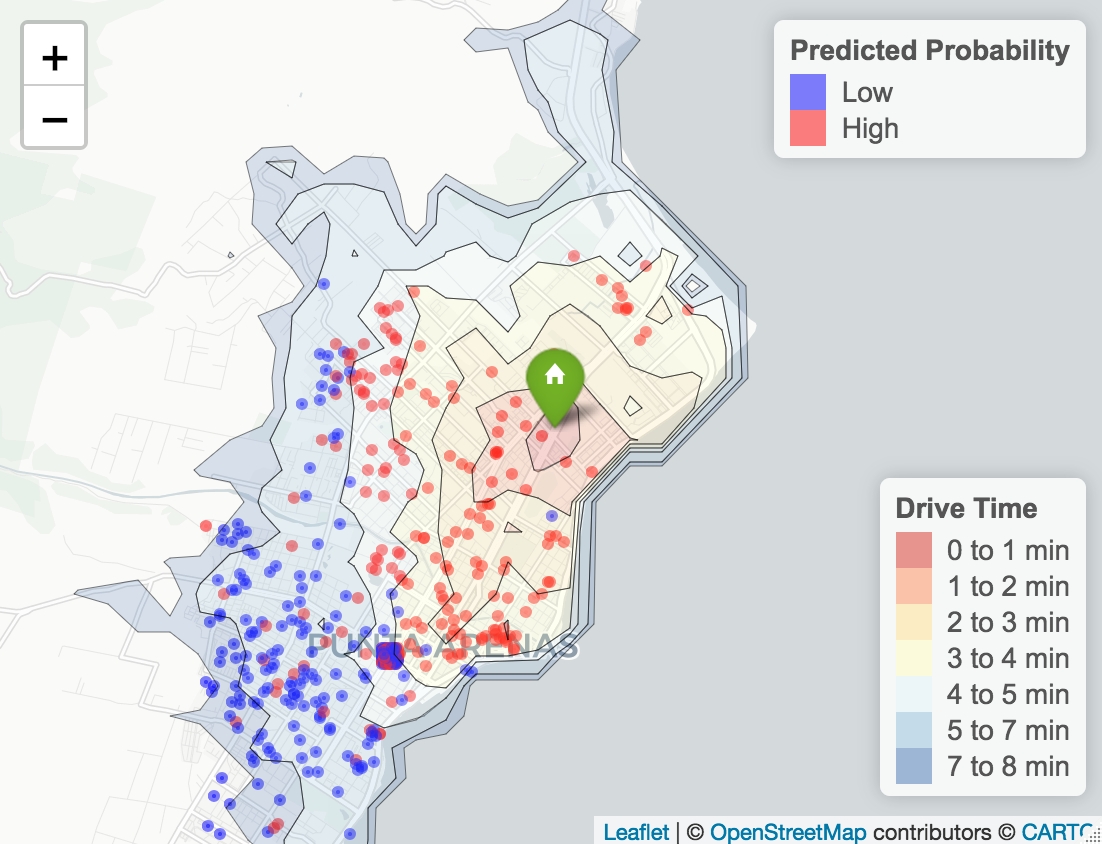}
    \begin{minipage}{17.0cm}
\footnotesize{
    {Notes: The probability of ranking this school with the median level of academic rigor is high (low) if the predicted probability is greater (lower) than the 75th percentile. }}
    \end{minipage}
\end{figure}

\begin{figure}[H] 
  \caption{Predicted probability of applying to top school by income}
  \centering
    \label{Rplot_latent_income} 
    \begin{minipage}[b]{0.3\linewidth}
    \includegraphics[width=\linewidth]{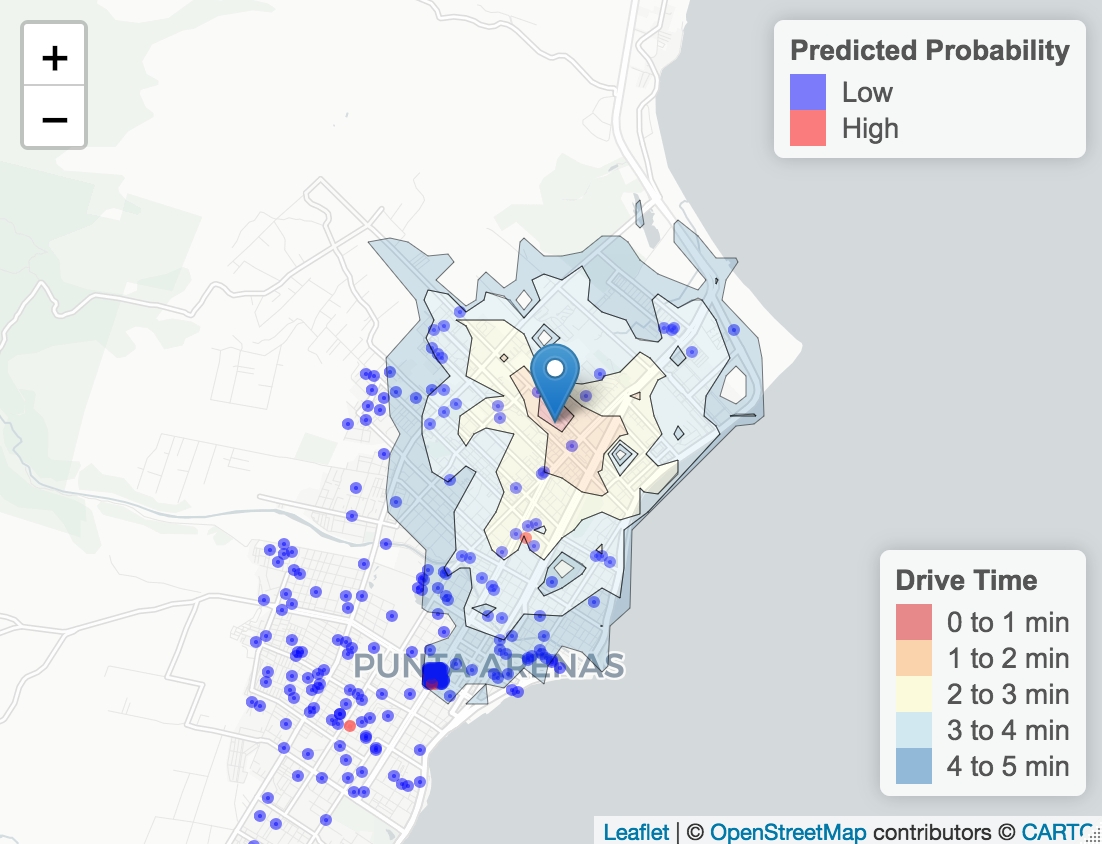}
    \caption*{A. Low income } 
  \end{minipage}%%
    \begin{minipage}[b]{0.3\linewidth}
    \includegraphics[width=\linewidth]{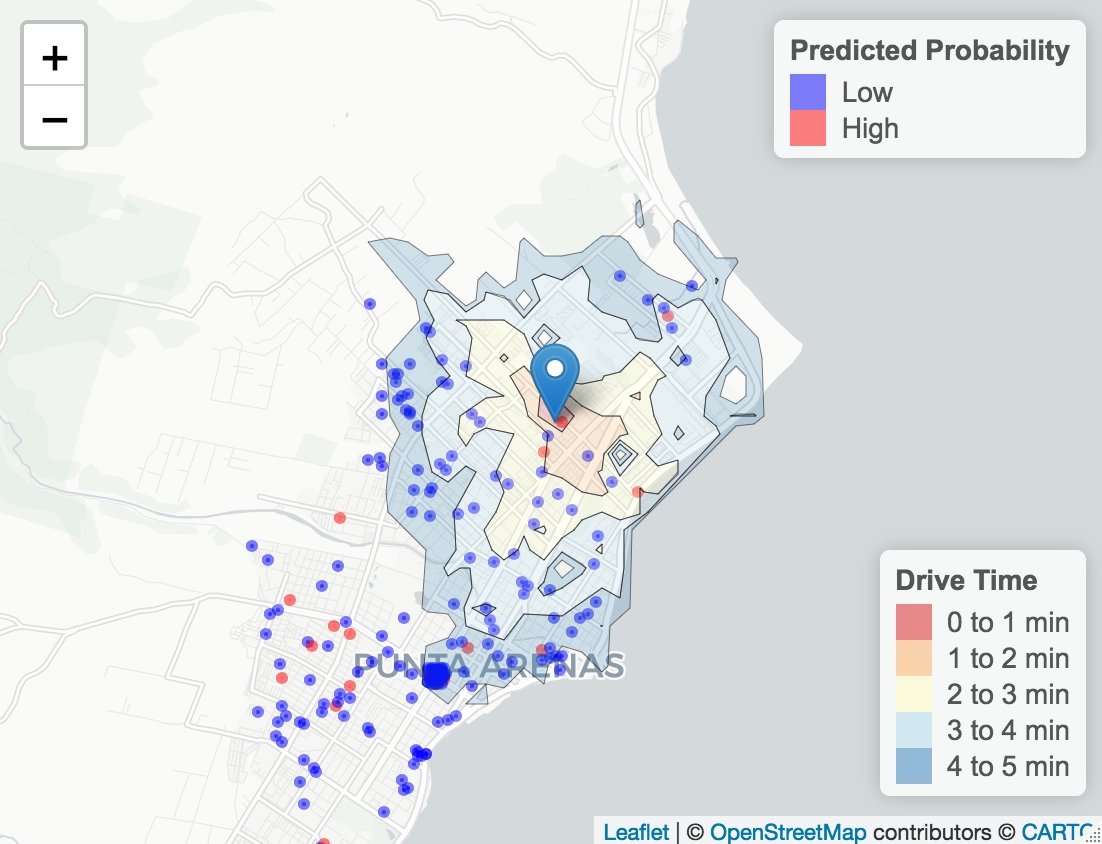}
    \caption*{B. High income } 
  \end{minipage}
  \begin{minipage}{17.0cm}
\footnotesize{
    {Notes: The sample consists of students who participated in DA for ninth grade admissions in 2016 in Magallanes and therefore started ninth grade in the allocated school in 2017. }}
    \end{minipage}
  \end{figure}

Next, I allow for unobserved heterogeneity in the above specification. Individual specific school invariant characteristics such as student test score as well as background characteristics parental income and education can be allowed to be correlated with unobserved cost. I display these estimates in Table \ref{resultstable2}. Column (1) illustrates the results with a restricted set of covariates. Additional functional forms for travel distance and academic ability are introduced in column (2). 
\begin{table}[H]
\centering
\captionsetup{width=15 cm}
\caption{School choice for ninth graders in Magallanes in 2016, allowing unobserved cost}
\label{resultstable2}
\begin{tabular}{p{7 cm}P{3.0 cm}P{3.0 cm}}\hline   \hline
Outcome variable:& \multicolumn{2}{c}{Rank ordered list}  \\ 
Variables& (1)  & (2)  \\ \hline
Travel distance (km)&	-0.444*&-0.088*\\
&[-0.918,-0.038]&[-1.118,-0.030]\\
Fee dummy&-0.797*&-1.301	\\
&[-1.945,-0.225]&[-1.538,0.306]\\
Student income&0.078&-0.137\\
&[-0.008,0.301]&[-0.646,1.436]\\
Student score&	0.203&0.008\\
&[-0.060,0.280]&[-0.196,0.124]\\
School score&0.131&0.318\\
&[-0.113,0.564]&[-0.047,0.605]\\
School score $\times$ Student score&&0.225*\\
&&[0.024,0.670]\\
Travel distance $\times$Student Income&&0.052\\
&&[-1.151,0.275]\\
Constant&-0.423&	-0.519\\
&[-0.916,0.101]&[-1.378,1.036]\\\\
Unobserved cost &\cmark &\cmark\\
J(Schools)&17&17\\
N(Students)&499&499\\\hline\hline
\multicolumn{3}{c}{ \begin{minipage}{15 cm}{\footnotesize{Notes: The 90\% bootstrap confidence intervals are provided in the square brackets. The sample for these specifications include the eighth grade students who applied for ninth-grade admission in 2016.}}
\end{minipage}} \\
  \end{tabular}
\end{table}
The coefficient on travel distance is negative across the two specifications, in line with the previous estimates (Baseline model without unobserved cost). Since the school choice model is highly non-linear, I cannot interpret the parameter estimates directly. Therefore, I display the marginal effects of travel distance on ROL in Figure \ref{distancelatent}. Here, I display the probability of ranking a school with an academic rigor close to the median for the set of schools present in the choice set. I plot the travel time isochrones using the OSRM API. The $x$ minute isochrone connects all the geo-coordinates around this school, which can be reached in $x$ minutes using a car. I want to highlight, though, that the isochrones are likely a lower limit on the actual commuting time as not all students have access to a car to commute to school. I observe that more the travel time increases, the likelihood of ranking this median school decreases. 
In addition to the median school, I show the predicted probability of ranking the best school and how it varies with the student's income. Figure \ref{Rplot_latent_income} displays that the likelihood of applying to the best school in this region is low for low-income students. The fraction of students who have a high likelihood of applying to this school is significantly higher for high-income students. Additionally, high-income students are less responsive to travel time to this school. 
Lastly, I examine the extent to which the match between student and school ability impacts school choice. I calculate the predicted ROL and examine the intersection between the predicted ROL and the set of best schools in this region. I divide my sample into two groups based on ability (High$>$Mean ability, Low$<$Mean ability). Figure \ref{abilitylatent} shows the extent of overlaps for two samples between the predicted top five ROL and the set of best schools. High student ability increases the likelihood of listing the best schools high up in the ROL. In fact, the distribution is completely shifted to the right for high-ability students compared to the low ability students. 

The school choice model for 2016 in Magallanes provides some meaningful predictions. However, since the new system was implemented only in one region in 2016, there are limitations in terms of the data variation that can be exploited to include the full set of covariates. I expand the covariates set for the school choice model in 2017 as then DA was implemented in five regions. 
\subsection{Results: 2017}
For this analysis, I am using the data on participants in 2017. The new system was implemented in 5 regions in 2017. Preliminary analysis for the 2017 participants suggests that 91.6\% of the students apply to schools within their residence province. I use province as a school market, and the choice set for student $i$ residing in province $p$ consists of the schools present in $p^{th}$ provinces. The students who applied for ninth-grade admission in 2017 switched to new schools in 2018. Table \ref{student2017summary} provides summary statistics for the eighth grade participants in 2017. 

The set of covariates in this model comprises of travel distance to school, school's academic score measured in 2017 (pre-DA), a dummy for fee as well as the socio-economic composition of the school, student pre-DA scores, student income, interactions between student income and school SES, student score and school scores as well as the interactions between distance and student score, distance and student income and triple interactions between student score, distance and student income. I account for the guaranteed school's academic score, and the school test score is used as a proxy variable for vacancies. Since a tiny fraction of students is eligible for a priority in the lotteries due to factors such as siblings enrolled in the same school, parents working in the same school or alumni, I include a priority indicator in the school choice model.

In table \ref{resultstable_allregions}, I provide the parameter estimates of the school choice model for the regions that got reallocated in 2017. The covariates set across the provinces in the five regions remain the same except the Coquimbo school fee variable. I do not observe any school fee variation for the schools that participated in DA in 2017 in Coquimbo with all the required information. 

\begin{figure}[h] 
  \caption{Predictive probability: Distance to mean score school}
    \centering
    \label{allregions_dist} 
  \begin{minipage}[b]{0.45\linewidth}
    %\centering
    \includegraphics[width=\linewidth]{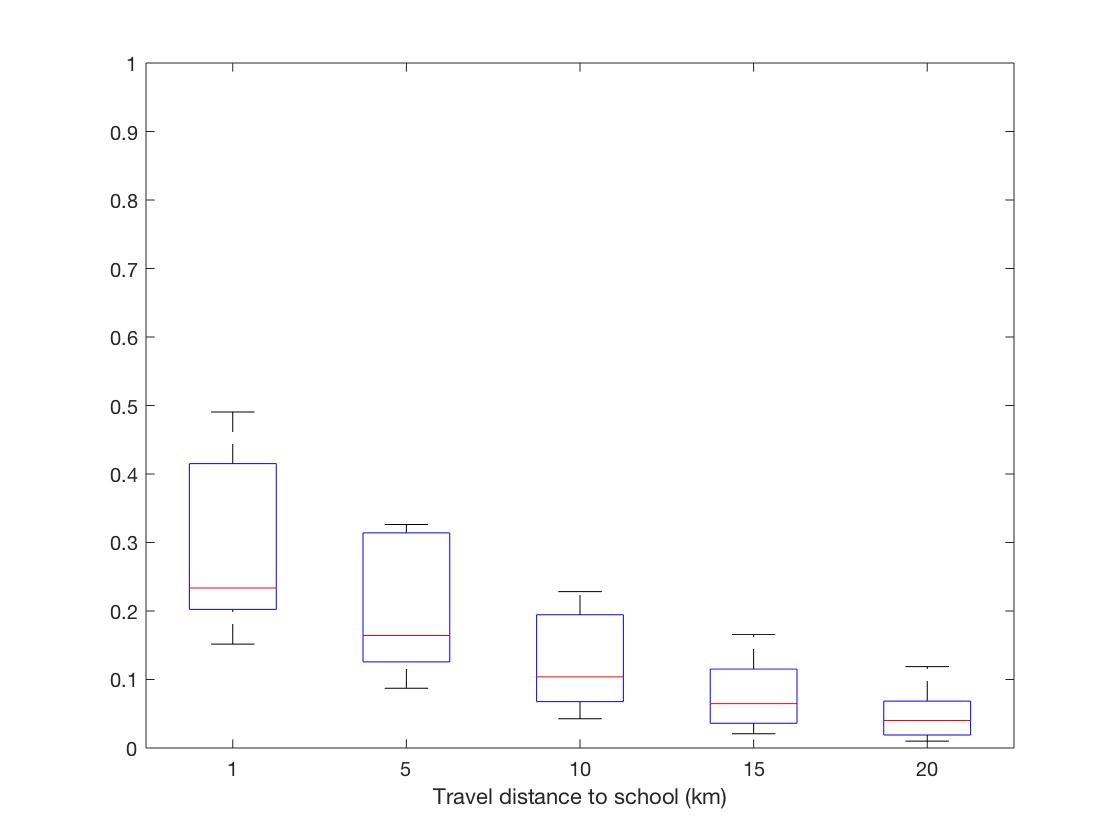} 
       \caption*{A. Student high income} 
           \vspace{0ex}
  \end{minipage} 
  \begin{minipage}[b]{0.45\linewidth}
    %\centering
      \includegraphics[width=\linewidth]{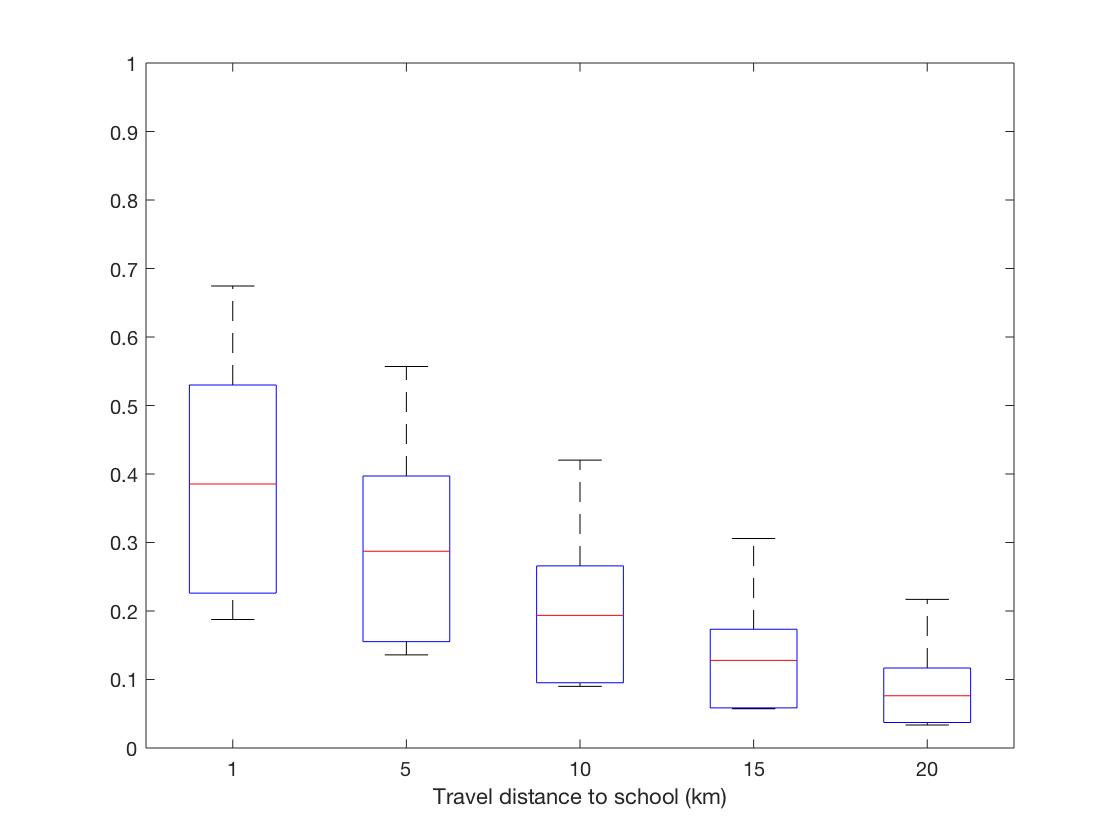}
       \caption*{B. Student low income} 
           \vspace{0ex}
  \end{minipage} 
  \begin{minipage}{17.5 cm}{\footnotesize{}}
\end{minipage} 
  \end{figure}

In order to unmask the impact of different covariates on the predicted probability of either listing the school or ranking it higher up in the ROL, I provide the marginal effects. First, I illustrate the response to travel distance to the mean school in the schooling market. I plot the predicted probability conditional on the average value of other covariates except for student income. I calculate the predictive margins for all the regions and display distribution of these marginal effects.
\begin{table}[H]
\centering
 \fontsize{9}{9}\selectfont
\captionsetup{width=15 cm}
\caption{School choice for ninth graders in 2017 (Parameter estimates)}
\label{resultstable_allregions}
\begin{tabular}{p{6.2 cm}P{2.0 cm}P{2.0 cm}P{2.0 cm}P{2.0 cm}P{2.0 cm}}\hline   \hline
Outcome variable:& \multicolumn{5}{c}{Rank ordered list}  \\ 
Region& Tarapaca  & Coquimbo & O'Higgins& Los Lagos& Magallanes \\
Province&Iquique &Choapa &Colchagua&Chiloe& Magallanes \\
Variables& (1)  & (2)&(3)&(4)&(5)  \\ \hline
Travel distance&-0.804*&-0.930*&-1.085*&-0.779*&-1.073*\\
&[-1.156,-0.509]&[-2.331,-0.702]&[-1.423,-0.678]&[-1.351,-0.470]&[-1.139,-0.558]\\
Students pre-DA score&0.306*&0.775*&-0.051&0.350*&0.261*\\
&[0.009,0.508]&[0.234,0.900]&[-0.234,0.034]&[0.241,0.473]&[0.019,0.318]\\
Student income&-0.217&-0.004&0.060&-1.776*&-0.102*\\
&[-0.479,0.005]&[-0.372,0.172]&[-0.053,0.108]&[-1.921,-1.664]&[-2.846,-0.070]\\
School's pre-DA score&0.408*&0.107&0.015&-0.136*&0.222\\
&[0.188,0.647]&[-0.292,0.292]&[-0.123,0.051]&[-0.263,-0.033]&[-0.039,0.278]\\
School SES&0.159&0.657*&0.021&-0.230*&-0.146\\
&[-0.109,0.387]&[0.397,0.794]&[-0.199,0.158]&[-0.353,-0.117]&[-0.645,0.327]\\
School fee&-1.024*&&-0.029&-0.558*&-0.917*\\
&[-1.206,-0.870]&&[-0.139,0.088]&[-0.685,-0.458]&[-1.005,-0.626]\\
Student score$\times$School score &0.094&
0.798*&0.028&0.421*&0.386*\\
&[-0.142,0.263]&[0.298,0.996
]&[-0.098,0.154]&[0.284,0.558]&[0.277,0.486]\\
School SES$\times$Student income&-0.350*&0.319*&0.146&0.885*&0.587*\\
&[-0.530,-0.185]&[0.121,0.487
]&[-0.032,0.246]&[0.748,1.025]&[0.336,4.459]\\
Distance$\times$Student score&-0.075&0.247&-0.854*&-0.285*&0.085\\
&[-0.329,0.181]&[-0.531,0.301
]&[-1.264,-0.431]&[-0.606,-0.020]&[-0.273,0.165]\\
Distance$\times$Student income&-0.743*&0.236&-0.281*&-0.229*&-0.215\\
&[-1.008,-0.526]&[-0.444,0.328
]&[-0.570,-0.158]&[-0.438,-0.081]&[-0.449,0.012]\\
Student score$\times$Student income&0.217&0.034&0.419*&0.058&-0.376*\\
&[-0.039,0.409]&[-0.206,0.242
]&[0.268,0.526]&[-0.083,0.209]&[-0.433,-0.078]\\
Distance$\times$Student score$\times$Student income&0.801*&1.089*&0.504*&0.373*&0.449*\\
&[0.567,0.969]&[0.062,1.197
]&[0.204,0.724]&[0.188,0.550]&[0.088,0.621]\\
Outside option value&-0.316*&-0.271
&-0.230*&-0.738*&-0.816*\\
&[-0.667,0.149]&[-0.373,0.229]&[-0.354,-0.059]&[-0.864,-0.598]&[-0.659,-0.215]\\
Constant&-1.462*&-0.629*&-0.314*&0.373*&-1.178*\\
&[-1.857,-1.143]&[-0.927,-0.474]&[-0.496,-0.195]&[0.259,0.522]&[-1.400,-0.570]\\
Priority dummy&0.052&-0.030&-0.096&0.859*&2.021*\\
&[-0.136,0.254]&[-0.214,0.895]&[
-0.213,0.008]&[0.763,1.003]&[1.219,2.253]\\
\hline\hline
\multicolumn{6}{c}{ \begin{minipage}{15 cm}{\footnotesize{ }}
\end{minipage}} \\
  \end{tabular}
\end{table}
In panel A in figure \ref{allregions_dist}, I display the responsiveness to the distance to average school for high-income students. There is a decline in the predicted probability of application by increasing distance to mean score school. However, compared to low-income students displayed in panel B, high-income students have lower likelihood of applying to the mean school than low-income students. In other words, probability of application drops more for high income students than low income students. 
In the Chilean context, there is variation in the value of the outside option. Student's whose pre-DA school offers ninth grade have a guaranteed seat if the algorithm fails to allocate the student to one of the listed schools in the ROL. If the pre-DA school does not offer ninth grade, then the student can be allocated to the nearest public school with vacancies. This nearest school is required to be within 17 km to the student's residence. Therefore, for students whose pre-DA school does not offer ninth grade, I extract set of public schools within 17 km using the travel distance and take the average school pre-DA test score. Figure \ref{allregions_dist1} shows that the predicted probability of application to the mean school drops with an increase in the value of outside option. 
 
I illustrated in section 2.3 that low-income students are under-represented in high score schools in Chile. Reducing segregation through DA was high up in the government's agenda. However, a better representation will depend on whether parents from low-income families end up listing the good schools in ROL. Parents are likely to care about the ability match between the student's ability and school academic rigor. Consequently, it is interesting to disentangle the effect of travel distance conditioned on both income and ability. Mainly, I next illustrate that student ability is compensating for income and resource constraints in Chile. I condition on low-income students and illustrate the probability of listing a high score school. Figures in panels A and B in \ref{allregions_dist1} showcase that high-ability low-income students are more likely to list a school with high test scores than low-ability low-income students. In panel C, I keep the distance constant at 1 km for low income category and calculate the probability of application to high score school by varying student ability. As student ability increases the likelihood of applying to good schools increase. 
 
\begin{figure}[h] 
  \caption{Marginal Effects: Distance, student score and income}
    \centering
    \label{marginalcoquimbo} 
    \begin{minipage}[b]{0.4\linewidth}
    \includegraphics[width=\linewidth]{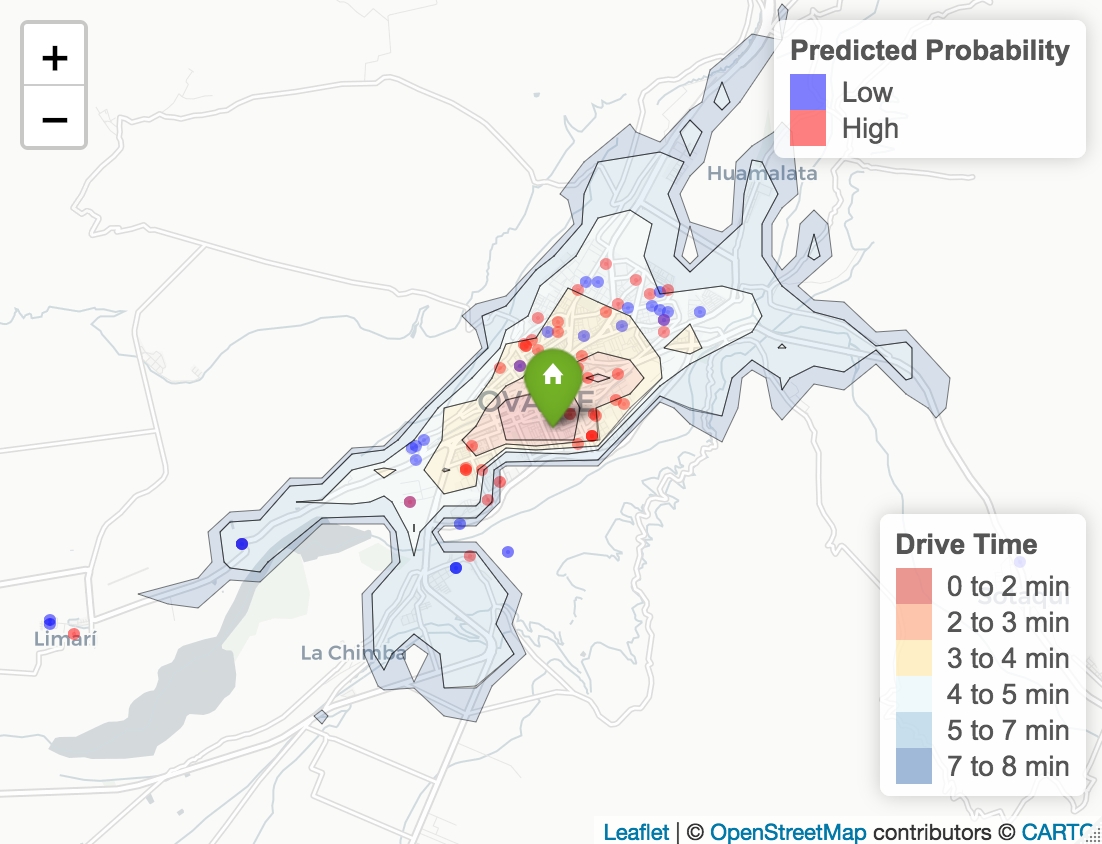}
    \caption*{\footnotesize{A. Low score high income}}
  \end{minipage}%%
    \begin{minipage}[b]{0.4\linewidth}
    \includegraphics[width=\linewidth]{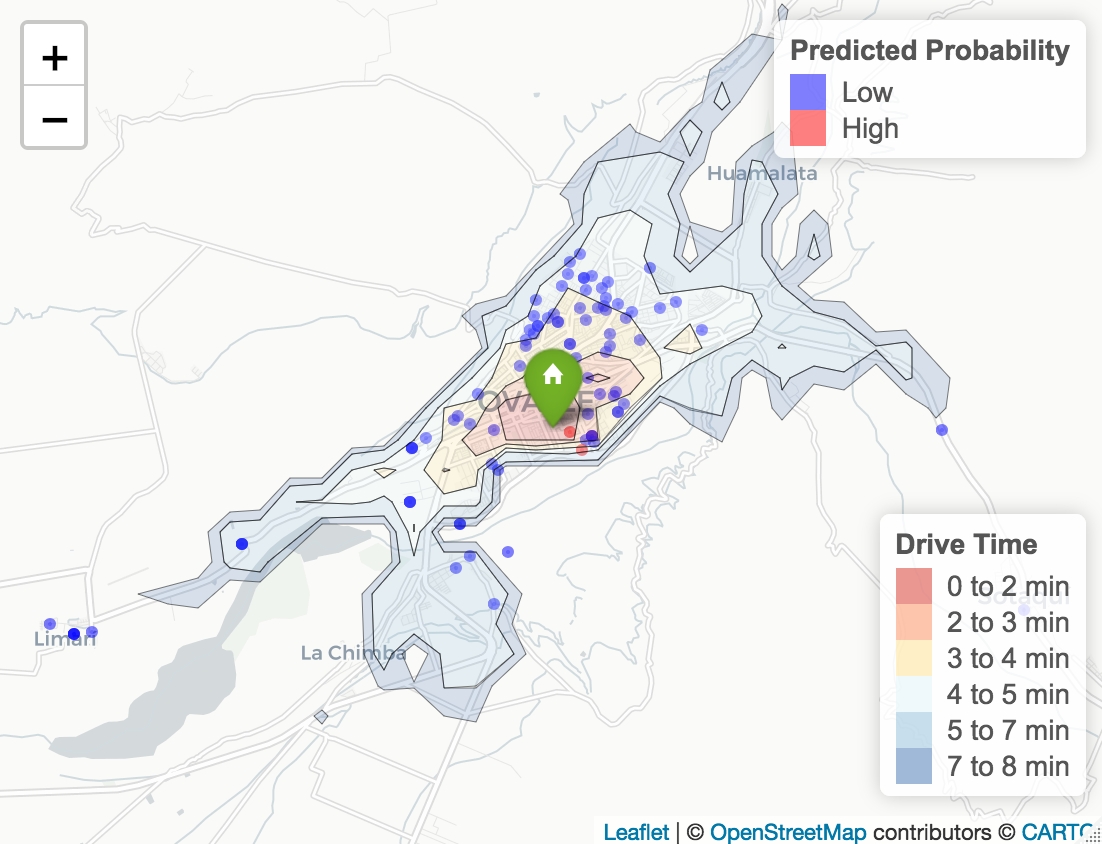}
    \caption*{\footnotesize{B. Low score low income}} 
  \end{minipage}
    \begin{minipage}[b]{0.4\linewidth}
    \includegraphics[width=\linewidth]{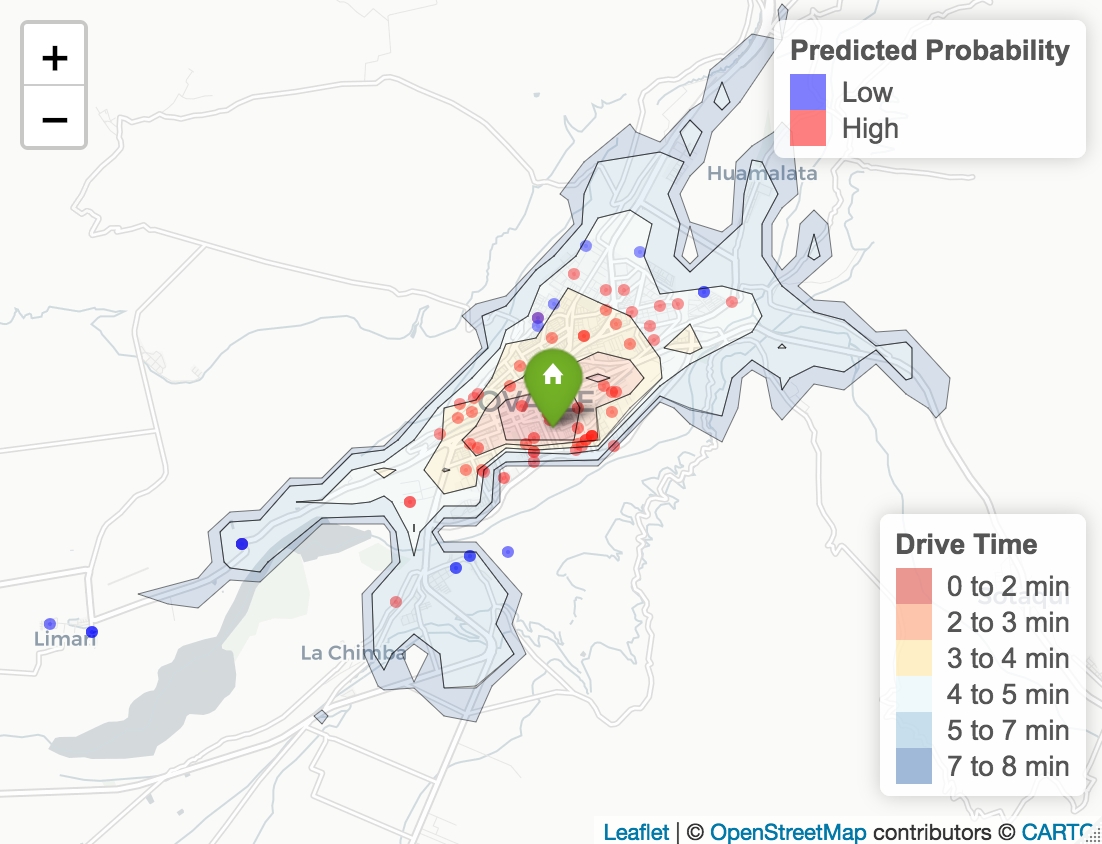}
    \caption*{\footnotesize{C. High score high income}}
  \end{minipage}%%
    \begin{minipage}[b]{0.4\linewidth}
    \includegraphics[width=\linewidth]{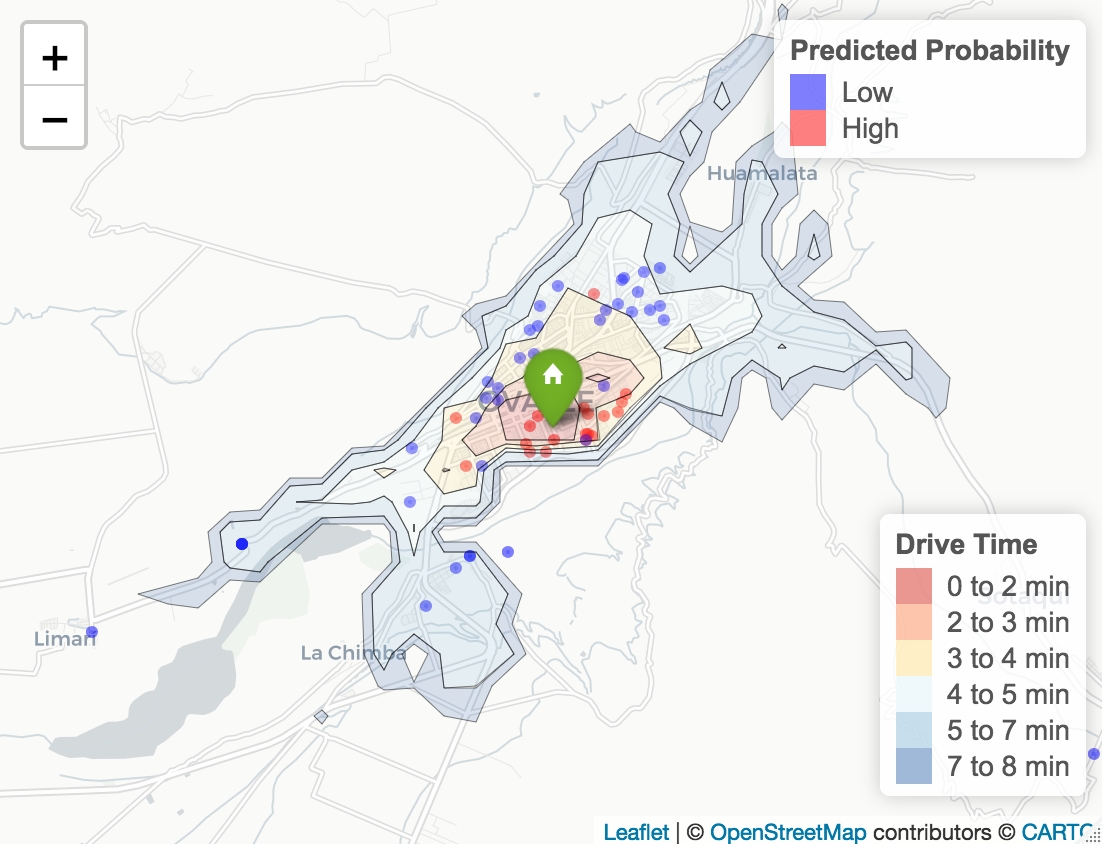}
    \caption*{\footnotesize{D. High score low income}}
  \end{minipage}
  \begin{minipage}{15.5 cm}{\footnotesize{Notes: These graphs display the travel time isocrones from the best school (\textcolor{green}{green} marker) in the Coquimbo region in Chile. A $x$ minute isocrone to the best school connects all points that are $x$ minutes away (travel time by car) from this school. The \textcolor{red}{red} (\textcolor{blue}{blue}) marker indicates a high (low) predicted probability of application to high score school. It displays conditioning on student score and income matters. Student score can compensate for travel costs for low income households.}}
\end{minipage} 
  \end{figure}

  These charts display the predicted probability of application to a high score school for low income students by varying student ability and travel distance. The other covariates are kept constant at the average values. In figure \ref{marginalcoquimbo}, I plot the unconditional marginal effect of application to the best school in Coquimbo region by student income and ability.   
I explore the angle between ability and school quality further by computing a rich array of marginal effects. I do this by calculating the likelihood of ranking a high score school above a low score school for low-income students (Figure \ref{allregions_ability2}). The figure illustrates that the predicted probability of observing such a ranking behavior is higher for high-ability students than low-ability students.

Lastly, I estimate the school choice model for all the areas together instead of estimating separately for each area. The summary statistics for this analysis is provided in \ref{resultstable_allregions}. I replicate the key predictions derived above using this joint analysis. Table \ref{allregions_abilityall} displays the predicted probability of ranking the mean school, low score and high score school by travel distance. 
\begin{table}[H]
 \fontsize{8}{8}\selectfont
\centering
\captionsetup{width=13cm}
\caption{Predictive Probability: Joint Analysis}
\label{allregions_abilityall}
\begin{tabular}{p{3.5 cm}P{2.1 cm}P{2.1 cm}P{2.1 cm}P{2.1 cm} P{2.1 cm}}\hline  \hline
Outcome Variable            &      \multicolumn{4}{c}{Predictive Margins}      \\
\hline
\textit{A. Travel distance (km)}&       1&        5&         10&         15&         20\\
Probability of application          &        &     &       &   &              \\
\hspace{3mm}High Score School     &0.695      &    0.550         &   0.384    & 0.255         &  0.164           \\
        & [0.681,0.716]       &  [0.536,0.571]       &    [0.373,0.402]       &   [0.247,0.268]         &    [0.158,0.173]         \\
\hspace{3mm}Mean Score School       &   0.592        &   0.452      &       0.307      &    0.199      &      0.126    \\
        & [0.578, 0.578]       &  [0.440, 0.440]       &    [0.297,    0.297]       &   [0.193,  0.193]         &    [0.122,0.122]         \\
\hspace{3mm}Low  Score School       &   0.492         &   0.365   & 0.241      &   0.155        &     0.097            \\
        & [0.479,0.512]       &  [0.355,0.382 ]       &    [0.234,0.254 ]       &   [0.149,0.163]         &    [0.093,0.102]         \\
\hline\hline
  \end{tabular}
\end{table}
\section{Discussion}\label{sec:future}
A key motivation behind the new reform for the government was to reduce the existing school segregation based on Chile's socioeconomic status. However, \cite{KNU2020} has analyzed the implications of DA for entire Chile and found no evidence of an unambiguous positive impact on segregation due to the new policy. The school choice model in illustrates that parents value travel costs and ability match while listing the ROL. Consequently, the combination of the school choice model and the government algorithm opens up the possibility of comparing the current policy's segregation outcomes with alternative counter-factual school networks. 

I begin by examining the participants in 2016. I use the Duncan index as the primary measure of school segregation. The main advantage of using the Duncan index for this analysis is that it makes the results easily comparable to existing studies on school segregation as it is the most common index used to measure school segregation. I use household income to construct the Duncan index. Income is a common measure of socioeconomic status in related work on Chile \citep{valenzuela2014socioeconomic,alves2015winners}.  
The Duncan index for Magallanes is defined as follows
\begin{equation}
D=\frac{1}{2}\sum_{i=1}^{J}|\frac{N_{i,ses=l}}{N_{ses=l}}-\frac{N_{i,ses=h}}{N_{ses=h}}|
\end{equation}
I compute the Duncan index for the public and voucher schools that participated in DA in 2016 in Magallanes for ninth-grade admissions. Figure \ref{segregation17} shows the changes in school segregation before and after DA. Overall, there has been a slight uptick in segregation levels post the reform.\footnote{I construct the Duncan index using student's family income and the income$>$200,000 Chilean pesos is used as the cut-off for low and high income students. } 

The school choice model in section \ref{sec:apply} illustrates that parents value the commuting distance to school. In my first simulation exercise, I use the Threshold Rank Order Model to generate alternative allocation under no travel cost. Such a simulation exercise can be interpreted as a thought experiment similar to busing. \footnote{I mimicked the government allocation rule for the applicants in Magallanes in 2016. Since, at the margin, there is a lottery, it is not possible to perfectly replicate the actual assignment. Overall, I could replicate the assignment for 82\% of the sample correctly. Moreover, students who had some priority at any applied school have relatively less uncertainty in admission than students without a priority as they face a random lottery with a higher probability. I observe that the replicated algorithm accurately allocates schools to 87\% of students who had indicated some priority in their application. The corresponding number for students without any priority is 79\%.} The Duncan index for 2017 is 0.372 under the actual allocation. However, if the low income students are provided busing which eliminates their travel cost then this index drops to 0.351 for 2017. The drop is marginal as not all students participate in DA and there are factors other than distance that are critical for submitted ROL. 

The school choice model for 2016 had limited set of covariates and this set has been expanded for 2017. The marginal effects suggest that although distance plays a critical role in school choice, student ability can compensate for such costs for low income students, specially the decision to list high score schools higher up in their ROL. Consequently, I address this interplay between costs and ability by conducting a simulation exercise which compensates for low ability and compute the probability of ranking high score schools under this alternative. 

  Figure \ref{allregions_simu} illustrates that students with mean or ability below mean often have a very low likelihood of ranking the high score school over a low score school. In other words, the match between student ability and school average test scores matters in school choice. Suppose a policy intervention can provide additional tutoring services to the students at or below the mean in the pre-DA test score distribution. In that case, that can go a long way in changing the DA ranking behavior of low-income low ability students. 
\section{Conclusion}
Centralized algorithms are increasingly used across the globe to assign students to schools and universities. A key motivation for policymakers to use this mechanism is an improvement in equity. However, the expansion in school choice need not necessarily translate into lower levels of segregation. The overall impact hinges critically on parental preferences, which is a crucial input for these algorithms. 

The urn model introduced by Plackett and Luce is the most commonly used model to estimate the determinants of the rank-ordered list (ROL). This model assumes that the decision to rank $n$ schools can be split into $n$ independent processes where each step can be thought as a multinomial choice by shrinking the choice set as one goes from ranking top to the bottom-most school. This assumption is unlikely to hold in ROL, where parents are expected to see the process of ranking $n$ schools as a one-step process. Modeling the ROL is computationally intensive. Consequently, I develop a recursive algorithm to estimate the likelihood of this scenario efficiently. I also allow for heterogeneity in the rank cut-off across individuals. This heterogeneity is handled using the Expectation-Maximization algorithm in the maximum likelihood estimator. 

I apply my new estimator to the centralized allocation in Chile. The Chilean government adopted a centralized system for student assignments starting in 2016. This new reform was sequentially implemented, and by 2019 all regions had a centralized assignment system for admissions to public and voucher schools in Chile. My analysis in Chile suggested that parents attach a high value to the school's distance in their ROL. Instead of crude distance proxies, I use the open street maps API and the entire road network of Chile to compute the actual travel distance to schools. I also find that the sensitivity to distance decreases with an improvement in student's socio-economic characteristics (SES) like income. Second, the results also suggest that parents care about the ability match between the student and the school while listing schools in ROL. Lastly, my simulation exercise shows policy interventions such as tutoring can help improve the representation of low income low ability students in high score schools. 
\newpage
\setcounter{table}{0}
\renewcommand{\thetable}{A\arabic{table}}
\setcounter{figure}{0}
\renewcommand{\thefigure}{A\arabic{figure}}

\section*{A. Appendix }\label{sec:appA}
\subsection*{Proof of Theorem 1}
\begin{proof}

Suppose, the individual ranks $n$ schools in their neighborhood where a rank of 1 corresponds to the top school and a rank of $n$ corresponds to the lowest rank among the schools that were ranked. I am not putting any restriction on the number of non-ranked schools. The integral that I want to evaluate is 
\begin{align}
P(u_{1}&-\frac{c_{i}}{p_{j}^{s}}>u_{2}-\frac{c_{i}}{p_{j}^{s}}>....>u_{K}-\frac{c_{i}}{p_{j}^{s}}>u_{0})=&\bigg[\int_{-V_{1}}^{\infty}\int_{-V_{2}}^{V_{1}+u_{1}-V_{2}}......\int_{-V_{n}}^{V_{n-1}+u_{n-1}-V_{n}}f(u_{n})du_{n}.....f(u_{1})du_{1}\bigg]
\end{align}
First, I evaluate the innermost integral over $u_{n}$
\begin{align*}
\int_{-V_{n}}^{V_{n-1}+u_{n-1}-V_{n}}f(u_{n})du_{n}=F(V_{n-1}+u_{n-1}-V_{n})-F(-V_{n})
\end{align*}
Putting this part back into (1) 
\small
\begin{align}
&P(U_{1}>U_{2}....>U_{n}>0)=\bigg[\int_{-V_{1}}^{\infty}\int_{-V_{2}}^{V_{1}+u_{1}-V_{2}}......\int_{-V_{n-1}}^{V_{n-2}+u_{n-2}-V_{n-1}}f(u_{n-1})du_{n-1}.....f(u_{1})du_{1}\bigg]-\bigg[F(-V_{n})\boldsymbol{I(n-1)}\bigg]
\end{align}
\normalsize
where $\boldsymbol{I(n-1)}=\int_{-V_{1}}^{\infty}\int_{-V_{2}}^{V_{1}+u_{1}-V_{2}}......\int_{-V_{n-1}}^{V_{n-2}+u_{n-2}-V_{n-1}}f(u_{n-1})du_{n-1}.....f(u_{1})du_{1}$. Next, I work with the innermost integral in for the first term in (2). I use integration by parts to solve the innermost integral in (2). 
\begin{align*}
\int_{-V_{n-1}}^{V_{n-2}+u_{n-2}-V_{n-1}}\underbrace{F(u_{n-1}+V_{n-1}-V_{n})}_{g}\underbrace{f(u_{n-1})du_{n-1}}_{\frac{dh}{dx}dx}
\end{align*}
The integration by parts formula for definite integrals is given below.
\begin{align*}
\int_{a}^{b}g\frac{dh}{dx}dx=[gh]_{a}^{b}-\int_{a}^{b}h\frac{dg}{dx}dx
\end{align*}
First, I compute $\frac{dg}{du_{n-1}},h$.
\begin{align*}
\frac{dg}{du_{n-1}}=f(u_{n-1}+V_{n-1}-V_{n}), h=F(u_{n-1})
\end{align*}
Using all the components derived so far, term 1 in integration by parts is 
\begin{align*}
[gh]_{a}^{b}&=[F(u_{n-1}+V_{n-1}-V_{n})F(u_{n-1})]_{-V_{n-1}}^{V_{n-2}+u_{n-2}-V_{n-1}}\\
&=F(V_{n-2}+u_{n-2}-V_{n})F(V_{n-2}+u_{n-2}-V_{n-1})-F(-V_{n})F(-V_{n-1})
\end{align*}
Next, using all the components derived so far the term two in integration by parts is 
\begin{align*}
\int_{a}^{b}h\frac{dg}{dx}dx&=\int_{-V_{n-1}}^{V_{n-2}+u_{n-2}-V_{n-1}}F(u_{n-1})f(u_{n-1}+V_{n-1}-V_{n})du_{n-1}
\end{align*}
I will apply result 2 derived below to term 2 of integration by parts. 
\begin{align*}
F(x)f(x+c)&=e^{-e^{-x}}e^{-(x+c)}e^{e^{-(x+c)}}\\
F(x+c)f(x)&=e^{-e^{-(x+c)}}e^{-x}e^{e^{-x}}\\
F(x)f(x+c)&=e^{-c}F(x+c)f(x)
\end{align*}
Using result 2, term 2 is 
\begin{align*}
\int_{a}^{b}h\frac{dg}{dx}dx&=\int_{-V_{n-1}}^{V_{n-2}+u_{n-2}-V_{n-1}}F(u_{n-1})f(u_{n-1}+V_{n-1}-V_{n})du_{n-1}\\
&=e^{-(V_{n-1}-V_{n})}\underbrace{\int_{-V_{n-1}}^{V_{n-2}+u_{n-2}-V_{n-1}}\underbrace{F(u_{n-1}+V_{n-1}-V_{n})}_{g}\underbrace{f(u_{n-1})du_{n-1}}_{\frac{dh}{dx}dx}}_{\text{LHS of integration by parts}}
\end{align*}
Bringing all the three terms together, I obtain 
\begin{align*}
\int_{-V_{n-1}}^{V_{n-2}+u_{n-2}-V_{n-1}}&F(u_{n-1}+V_{n-1}-V_{n})f(u_{n-1})du_{n-1}\\
&=\bigg[F(V_{n-2}+u_{n-2}-V_{n})F(V_{n-2}+u_{n-2}-V_{n-1})-F(-V_{n})F(-V_{n-1})\\
&-e^{-(V_{n-1}-V_{n})}\int_{-V_{n-1}}^{V_{n-2}+u_{n-2}-V_{n-1}}F(u_{n-1}+V_{n-1}-V_{n})f(u_{n-1})du_{n-1}\bigg]\\
(1+e^{-(V_{n-1}-V_{n})})&\int_{-V_{n-1}}^{V_{n-2}+u_{n-2}-V_{n-1}}F(u_{n-1}+V_{n-1}-V_{n})f(u_{n-1})du_{n-1}\\
&=F(V_{n-2}+u_{n-2}-V_{n})F(V_{n-2}+u_{n-2}-V_{n-1})-F(-V_{n})F(-V_{n-1})\\
\int_{-V_{n-1}}^{V_{n-2}+u_{n-2}-V_{n-1}}&F(u_{n-1}+V_{n-1}-V_{n})f(u_{n-1})du_{n-1}\\
&=\kappa_{n-1}F(V_{n-2}+u_{n-2}-V_{n})F(V_{n-2}+u_{n-2}-V_{n-1})-\kappa_{n-1}F(-V_{n})F(-V_{n-1})
\end{align*}
where $\kappa_{n-1}=\frac{1}{1+e^{-(V_{n-1}-V_{n})}}$
Next, I put this term back into the integral and obtain the following
\begin{align*}
P(U_{1}&>U_{2}....>U_{n}>0)=\\
&\bigg[\kappa_{n-1}\int_{-V_{1}}^{\infty}...\int_{-V_{n-2}}^{V_{n-3}+u_{n-3}-V_{n-2}}F(V_{n-2}+u_{n-2}-V_{n})F(V_{n-2}+u_{n-2}-V_{n-1})f(u_{n-2})du_{n-2}.....f(u_{1})du_{1}\bigg]-\\&\bigg[\kappa_{n-1}F(-V_{n})F(-V_{n-1})\boldsymbol{I(n-2)}\bigg]-\bigg[F(-V_{n})\boldsymbol{I(n-1)}\bigg]
\end{align*}
Again I repeat the same steps of integration by parts to solve for the innermost integral. I redefine $g=F(V_{n-2}+u_{n-2}-V_{n})F(V_{n-2}+u_{n-2}-V_{n-1})$ and $\frac{dh}{dx}=f(u_{n-2})$. Repeating the similar scaling transformation in Integration by parts I obtain the following.  Here,$g=F(V_{n-2}+u_{n-2}-V_{n})F(V_{n-2}+u_{n-2}-V_{n-1})$ and $\frac{dh}{dx}=f(u_{n-2})$ which will result in the following two components
\begin{align*}
[gh]_{a}^{b}&=F(V_{n-2}+u_{n-2}-V_{n})F(V_{n-2}+u_{n-2}-V_{n-1})F(u_{n-2})|^{V_{n-3}+u_{n-3}-V_{n-2}}_{-V_{n-2}}\\
&=F(V_{n-3}+u_{n-3}-V_{n})F(V_{n-3}+u_{n-3}-V_{n-1})F(V_{n-3}+u_{n-3}-V_{n-2})-F(-V_{n})F(-V_{n-1})F(-V_{n-2})
\end{align*}
The second component in integration by parts is 
\begin{align*}
\int_{a}^{b}h\frac{dg}{dx}dx=\int_{V_{n-3}+u_{n-3}-V_{n-2}}^{-V_{n-2}}f(V_{n-2}+u_{n-2}-V_{n})F(V_{n-2}+u_{n-2}-V_{n-1})F(u_{n-2})du_{n-2}\\
+\int_{V_{n-3}+u_{n-3}-V_{n-2}}^{-V_{n-2}}f(V_{n-2}+u_{n-2}-V_{n-1})F(V_{n-2}+u_{n-2}-V_{n})F(u_{n-2})du_{n-2}
\end{align*}
Applying the same transformation as before, I obtain result a and result b\\
\textbf{Result a}
\begin{align*}
F(x)F(x+c_{2})f(x+c_{1})&=e^{e^{-x}}e^{-e^{-(x+c_{2})}}e^{-e^{-(x+c_{1})}}e^{e^{-(x+c_{1})}}\\
&=e^{-c_{1}}F(x+c_{1})F(x+c_{2})f(x)
\end{align*}
\textbf{Result b}
\begin{align*}
F(x)F(x+c_{1})f(x+c_{2})=e^{e^{-x}}e^{-e^{-(x+c_{1})}}e^{-e^{-(x+c_{2})}}e^{e^{-(x+c_{2})}}\\
=e^{-c_{2}}F(x+c_{1})F(x+c_{2})f(x)
\end{align*}
In my setting $c_{1}=V_{n-2}-V_{n}$ and $c_{2}=V_{n-2}-V_{n-1}$. Therefore, I write term 2 of integration by parts as 
\begin{align*}
\text{term 2}=e^{-c_{1}}LHS+e^{-c_{2}}LHS\\
\text{LHS}=\int^{V_{n-3}+u_{n-3}-V_{n-2}}_{-V_{n-2}}F(V_{n-2}+u_{n-2}-V_{n})F(V_{n-2}+u_{n-2}-V_{n-1})f(u_{n-2})du_{n-2}
\end{align*}
\small
\begin{align*}
&P(U_{1}>U_{2}....>U_{n}>0)=\\
&\bigg[\kappa_{n-2}\kappa_{n-1}\int_{-V_{1}}^{\infty}..\int_{-V_{n-3}}^{V_{n-4}+u_{n-4}-V_{n-3}}\prod_{j=0,1,2}F(V_{n-3}+u_{n-3}-V_{n-j})f(u_{n-3})du_{n-3}..f(u_{1})du_{1}\bigg]-\\&\bigg[\kappa_{n-2}\kappa_{n-1}F(-V_{n})F(-V_{n-1})F(-V_{n-3})\boldsymbol{I(n-3)}\bigg]-\bigg[\kappa_{n-1}F(-V_{n})F(-V_{n-1})\boldsymbol{I(n-2)}\bigg]-\bigg[F(-V_{n})\boldsymbol{I(n-1)}\bigg]
\end{align*}
\normalsize
where $\kappa_{n-2}=\frac{1}{1+e^{-(V_{n-2}-V_{n})}+e^{-(V_{n-2}-V_{n-1})}}=\frac{1}{\sum_{j=n-2}^{n}e^{-(V_{n-2}-V_{j})}}$. I keep on iterating backward to get the final solution stated as result (3). For completeness of the proof, I next derive the last step for a general $m$ where $1<m<n$. First, the last result if I have iterated backward m times is 
\begin{align*}
&P(U_{1}>U_{2}....>U_{n}>0)=\\
&\bigg[\kappa_{n-(m+1)}..\kappa_{n-1}\int_{-V_{1}}^{\infty}..\int_{-V_{n-m}}^{V_{n-(m+1)}+u_{n-(m+1)}-V_{n-m}}\prod_{j=0,1,..,m-1}F(V_{n-m}+u_{n-m}-V_{n-j})f(u_{n-m})du_{n-m}..f(u_{1})du_{1}\bigg]\\
&-\bigg[\kappa_{n-(m+1)}..\kappa_{n-1}F(-V_{n})...F(-V_{n-m})\boldsymbol{I(n-m)}\bigg]-....-\bigg[\kappa_{n-1}F(-V_{n})F(-V_{n-1})\boldsymbol{I(n-2)}\bigg]\\
&-\bigg[F(-V_{n})\boldsymbol{I(n-1)}\bigg]
\end{align*}
Second, I prove that the result for having iterated $m$ times is what I derived above. I begin with $m-1$ and then derive the solution for the $m^{th}$ iteration. 
\small
\begin{align*}
&P(U_{1}>U_{2}....>U_{n}>0)=\\
&\bigg[\kappa_{n-m}..\kappa_{n-1}\int_{-V_{1}}^{\infty}..\int_{-V_{n-(m-1)}}^{V_{n-m}+u_{n-m}-V_{n-(m-1)}}\underbrace{\prod_{j=0,1,..,m-2}F(V_{n-(m-1)}+u_{n-(m-1)}-V_{n-j})}_{g}\underbrace{f(u_{n-(m-1)})du_{n-(m-1)}}_{\frac{dh}{dx}}..f(u_{1})du_{1}\bigg]-\\&\bigg[\kappa_{n-m}..\kappa_{n-1}F(-V_{n})F(-V_{n-1})F(-V_{n-3})\boldsymbol{I(n-(m-1))}\bigg]-..-\bigg[F(-V_{n})\boldsymbol{I(n-1)}\bigg]
\end{align*}
\normalsize
Using integration by parts for the innermost integral. 
\begin{align*}
[gh]_{a}^{b}&=[\prod_{j=0,1,..,m-2}F(V_{n-(m-1)}+u_{n-(m-1)}-V_{n-j})\times F(u_{n-(m-1)})]_{-V_{n-(m+1)}}^{V_{n-m}+u_{n-m}-V_{n-(m-1)}}\\
&=[\prod_{j=0,1,..,m-1}F(V_{n-m}+u_{n-m}-V_{n-j})-F(-V_{n-(m-1)})..F(-V_{n})]
\end{align*}
Now for the second term in integration by parts I use the result $\prod_{j=2,...,b}F(x)F(x+c_{j})f(x+c_{1})=e^{-c_{1}}\prod_{j=1,...,b}F(x+c_{j})f(x)$
\small
\begin{align*}
\int_{a}^{b}h\frac{dg}{dx}dx&=(\sum_{j}e^{-c_{j}})\int_{-V_{n-(m-1)}}^{V_{n-m}+u_{n-m}-V_{n-(m-1)}}\prod_{j=0,1,..,m-2}F(V_{n-(m-1)}+u_{n-(m-1)}-V_{n-j})f(u_{n-(m-1)})du_{n-(m-1)}\\
(1+(\sum_{j}e^{-c_{j}}))&\int_{-V_{n-(m-1)}}^{V_{n-m}+u_{n-m}-V_{n-(m-1)}}\prod_{j=0,1,..,m-2}F(V_{n-(m-1)}+u_{n-(m-1)}-V_{n-j})f(u_{n-(m-1)})du_{n-(m-1)}=\\
&[\prod_{j=0,1,..,m-1}F(V_{n-m}+u_{n-m}-V_{n-j})-F(-V_{n-(m-1)})..F(-V_{n})]\\
\int_{-V_{n-(m-1)}}^{V_{n-m}+u_{n-m}-V_{n-(m-1)}}&\prod_{j=0,1,..,m-2}F(V_{n-(m-1)}+u_{n-(m-1)}-V_{n-j})f(u_{n-(m-1)})du_{n-(m-1)}=\\
&\kappa_{n-(m+1)}[\prod_{j=0,1,..,m-1}F(V_{n-m}+u_{n-m}-V_{n-j})-F(-V_{n-(m-1)})..F(-V_{n})]
\end{align*}
\normalsize
Putting this term back into the integral, I obtain the following expression
\small
\begin{align*}
&P(U_{1}>U_{2}....>U_{n}>0)=\\
&\bigg[\kappa_{n-(m+1)}..\kappa_{n-1}\int_{-V_{1}}^{\infty}..\int_{-V_{n-m}}^{V_{n-(m+1)}+u_{n-(m+1)}-V_{n-m}}\prod_{j=0,1,..,m-1}F(V_{n-m}+u_{n-m}-V_{n-j})f(u_{n-m})du_{n-m}..f(u_{1})du_{1}\bigg]\\
&-\bigg[\kappa_{n-(m+1)}..\kappa_{n-1}F(-V_{n})...F(-V_{n-m})\boldsymbol{I(n-m)}\bigg]-....-\bigg[\kappa_{n-1}F(-V_{n})F(-V_{n-1})\boldsymbol{I(n-2)}\bigg]\\
&-\bigg[F(-V_{n})\boldsymbol{I(n-1)}\bigg]
\end{align*}
\normalsize
The last component left is to solve for the topmost integral over the highest ranked school which is 1 in my setting. 
\small
\begin{align*}
&P(U_{1}>U_{2}....>U_{n}>0)=\kappa_{2}...\kappa_{n-1}\int_{-V_{1}}^{\infty}\prod_{j\neq 1}e^{-e^{-(u_{1}+V_{1}-V_{j})}}e^{-u_{1}}e^{-e^{-u_{1}}}du_{1}-\bigg[\kappa_{2}..\kappa_{n-1}F(-V_{n})..F(-V_{2})\boldsymbol{I(1)}\bigg]-\\
&\bigg[\kappa_{3}..\kappa_{n-1}F(-V_{n})F(-V_{n-1})..F(-V_{2})\boldsymbol{I(2)}\bigg]-......-
\bigg[\kappa_{n-1}F(-V_{n})F(-V_{n-1})\boldsymbol{I(n-2)}\bigg]-\bigg[F(-V_{n})\boldsymbol{I(n-1)}\bigg]\\
\end{align*}
\normalsize
I solve for the first term in the above integral to complete the proof. The derivation of the closed form solution for $\kappa_{2}...\kappa_{n-1}\int_{-V_{1}}^{\infty}\prod_{j\neq 1}e^{-e^{-(u_{1}+V_{1}-V_{j})}}e^{-u_{1}}e^{-e^{-u_{1}}}du_{1}$ is given below
\small
\begin{align*}
\kappa_{2}...\kappa_{n-1}\int_{-V_{1}}^{\infty}\prod_{j\neq 1}e^{-e^{-(u_{1}+V_{1}-V_{j})}}e^{-u_{1}}e^{-e^{-u_{1}}}du_{1}&=\kappa_{2}...\kappa_{n-1}\int_{-V_{1}}^{\infty}\prod_{j=1}^{n}e^{-e^{-(u_{1}+V_{1}-V_{j})}}e^{-u_{1}}du_{1}\\
&=\kappa_{2}...\kappa_{n-1}\int_{-V_{1}}^{\infty}exp(-\sum_{j=1}^{n}-e^{-(u_{1}+V_{1}-V_{j})})e^{-u_{1}}du_{1}\\
&=\kappa_{2}...\kappa_{n-1}\int_{-V_{1}}^{\infty}exp(-e^{-u_{1}}\sum_{j=1}^{n}-e^{-(V_{1}-V_{j})})e^{-u_{1}}du_{1}\\
\text{I substitute $t=e^{-u_{1}}$ and obtain the following expression.}\\
&=\int_{e^{V_{1}}}^{0}exp(-t\sum_{j}e^{-(V_{1}-V_{j})})-dt\\
&=\int^{e^{V_{1}}}_{0}exp(-t\sum_{j}e^{-(V_{1}-V_{j})})dt\\
&=\frac{exp(-t\sum_{j}e^{-(V_{1}-V_{j})})}{\sum_{j}e^{-(V_{1}-V_{j})}}\bigg|^{e^V_{1}}_{0}\\
&=\frac{1}{\sum_{j}e^{-(V_{1}-V_{j})}}\bigg[1-F(-V_{1})F(-V_{2})....F(-V_{n})\bigg]
\end{align*}
\normalsize
Substituting for the closed form solution for term 1 as derived above, I obtain the final recursive solution for the likelihood. 
\small
\begin{equation}
\begin{aligned}
&P(U_{1}>U_{2}....>U_{n}>0)=\kappa_{1}\kappa_{2}...\kappa_{n-2}\kappa_{n-1}(1-F(-V_{1})...F(-V_{n}))-\bigg[\kappa_{2}..\kappa_{n-1}F(-V_{n})F(-V_{n-1})..F(-V_{2})\boldsymbol{I(1)}\bigg]\\
&-......-
\bigg[\kappa_{n-1}F(-V_{n})F(-V_{n-1})\boldsymbol{I(n-2)}\bigg]-\bigg[F(-V_{n})\boldsymbol{I(n-1)}\bigg]
\end{aligned}
\end{equation}
\end{proof}
%\newpage 
\subsection*{Proof of Step 1, solving for $\hat{\sigma}^{2}$}
\begin{proof}
The score function is $S_{i}(\sigma^2)=-\frac{1}{2\sigma^2}+\frac{(c_{i}-z_{i}\gamma)^{2}}{2(\sigma^{2})^2}$. Solving for $\sum_{i=1}^{N}E_{h(c|R)}S_{i}(\sigma^{2})=0$
\begin{equation*}
  \begin{aligned}
&\sum_{i=1}^{N}E_{h(c|R)}S_{i}(\sigma^{2})=0\\
&\sum_{i=1}^{N}\int [-\frac{1}{2\sigma^2}+\frac{(c_{i}-z_{i}\gamma)^{2}}{2(\sigma^{2})^2} ]h(c_{i}|R_{i};\gamma,\sigma^2)dc=0\\
&-\frac{n}{2\sigma^2}+\sum_{i=1}^{N}\int [\frac{(c_{i}-z_{i}\gamma)^{2}}{2(\sigma^{2})^2} ]h(c_{i}|R_{i};\gamma,\sigma^2)dc=0\\
&\hat{\sigma}^{2}=\frac{1}{n}\sum_{i=1}^{N}\int (c_{i}-z_{i}\gamma)^{2}h(c_{i}|R_{i};\gamma,\sigma^2)dc\\
&\hat{\sigma}^{2}=\frac{1}{n}\sum_{i=1}^{N}\bigg[\frac{\sum_{t=1}^{T}  (c_{i}-z_{i}\gamma)^{2}f(R_{i}|c;\beta)}{\sum_{t=1}^{T} f(R_{i}|c;\beta)}\bigg] \\
 \end{aligned}
\end{equation*}
\end{proof}
\normalsize

\newpage

\setcounter{table}{0}
\renewcommand{\thetable}{B\arabic{table}}
\setcounter{figure}{0}
\renewcommand{\thefigure}{B\arabic{figure}}
\section*{B. Appendix}
\subsection*{Tables}

\begin{table}[H]
 \fontsize{10}{10}\selectfont
\centering
\captionsetup{width=13cm}
\caption{Student and school participation in DA by region}
\label{datatable1}
\begin{tabular}{p{5.0 cm}p{2.0 cm}p{2.0 cm}p{2.0 cm}p{2.0 cm}p{2.0 cm}}\hline  \hline
 & \multicolumn{3}{c}{School}  &\multicolumn{2}{c}{Student} \\ 
Regions& N  & \% public  & \% voucher  & N  & \% enroll  \\ \hline  
A. 2016&&&&&\\
\hspace{3 mm}Magallanes &        24 &     45.83 &     54.17 &      1040 &     46.19 \\ \\
B. 2017&&&&&\\
\hspace{3 mm}Tarapac\'a &        57 &     28.07 &     71.93 &      1757 &     32.78 \\ 
\hspace{3 mm}Coquimbo &       129 &     25.78 &     66.41 &      4824 &     43.14 \\ 
\hspace{3 mm}O'Higgins &       132 &     43.51 &     56.49 &      7872 &     59.47 \\ 
\hspace{3 mm}Los Lagos &       145 &     47.22 &     52.78 &      7439 &     57.86 \\ 
\hspace{3 mm}Magallanes &        25 &     48.00 &     52.00 &      1041 &     46.61 \\ 
\\
C. 2018&&&&&\\
\hspace{3 mm}Tarapac\'a &        56 &     27.27 &     72.73 &      1804 &     35.33 \\ 
\hspace{3 mm}Antofagasta &        67 &     44.78 &     55.22 &      4459 &     51.79 \\ 
\hspace{3 mm}Atacama &        33 &     54.55 &     45.45 &      2856 &     67.41 \\  
\hspace{3 mm}Coquimbo &       124 &     33.06 &     66.94 &      4517 &     41.67 \\  
\hspace{3 mm}Valpara\'iso &       321 &     31.35 &     68.65 &      9909 &     42.29 \\ 
\hspace{3 mm}O'Higgins &       132 &     43.18 &     56.82 &      7517 &     58.22 \\ 
\hspace{3 mm}Maule &       153 &     44.08 &     55.92 &      8794 &     60.75 \\ 
\hspace{3 mm}Biob\'io &       309 &     47.23 &     52.77 &     15127 &     53.56 \\ 
\hspace{3 mm}Araucan\'ia &       166 &     36.97 &     63.03 &      8502 &     60.20 \\ 
\hspace{3 mm}Los Lagos &       142 &     47.18 &     52.82 &      7290 &     58.89 \\  
\hspace{3 mm}Ays\'en &        25 &     52.00 &     48.00 &       469 &     28.71 \\ 
\hspace{3 mm}Magallanes &        25 &     48.00 &     52.00 &      1012 &     46.36 \\ 
\hspace{3 mm}Los R\'ios &        74 &     41.10 &     58.90 &      2990 &     56.49 \\ 
\hspace{3 mm}Arica and Parinacota &        31 &     35.48 &     64.52 &      1655 &     49.34 \\ \hline\hline
\multicolumn{6}{c}{ \begin{minipage}{17.5 cm}{\footnotesize{Notes: In this table I illustrate the distribution of school participation in DA by its type. Public and private voucher schools participated in DA. Private non-voucher schools did not participate in DA. The participation of students corresponds to those who applied for ninth-grade admissions. }}
\end{minipage}} \\
  \end{tabular}
\end{table}

\begin{table}[H]
\centering
 \fontsize{10}{10}\selectfont
\captionsetup{width=13cm}
\caption{School variables for tenth grade}
\label{dataschooltenth}
\begin{tabular}{p{5 cm}p{2.0 cm}p{2.0 cm}p{2.0 cm}p{2.0 cm} p{2.0 cm}}\hline  \hline
Variables& N  & mean  & sd  & min  & max  \\ \hline 
\textit{Tenth grade (2015)}&&&&&\\ 
Language            &        2856&      247.42&       30.55&         171&         340\\
mathematics         &        2859&      261.23&       46.74&         136&         387\\
\% public              &        2860&        0.30&        --&           0&           1\\
\% private voucher     &        2860&        0.57&        --&           0&           1\\
\% private nonvoucher  &        2860&        0.13&        --&           0&           1\\
\% rural     &        2860&        0.06&        --&           0&           1\\\\
\textit{Tenth grade (2016)}&&&&&\\ 
Language            &        2881&      247.87&       30.28&         176&         340\\
Mathematics         &        2881&      264.04&       46.72&         138&         391\\
\% public              &        2884&        0.30&        --&           0&           1\\
\% private voucher     &        2884&        0.56&        --&           0&           1\\
\% private nonvoucher  &        2884&        0.13&        --&           0&           1\\
\% rural     &        2884&        0.06&        --&           0&           1\\\\
\textit{Tenth grade (2017)}&&&&&\\ 
Language            &        2901&      251.89&       28.82&         169&         338\\
mathematics         &        2900&      264.71&       45.13&         149&         393\\
\% public              &        2901&        0.31&        --&           0&           1\\
\% private voucher     &        2901&        0.56&        --&           0&           1\\
\% private nonvoucher  &        2901&        0.14&        --&           0&           1\\
\% rural     &        2901&        0.06&        --&           0&           1\\
\hline\hline
\multicolumn{6}{c}{ \begin{minipage}{17.5 cm}{\footnotesize{Notes: I use tenth grade SIMCE for 2015, 2016 and 2017 and the enrolment files for these years to obtain the school variables for the school choice model.}}
\end{minipage}} \\
  \end{tabular}
\end{table}

\begin{table}[H]
\centering
 \fontsize{10}{10}\selectfont
\captionsetup{width=12 cm}
\caption{Distribution of school type and \% charging copayment}
\label{datafee}
\begin{tabular}{p{4 cm}p{2.0 cm}p{2.0 cm}p{3.0 cm}}\hline  \hline
Variables& N  & \% in total  & \% fee  \\
&(1)&(2)&(3)\\ \hline 
\textit{2015}&&&\\ 
Public &     877.0 &      30.0 &       7.3 \\ 
Private Voucher&    1671.0 &      57.1 &      77.7 \\ 
Private Non-Voucher &     380.0 &      13.0 &     100.0 \\ 
Total  &    2928.0 &     100.0 &         -- \\ \\
\textit{2016}&&&\\ 
Public&     887.0 &      30.2 &       0.6 \\  
Private  Voucher&    1657.0 &      56.4 &      57.3 \\ 
Private  Non-Voucher&     393.0 &      13.4 &     100.0 \\ 
Total &    2937.0 &     100.0 &         -- \\ \\
\textit{2017}&&&\\ 
Public &     902.0 &      30.5 &       0.4 \\ 
Private Voucher&    1659.0 &      56.0 &      53.3 \\ 
 Private Non-Voucher &     401.0 &      13.5 &     100.0 \\  
Total &    2962.0 &     100.0 &         -- \\ \\
\textit{2018}&&&\\ 
Public&     916.0 &      30.7 &       0.4 \\ 
Private Voucher&    1621.0 &      54.3 &      47.3 \\ 
Private Non-Voucher &     446.0 &      15.0 &     100.0 \\ 
Total &    2983.0 &     100.0 &         -- \\ \hline\hline
\multicolumn{4}{c}{ \begin{minipage}{12 cm}{\footnotesize{Notes: I use files on the details of schools in Chile to obtain the complete set of schools. Second, I construct the fee variable using the ficom data file that reports the fee structure of public and private non-voucher schools in Chile. The private non-voucher schools are allowed to charge any amount of fee. They are not required to give details of the fee to the Ministry of Education.}}
\end{minipage}} \\
  \end{tabular}
\end{table}

\begin{table}[H]
\centering
 \fontsize{10}{10}\selectfont
\captionsetup{width=9 cm}
\caption*{Table 1: Regular stage}
\label{pro124table1}
\begin{tabular}{p{3.5 cm}P{2.5 cm}P{2.0 cm}}\hline  \hline
 Region& Participants (N)  & Enrollment$==$Assignment (\% ) \\ \hline  
 \textit{2017}&&\\
 Magallanes &      1029 &     74.34 \\ \\
 \textit{2018} &&\\
tarapaca &      1664 &     76.62 \\ 
Coquimbo &      4655 &     75.32 \\ 
Ohiggins &      7670 &     74.81 \\ 
loslagos &      7182 &     76.65 \\ 
Magallanes &      1010 &     75.94 \\  \\
 \textit{2019}&&\\
tarapaca &      1827 &     71.81 \\ 
antofagasta &      4446 &     74.85 \\ 
atacama &      2830 &     75.09 \\  
coquimbo &      4548 &     71.92 \\  
valparaiso &     10008 &     75.70 \\  
ohiggins &      7492 &     73.75 \\ 
maule &      8851 &     78.93 \\ 
biobio &     11597 &     79.34 \\  
araucania &      8550 &     81.78 \\  
loslagos &      7242 &     70.52 \\ 
aysen &       471 &     74.73 \\  
magallanes &       996 &     75.40 \\  
rios &      3055 &     79.18 \\  
ayp &      1651 &     77.83 \\
 \hline\hline
\multicolumn{3}{c}{ \begin{minipage}{10 cm}{\footnotesize{Notes: The program was sequentially implemented in Chile.}}
\end{minipage}} \\
  \end{tabular}
\end{table}

\newpage
\begin{table}[H]
\centering
\captionsetup{width=13cm}
\caption{Descriptive Statistics: Magallanes in 2016}
\label{magallanes1}
\begin{tabular}{p{7 cm}p{1.5 cm}p{1.5 cm}p{1.5 cm}p{1.5 cm} p{1.5 cm}}\hline  \hline
Variables            &       N&        Mean&         Std. dev.&         Min.&         Max.\\
\hline
\textit{A. Student characteristics}&&&&&\\
Math score          &         499&      235.13&       42.65&         134&         368\\
Language score      &         499&      231.76&       49.09&         112&         363\\
Mother's education  &         499&       11.57&        3.11&           0&          19\\
Father's education  &         499&       11.41&        3.03&           3&          19\\
Household income index&         499&        4.74&        2.31&           1&          15\\
 Number of ranked schools& 499   &   4.10 &  1.54 &         2  &       11\\\\
\textit{B. Top choice school characteristics}&&&&&\\
Math score          &         499&      252.19&       40.50&         191&         328\\
Language score      &         499&      242.70&       29.47&         198&         299\\
Household income index&         499&        5.66&        1.27&           4&           9\\
Mother's education  &         499&       11.89&        1.40&          10&          15\\
Father's education  &         499&       11.87&        1.55&          10&          16\\
Travel distance (kms)&         499&        2.41&        1.36&           0&           7\\
\hline\hline
\multicolumn{6}{c}{ \begin{minipage}{17.5 cm}{\footnotesize{Notes: This table provides information on eighth-graders who applied for ninth-grade admission in 2016 in Magallanes. I obtain their background variables and test scores from sixth grade SIMCE. Travel distance has been computed using OSRM API.}}
\end{minipage}} \\
  \end{tabular}
\end{table}

\newpage

\begin{table}[H]
 \fontsize{10}{10}\selectfont
\centering
\captionsetup{width=12 cm}
\caption{Descriptive Statistics by Region}
\label{student2017summary}
\begin{tabular}{p{9.5 cm}P{0.5 cm}P{1.0 cm}p{1.0 cm}P{1.0 cm}P{1.0 cm}}\hline  \hline
                    &      N&        Mean&          Std. dev.&         Min&         Max\\
  \hline 
 \textit{Tarapaca}&&&&&\\ 
Pre-reform Math (Standardized)&         858&       -0.09&        1.05&   -2.41&           3\\
Pre-reform Language (Standardized)&         858&       -0.08&        1.03&   -2.58&           3\\
Mother's education  &         858&       11.30&        3.31&           0&          22\\
Father's education  &         858&       11.27&        3.25&           0&          19\\
Income Index              &         858&        4.87&        2.59&           1&          15\\
Avg. academic score of outside option (Standardized)&         858&       -0.74&        0.52&    -1.76&           1\\
 \textit{Coquimbo}&&&&&\\ 
Pre-reform Math (Standardized)&        2693&       -0.06&        1.00&   -2.44&           3\\
Pre-reform Language (Standardized)&        2693&       -0.01&        1.00&   -2.63&           3\\
Mother's education  &        2693&       10.77&        3.33&           0&          22\\
Father's education  &        2693&       10.54&        3.41&           0&          22\\
Income Index              &        2693&        4.06&        2.23&           1&          15\\
Avg. academic score of outside option (Standardized)&        2693&       -0.43&        0.60&   -1.60&           2\\
 \textit{O'Higgins}&&&&&\\ 
Pre-reform Math (Standardized)&        4765&        0.03&        0.99&   -2.45&           3\\
Pre-reform Language (Standardized)&        4765&       -0.01&        1.01&   -2.86&           3\\
Mother's education  &        4765&       10.55&        3.34&           0&          19\\
Father's education  &        4765&       10.28&        3.44&           0&          22\\
Income Index            &        4765&        4.08&        2.03&           1&          15\\
Avg. academic score of outside option (Standardized)&        4765&       -0.59&        0.43&   -1.75&           2\\
 \textit{Los Lagos}&&&&&\\ 
Pre-reform Math (Standardized)&        4203&        0.04&        1.01&   -2.48&           3\\
Pre-reform Language (Standardized)&        4203&        0.03&        0.99&   -2.84&           3\\
Mother's education  &        4203&       10.15&        3.59&           0&          22\\
Father's education  &        4203&        9.95&        3.41&           0&          22\\
Income Index              &        4203&        3.92&        2.09&           1&          15\\
Avg. academic score of outside option (Standardized)&        4203&       -0.58&        0.49&   -1.49&           2\\
 \textit{Magallanes}&&&&&\\ 
Pre-reform Math (Standardized)&         544&       -0.03&        0.97&   -2.27&           3\\
Pre-reform Language (Standardized)&         544&        0.01&        0.97&   -2.46&           3\\
Mother's education  &         544&       11.67&        3.11&           1&          19\\
Father's education  &         544&       11.52&        3.18&           0&          17\\
Income Index              &         544&        5.67&        2.88&           1&          15\\
Avg. academic score of outside option  (Standardized)&         544&       -0.66&        0.40&   -1.00&           2\\
 \hline\hline
\multicolumn{6}{c}{ \begin{minipage}{15 cm}{\footnotesize{Notes: The program was sequentially implemented in Chile. In 2017 the program was introduced in a total of 5 regions.}}
\end{minipage}} \\
  \end{tabular}
\end{table}
\newpage
\begin{table}[H]
\centering
 \fontsize{8}{8}\selectfont
\captionsetup{width=15 cm}
\caption{Joint Analysis: Summary Statistics}
\label{resultstable_allregions}
\begin{tabular}{p{4.4 cm}P{1.0 cm}P{1.0 cm}P{1.0 cm}P{1.0 cm}P{1.0 cm}}\hline   \hline
                   &      N&        Mean&          Std. dev.&         Min&         Max\\
\hline
Pre-reform Math (Standardized)&        7030&        0.02&        1.01&   -2.44&           3\\
Pre-reform Language (Standardized)&        7030&        0.03&        0.99&   -2.84&           3\\
Mother's education  &        7030&       10.64&        3.42&           0&          22\\
Father's education  &        7030&       10.39&        3.40&           0&          22\\
income              &        7030&        4.14&        2.28&           1&          15\\
Avg. academic score of outside option school (Standardized)&        7030&       -0.49&        0.53&    -1.76&           2\\
N       &        7030&            &            &            &            \\
J       &        286&            &            &            &            \\
\hline\hline
\multicolumn{6}{c}{ \begin{minipage}{10 cm}{\footnotesize{}}
\end{minipage}} \\
  \end{tabular}
\end{table}
\newpage 
\subsection*{Figures}
\begin{figure}[H] 
  \caption{Simulation results: Multiple covariates without individual heterogeneity}
  \centering
    \label{simulationnolatent} 
  \begin{minipage}[b]{0.33\linewidth}
    \centering
    \includegraphics[width=\linewidth]{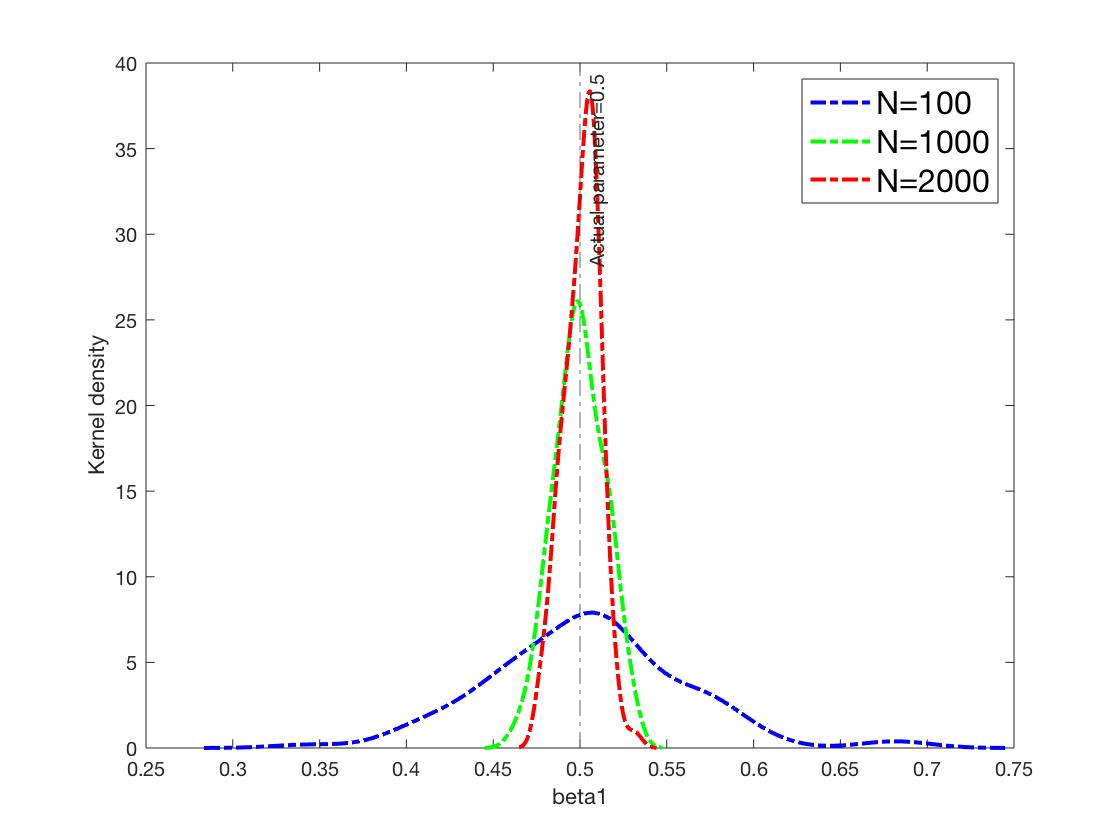} 
        \caption*{A. $\beta_{4}=0.5$}
    \vspace{0ex}
  \end{minipage}%%
  \begin{minipage}[b]{0.33\linewidth}
    \centering
    \includegraphics[width=\linewidth]{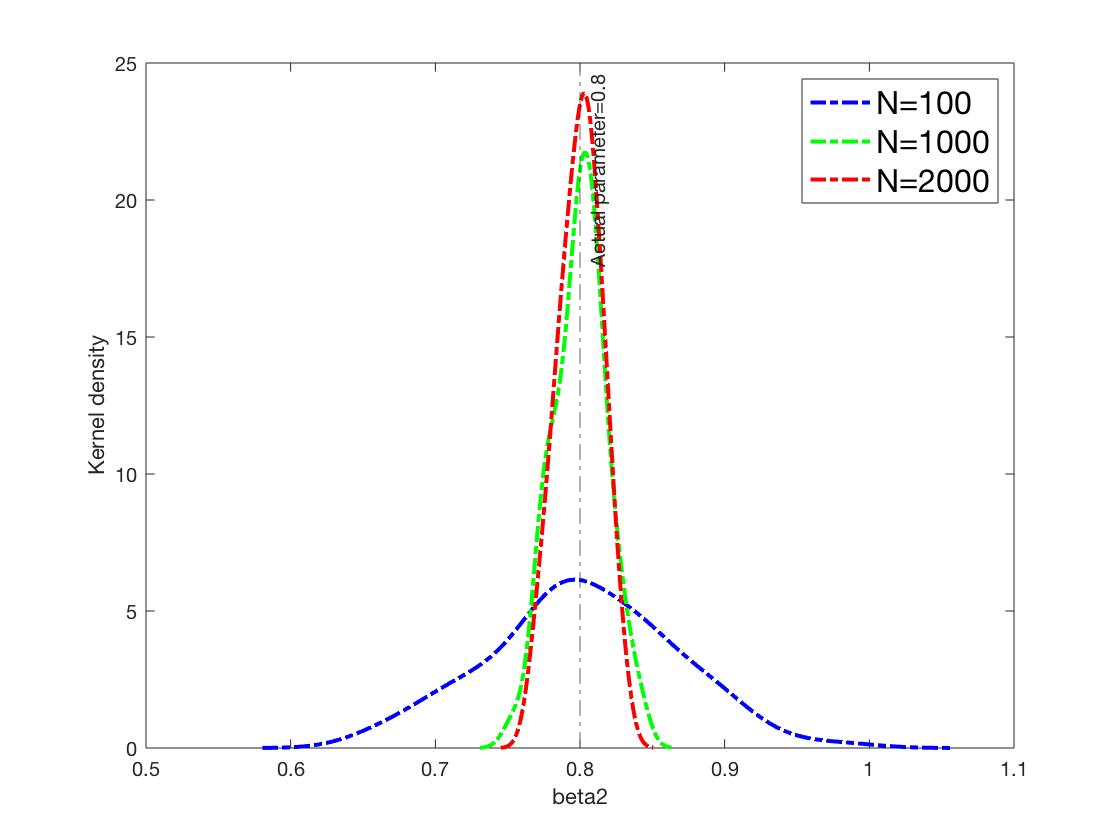} 
    \caption*{B. $\beta_{2}=0.8$}
    \vspace{0ex}
  \end{minipage} 
  \begin{minipage}[b]{0.33\linewidth}
    \centering
    \includegraphics[width=\linewidth]{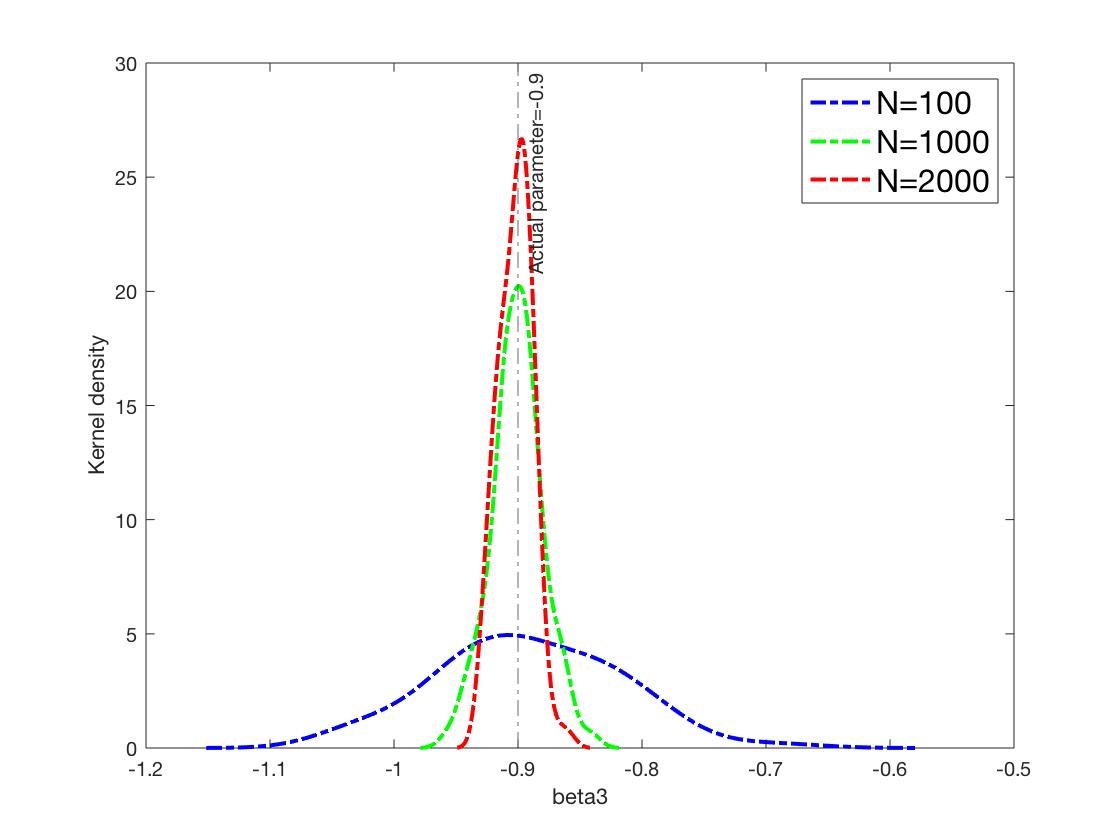} 
        \caption*{C. $\beta_{3}=-0.9$}
    \vspace{0ex}
  \end{minipage}%%
  \begin{minipage}[b]{0.33\linewidth}
    \centering
    \includegraphics[width=\linewidth]{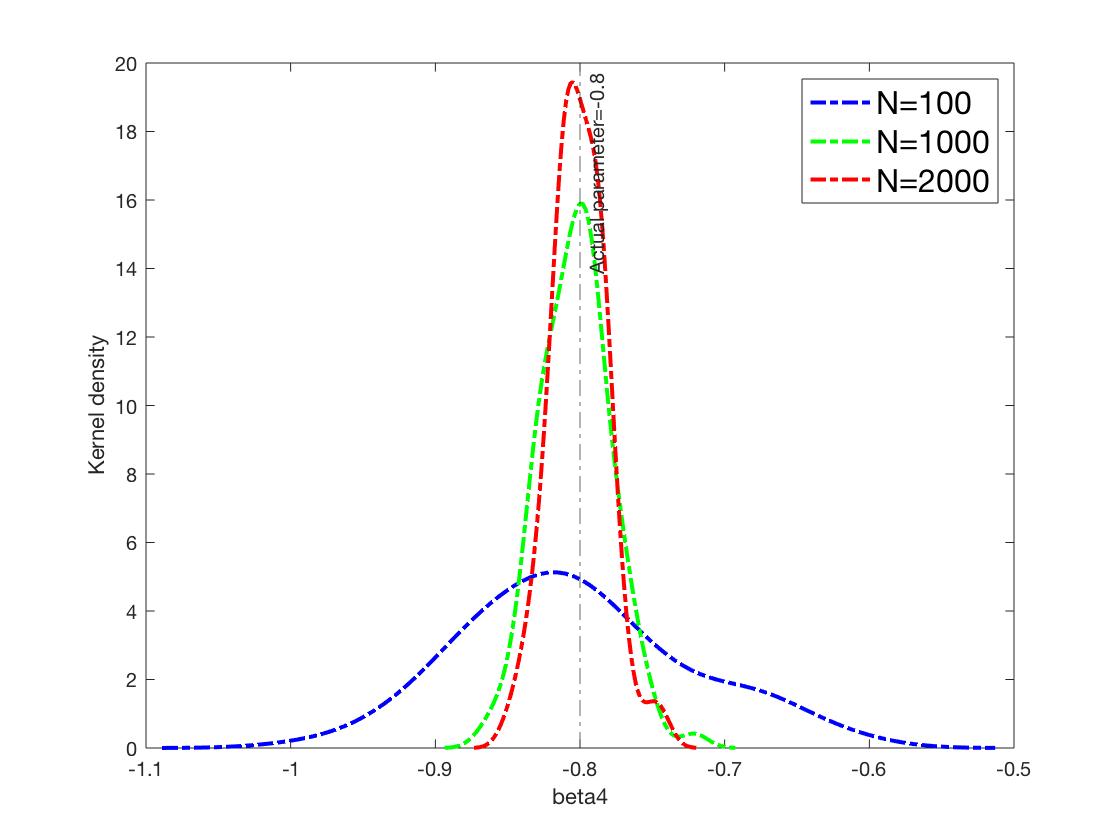} 
            \caption*{D. $\beta_{4}=-0.8$}
    \vspace{0ex}
  \end{minipage}
      \begin{minipage}[b]{0.33\linewidth}
    \centering
    \includegraphics[width=1.0\linewidth]{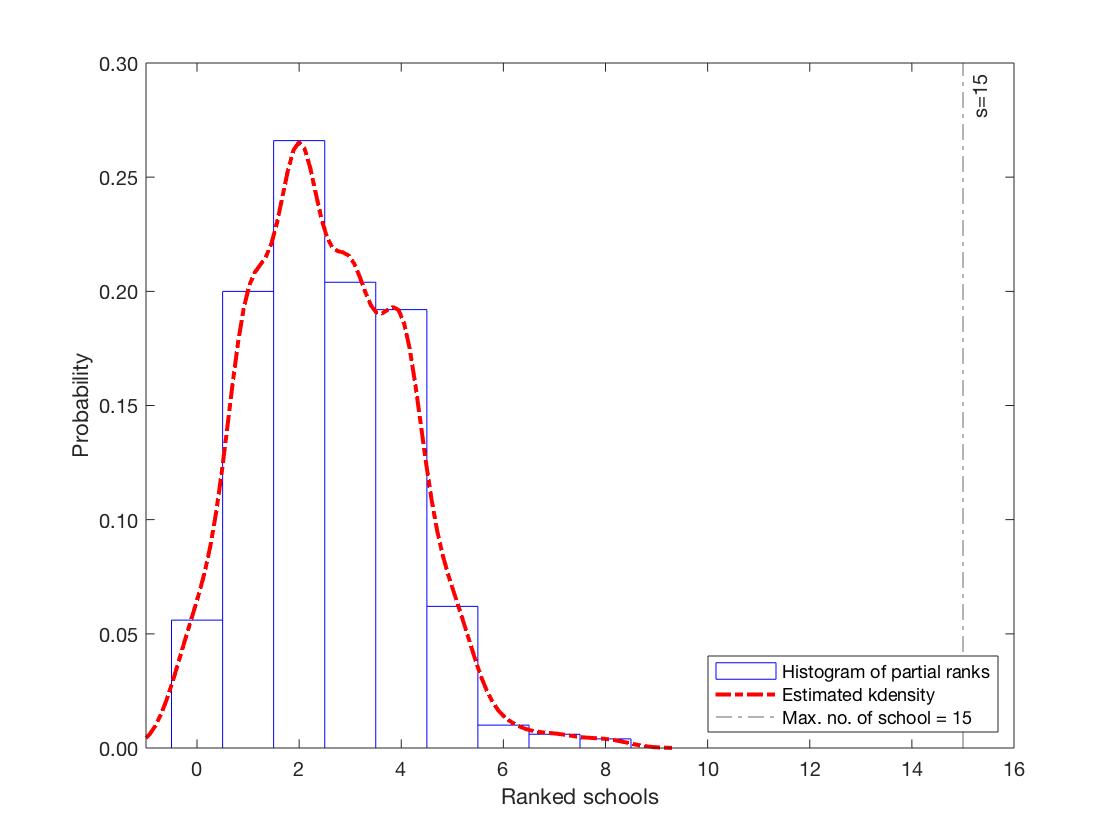} 
            \caption*{E. Rank}
    \vspace{0ex}
  \end{minipage}
  \begin{minipage}{17.5 cm}{\footnotesize{Notes: These graphs display the smoothed kernel density plots of the empirical distribution of the parameter estimates in repeated MC samples. The number of schools for this analysis is fixed at 15. }}
\end{minipage} 
\end{figure}

\newpage

\begin{figure}[H] 
  \caption{Simulation results: Multiple covariates without individual heterogeneity}
  \centering
    \label{simulationlatent} 
  \begin{minipage}[b]{0.33\linewidth}
    \centering
    \includegraphics[width=\linewidth]{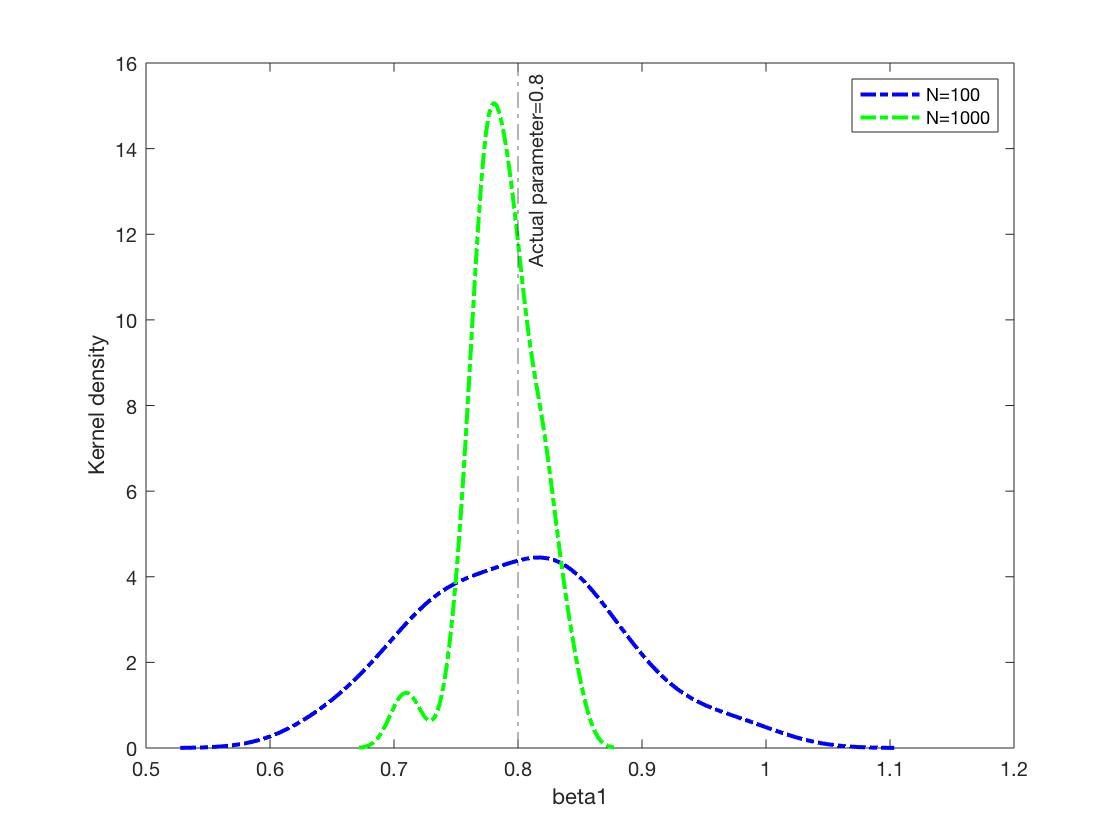} 
        \caption*{A. $\beta_{4}=0.8$}
    \vspace{0ex}
  \end{minipage}%%
  \begin{minipage}[b]{0.33\linewidth}
    \centering
    \includegraphics[width=\linewidth]{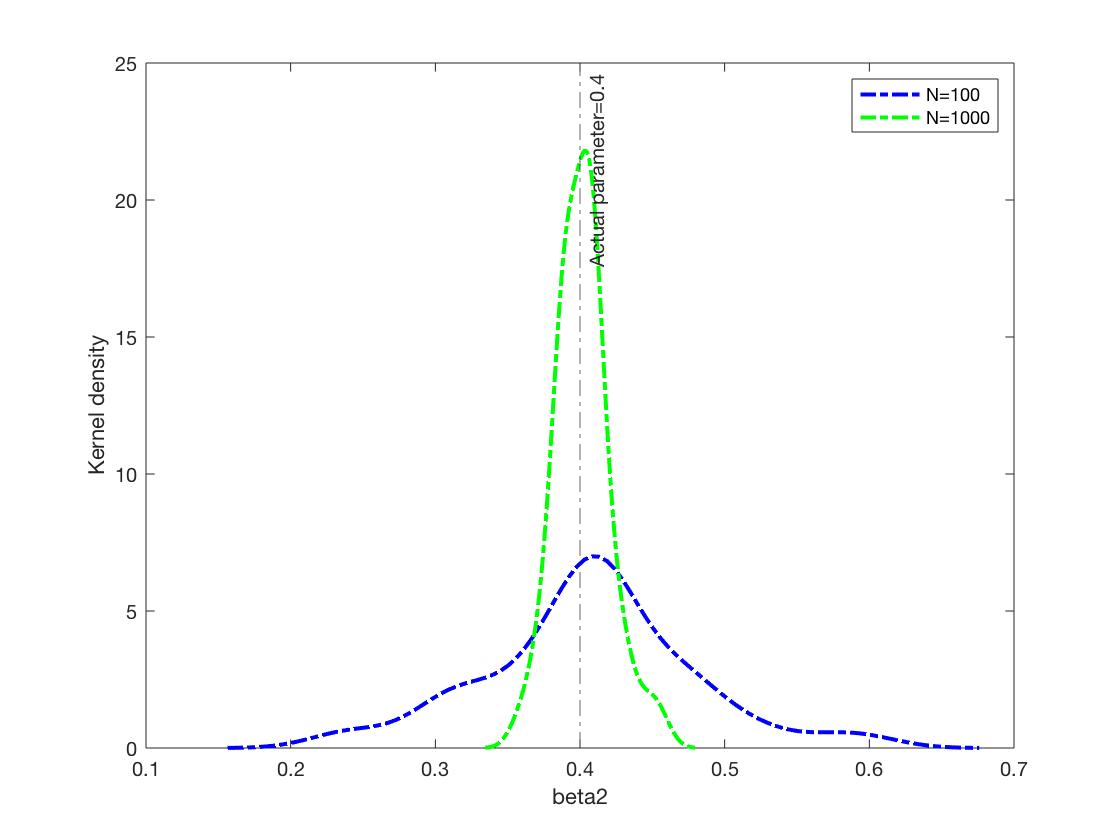} 
    \caption*{B. $\beta_{2}=0.4$}
    \vspace{0ex}
  \end{minipage} 
  \begin{minipage}[b]{0.33\linewidth}
    \centering
    \includegraphics[width=\linewidth]{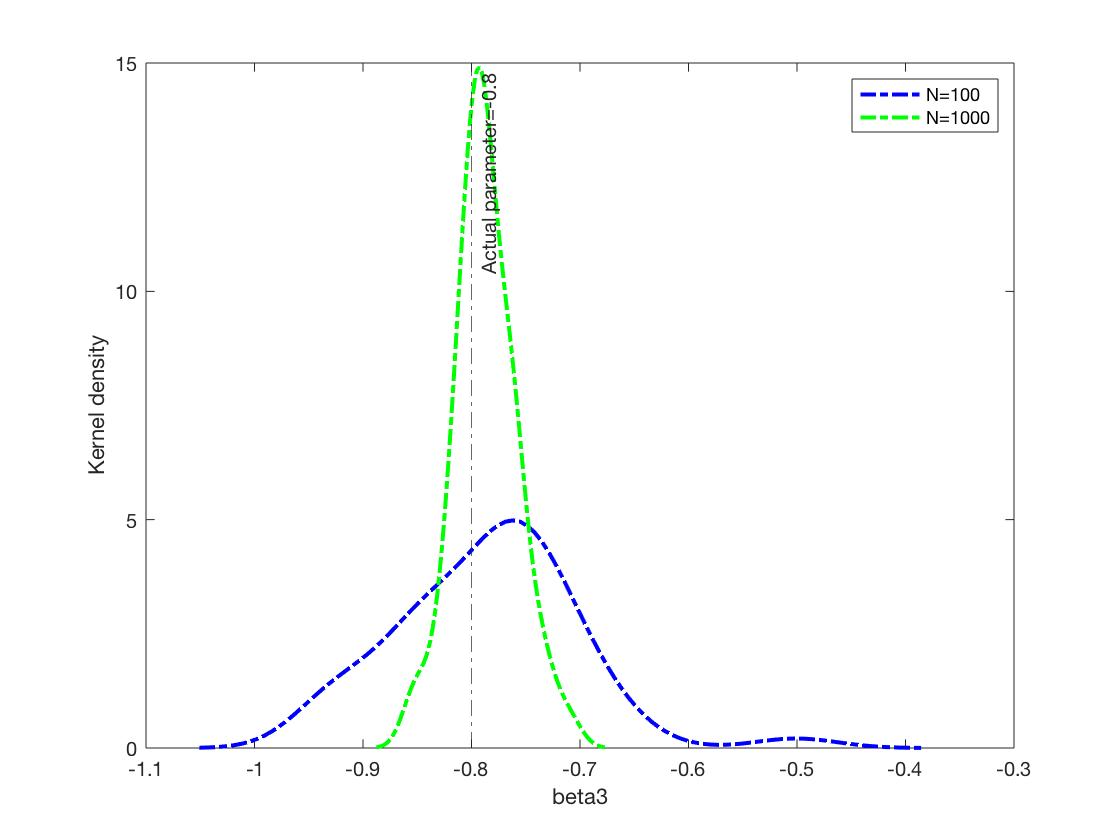} 
        \caption*{C. $\beta_{3}=-0.8$}
    \vspace{0ex}
  \end{minipage}
  \begin{minipage}[b]{0.33\linewidth}
    \centering
    \includegraphics[width=\linewidth]{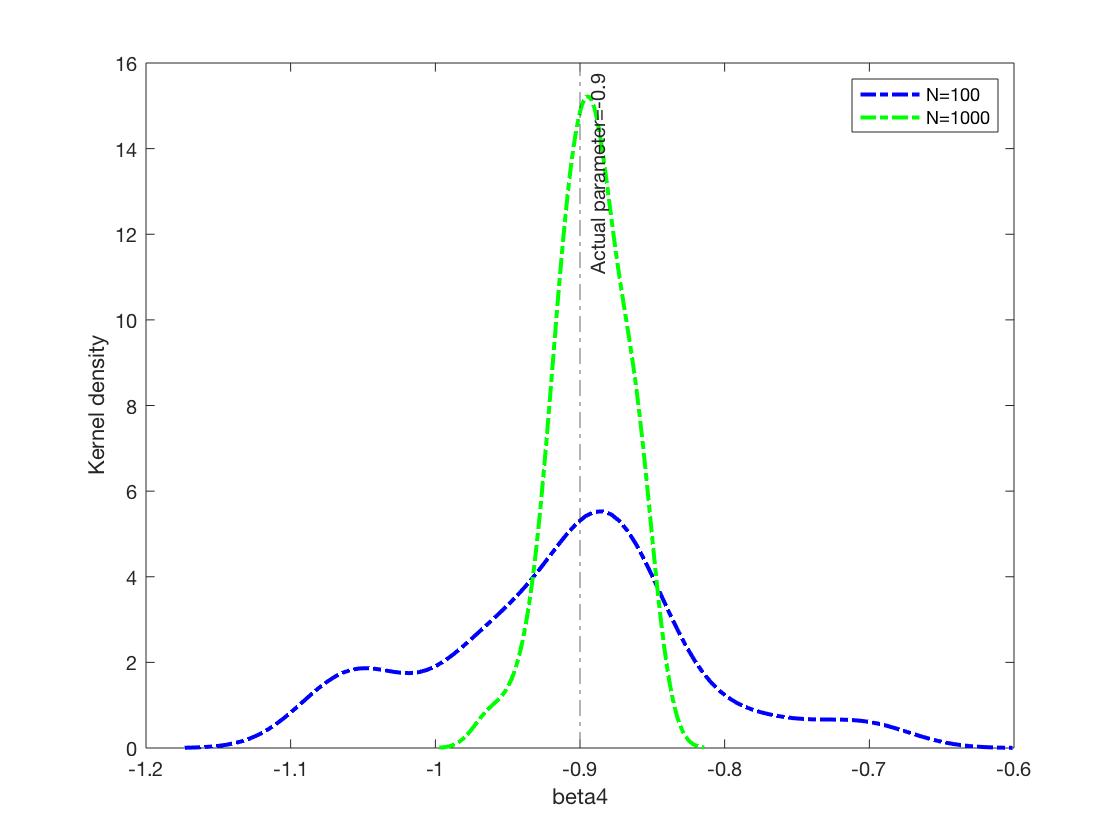} 
            \caption*{D. $\beta_{4}=-0.9$}
    \vspace{0ex}
  \end{minipage}
    \begin{minipage}[b]{0.33\linewidth}
    \centering
    \includegraphics[width=\linewidth]{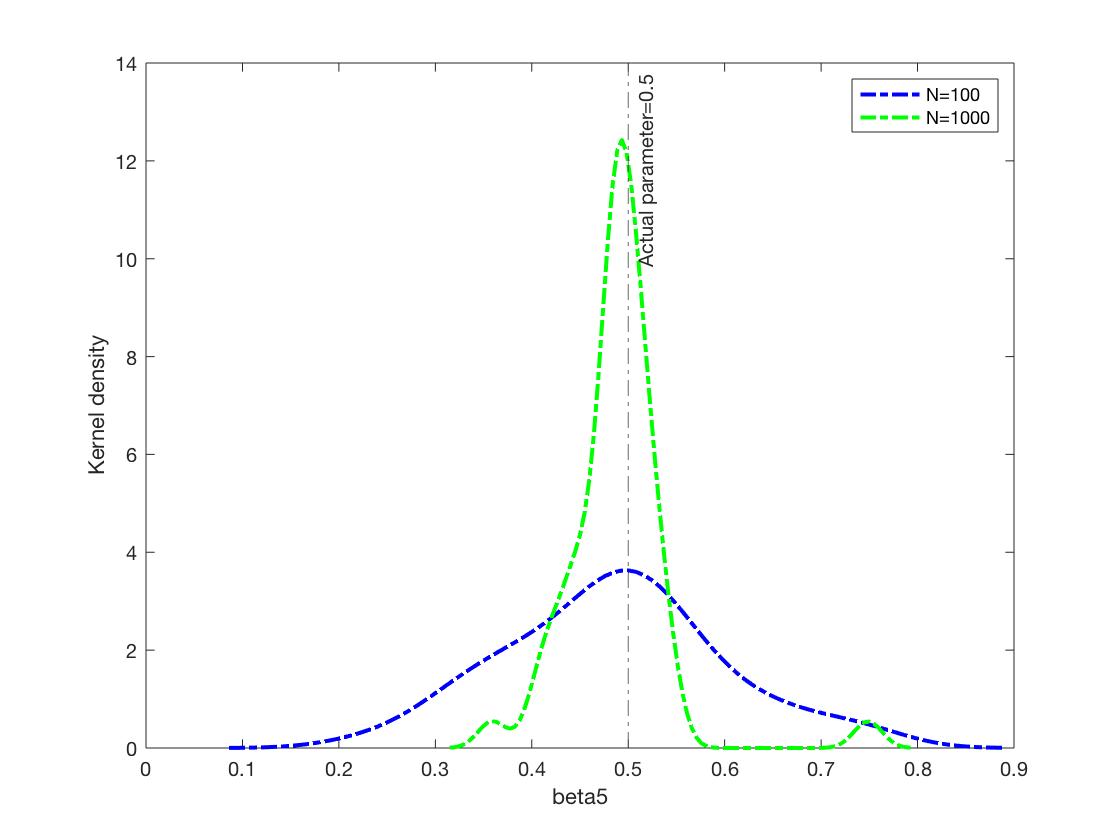} 
            \caption*{E. $\gamma=0.5$}
    \vspace{0ex}
  \end{minipage}
  \begin{minipage}{17.5 cm}{\footnotesize{Notes: These graphs display the smoothed kernel density plots of the empirical distribution of the parameter estimates in repeated MC samples.}}
\end{minipage} 
\end{figure}

\newpage

\begin{figure}[H] 
  \caption{Comparison of threshold rank model with urn model}
    \centering
    \label{dataurn} 
  \begin{minipage}[b]{0.33\linewidth}
    %\centering
    \includegraphics[width=\linewidth]{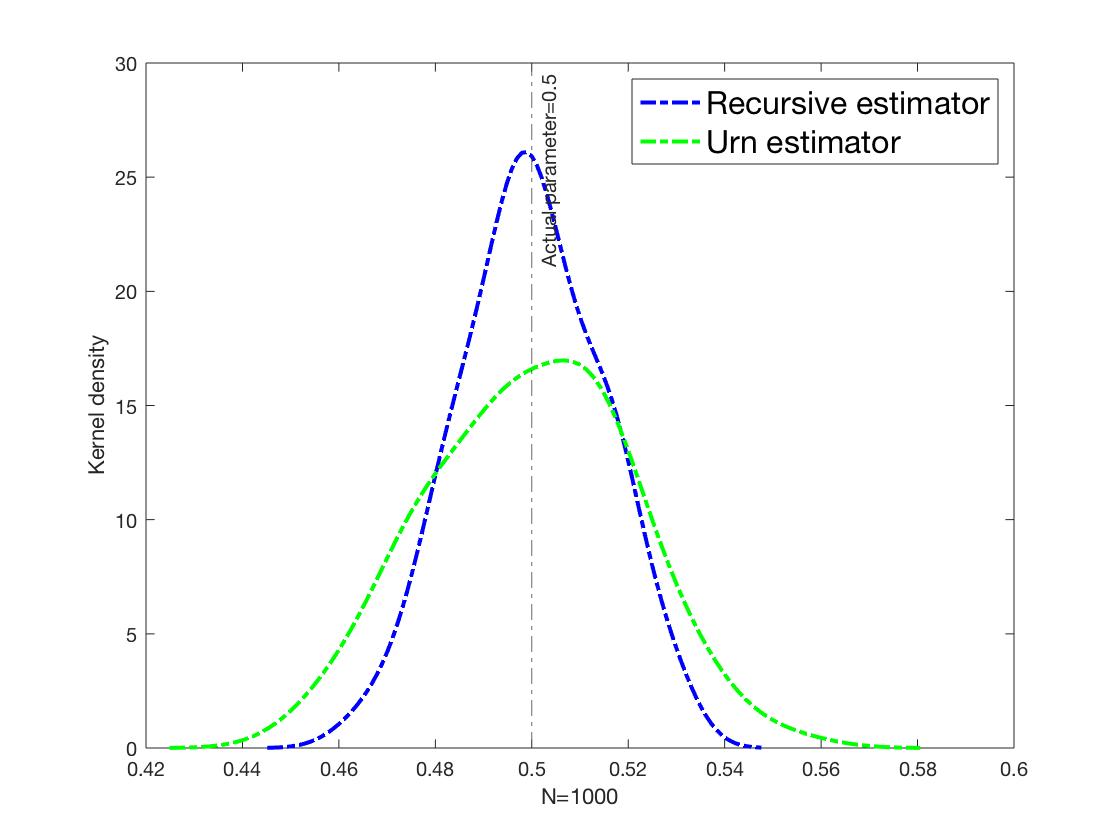}
    \caption*{A. $\beta_{1}=0.5$} 
        \vspace{0ex}
  \end{minipage}%%
  \begin{minipage}[b]{0.33\linewidth}
    %\centering
    \includegraphics[width=\linewidth]{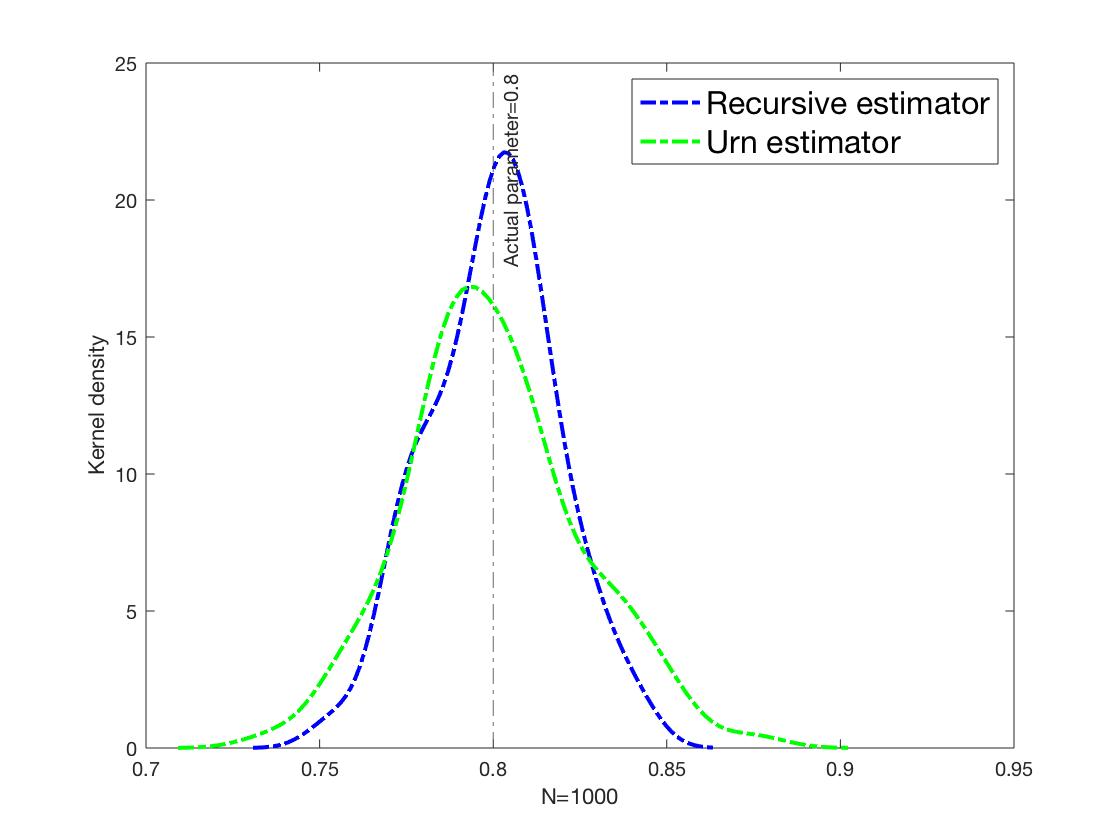} 
       \caption*{B. $\beta_{2}=0.8$} 
           \vspace{0ex}
  \end{minipage} 
  \begin{minipage}[b]{0.33\linewidth}
    %\centering
      \includegraphics[width=\linewidth]{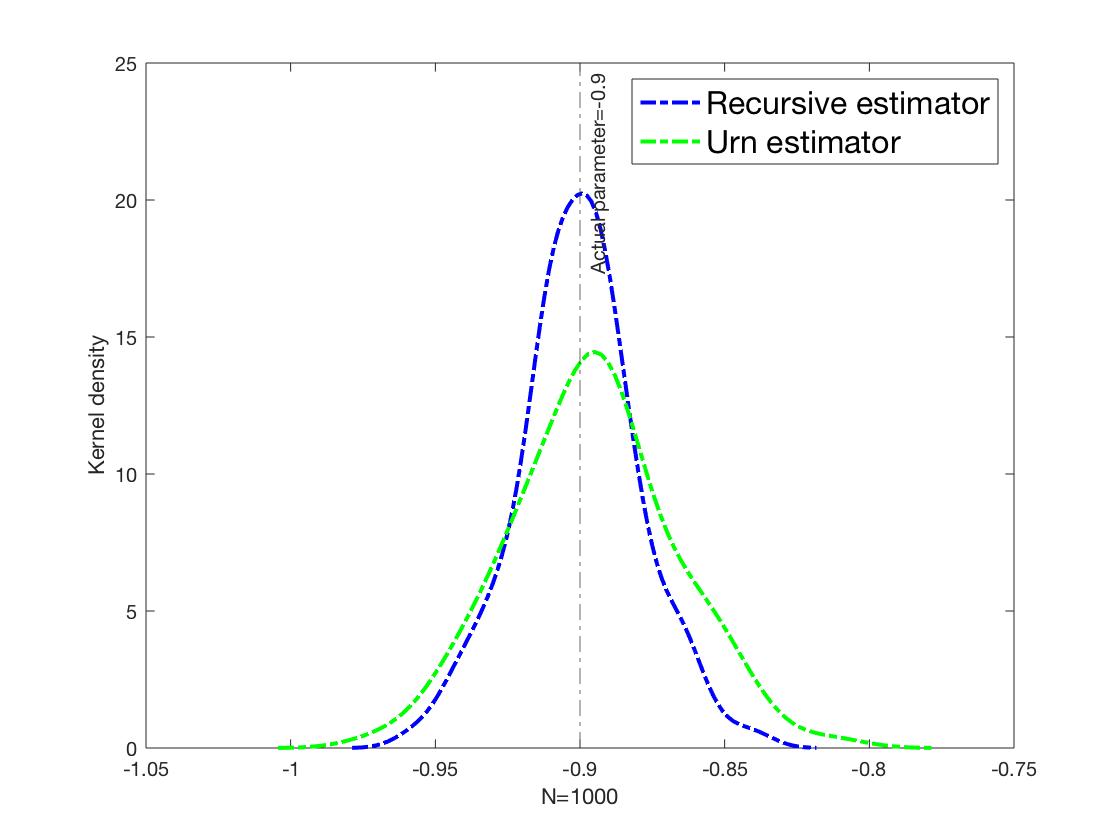}
       \caption*{C. $\beta_{3}=-0.9$} 
           \vspace{0ex}
  \end{minipage} 
    \begin{minipage}[b]{0.33\linewidth}
    %\centering
      \includegraphics[width=\linewidth]{urn3.jpg}
       \caption*{D. $\beta_{4}=-0.8$} 
           \vspace{0ex}
  \end{minipage} 
  \begin{minipage}{17.5 cm}{\footnotesize{Notes: These graphs display the smoothed kernel density plots of the empirical distribution of the parameter estimates in repeated MC samples.}}
\end{minipage} 
  \end{figure}
\newpage 
\begin{figure}[H] 
  \caption{Segregation (Income) by school type}
  \centering
    \label{seg20152016} 
  \begin{minipage}[b]{0.35\linewidth}
    %\centering
    \includegraphics[width=1.0\linewidth]{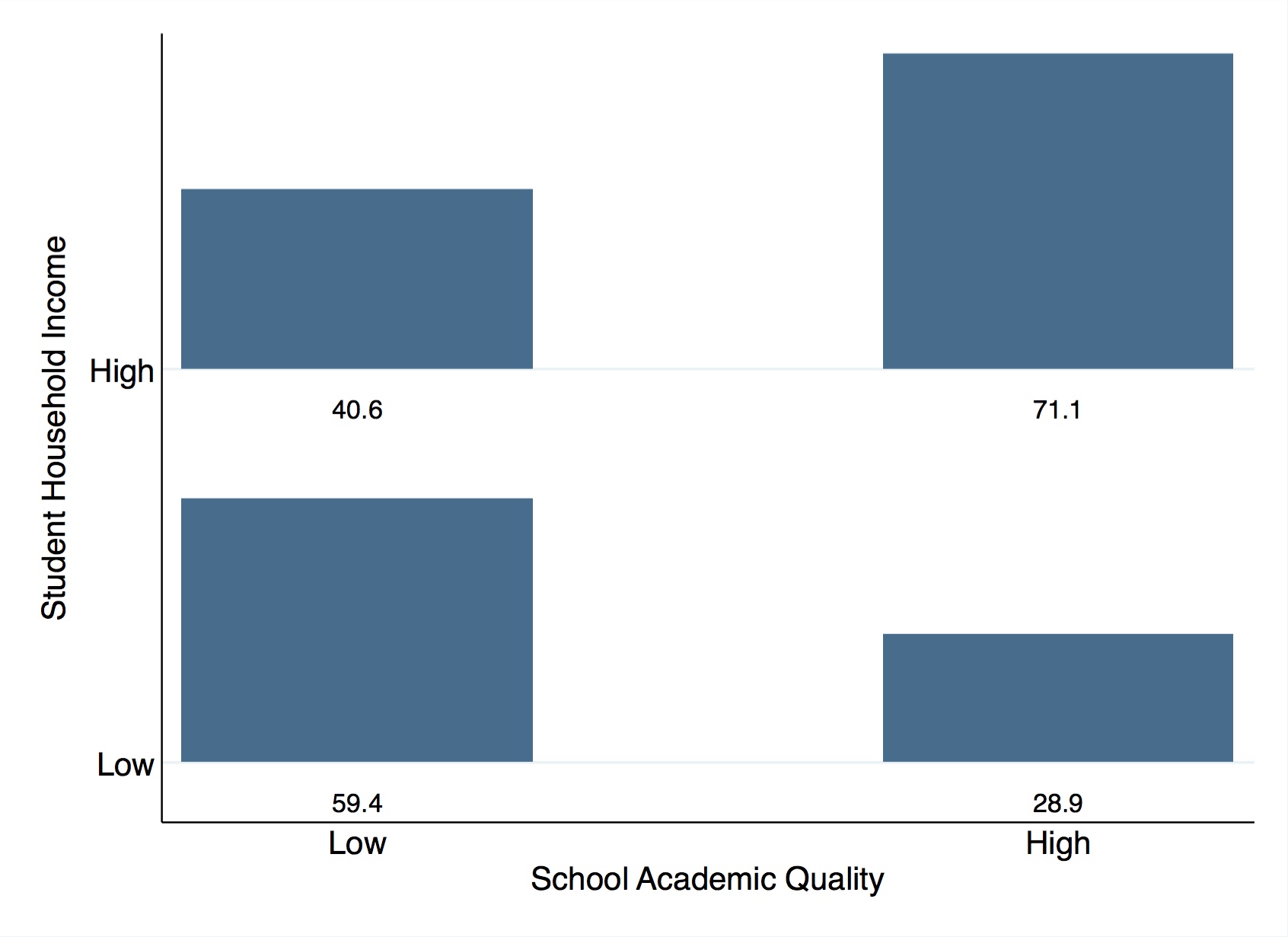}
    \caption*{A. 2015 (Before)} 
  \end{minipage}%%
  \begin{minipage}[b]{0.35\linewidth}
    %\centering
    \includegraphics[width=1.0\linewidth]{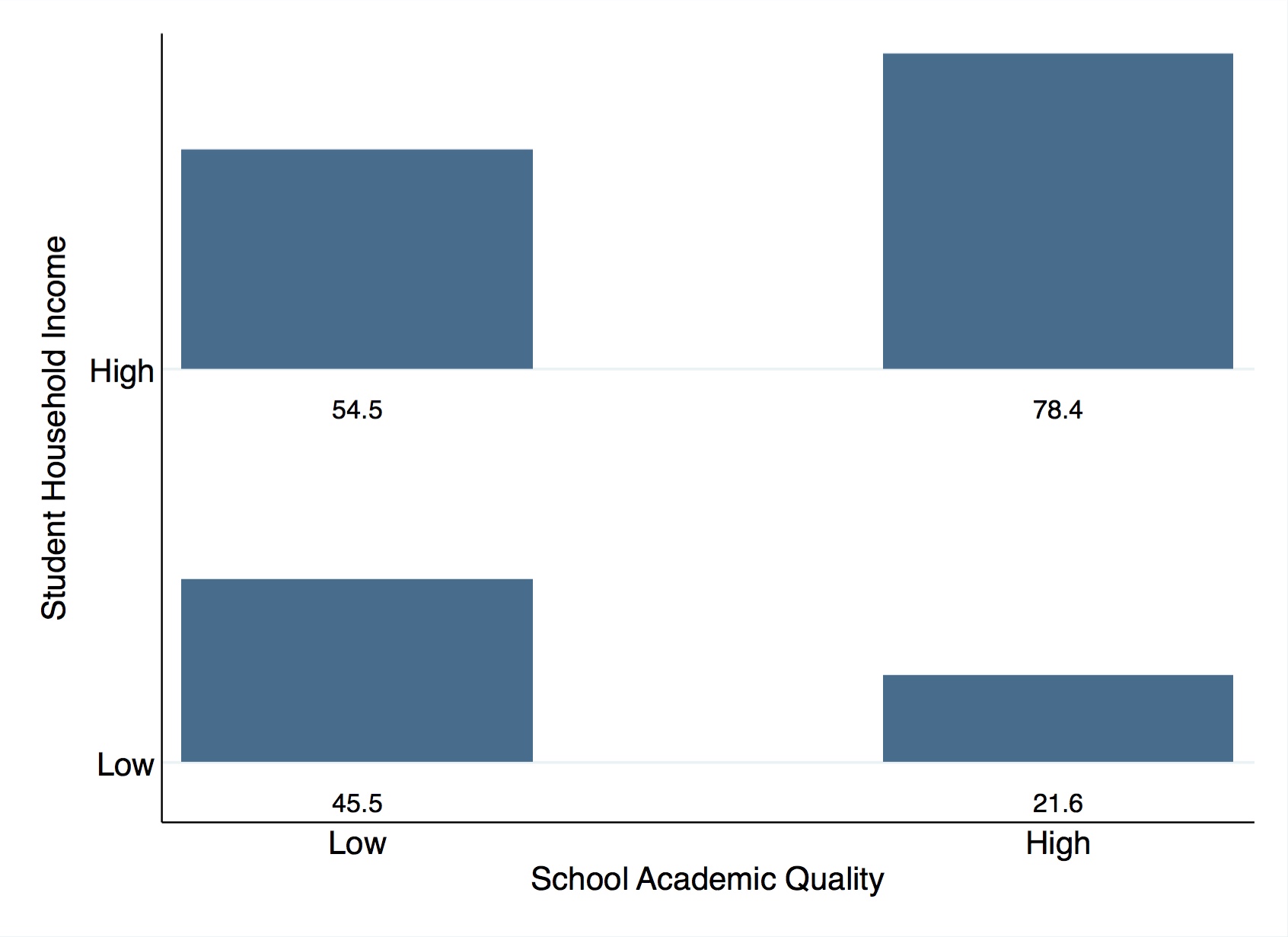} 
        \caption*{D. 2018 (After)} 
  \end{minipage} 
    \begin{minipage}{17.5 cm}{\footnotesize{}}
\end{minipage} 
\begin{minipage}{14.5 cm}{\footnotesize{Notes: These graphs display the composition of low income and high income students in high and low ability school types. For each school type I calculate the \% of students who are low income and the \% students who are high income. Students with low income have a monthly income of less than CLP 300,000. Low income students have a much lower representation in high score schools compared to their representation in low score schools.}}
\end{minipage} 
\end{figure}

\newpage

\begin{figure}[H] 
    \centering

  \caption{Attendance and school dummies: OLS}
    \label{attendanceschooltype} 
    \begin{minipage}[b]{0.43\linewidth}
    \includegraphics[width=\linewidth]{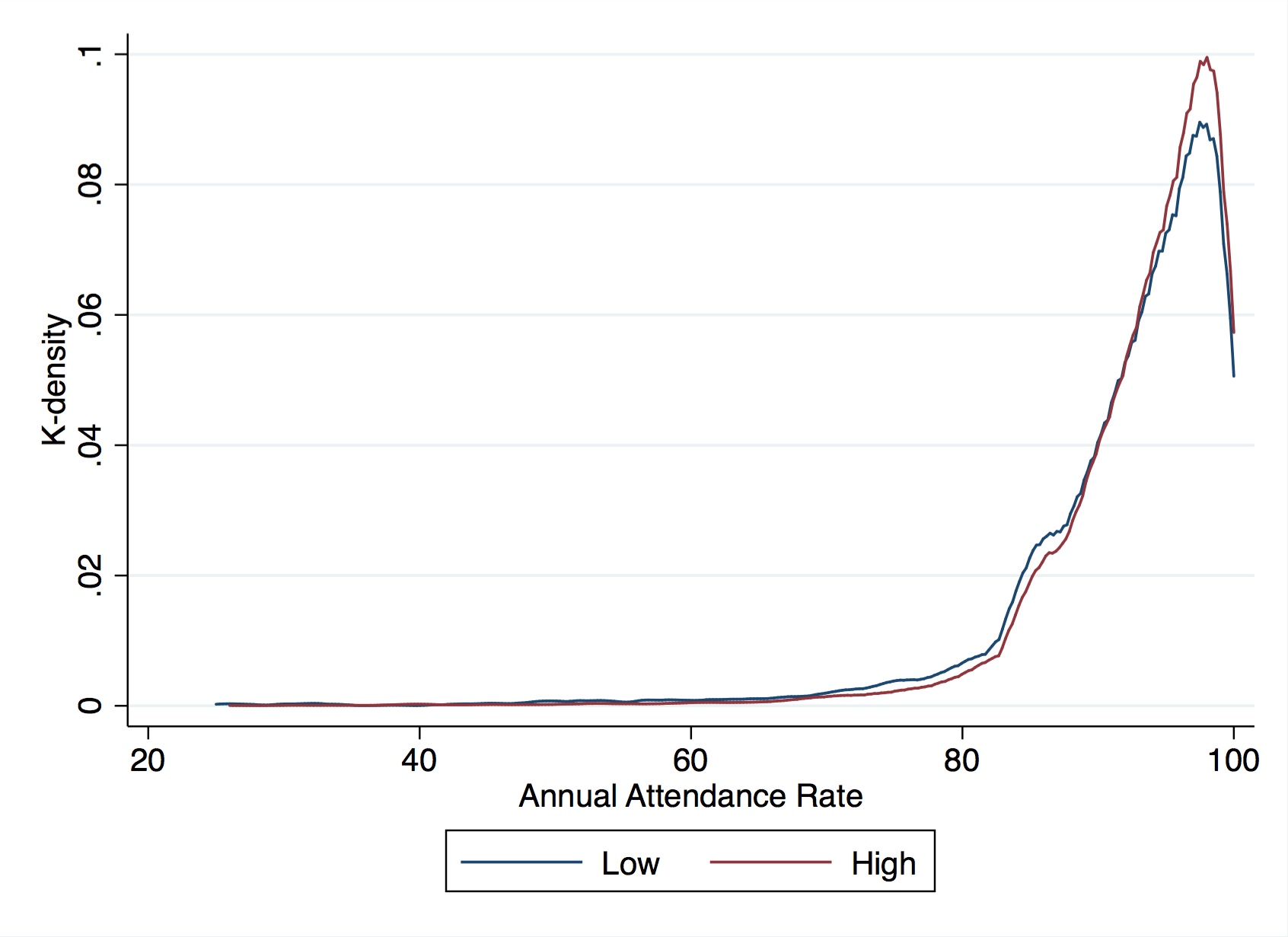} 
       \caption*{A. Density by student income} 
              \end{minipage} 
  \begin{minipage}[b]{0.43\linewidth}
    \includegraphics[width=\linewidth]{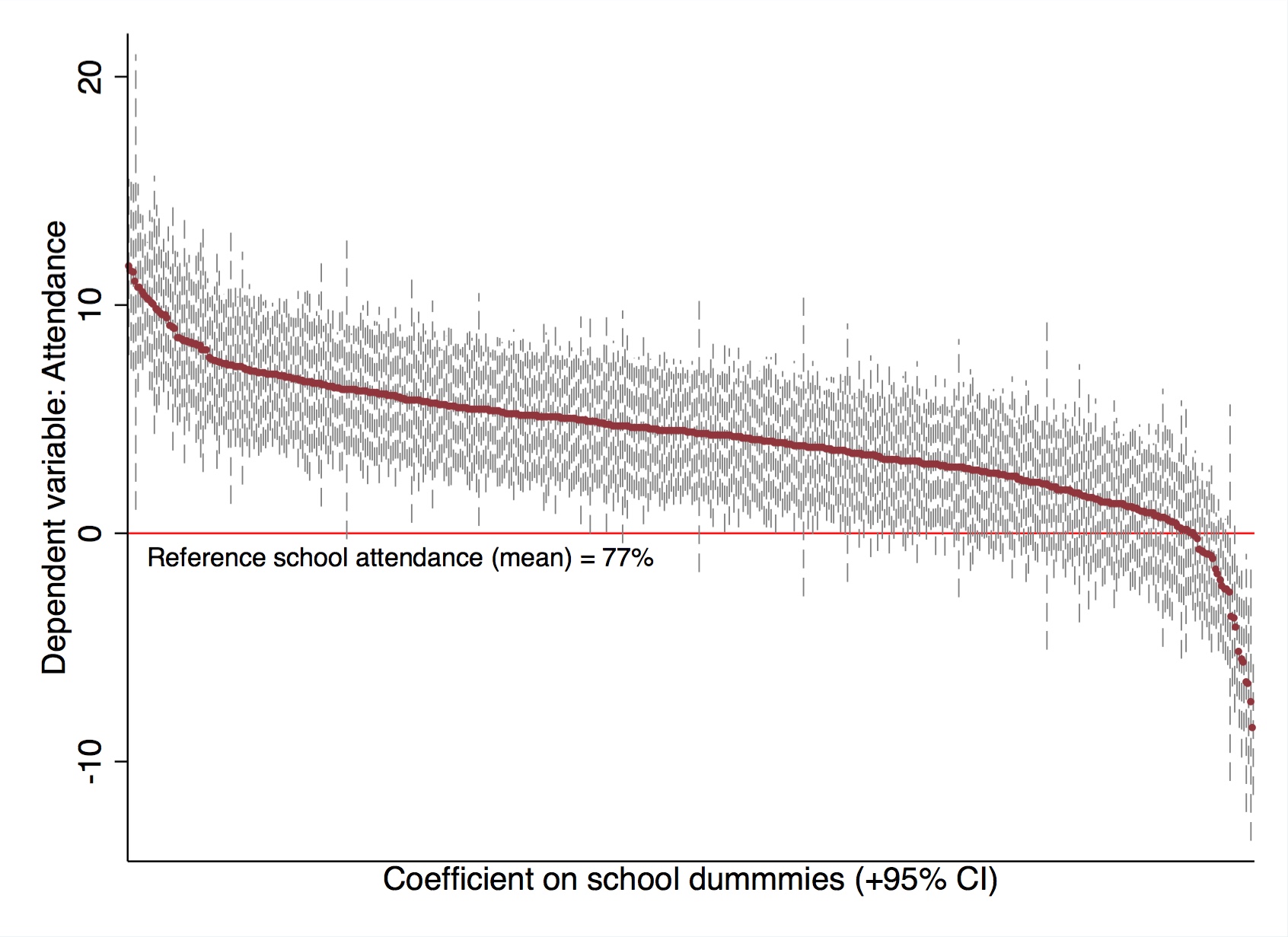} 
       \caption*{B. School dummies and attendance} 
              \end{minipage} 
   \begin{minipage}[b]{0.43\linewidth}
    \includegraphics[width=\linewidth]{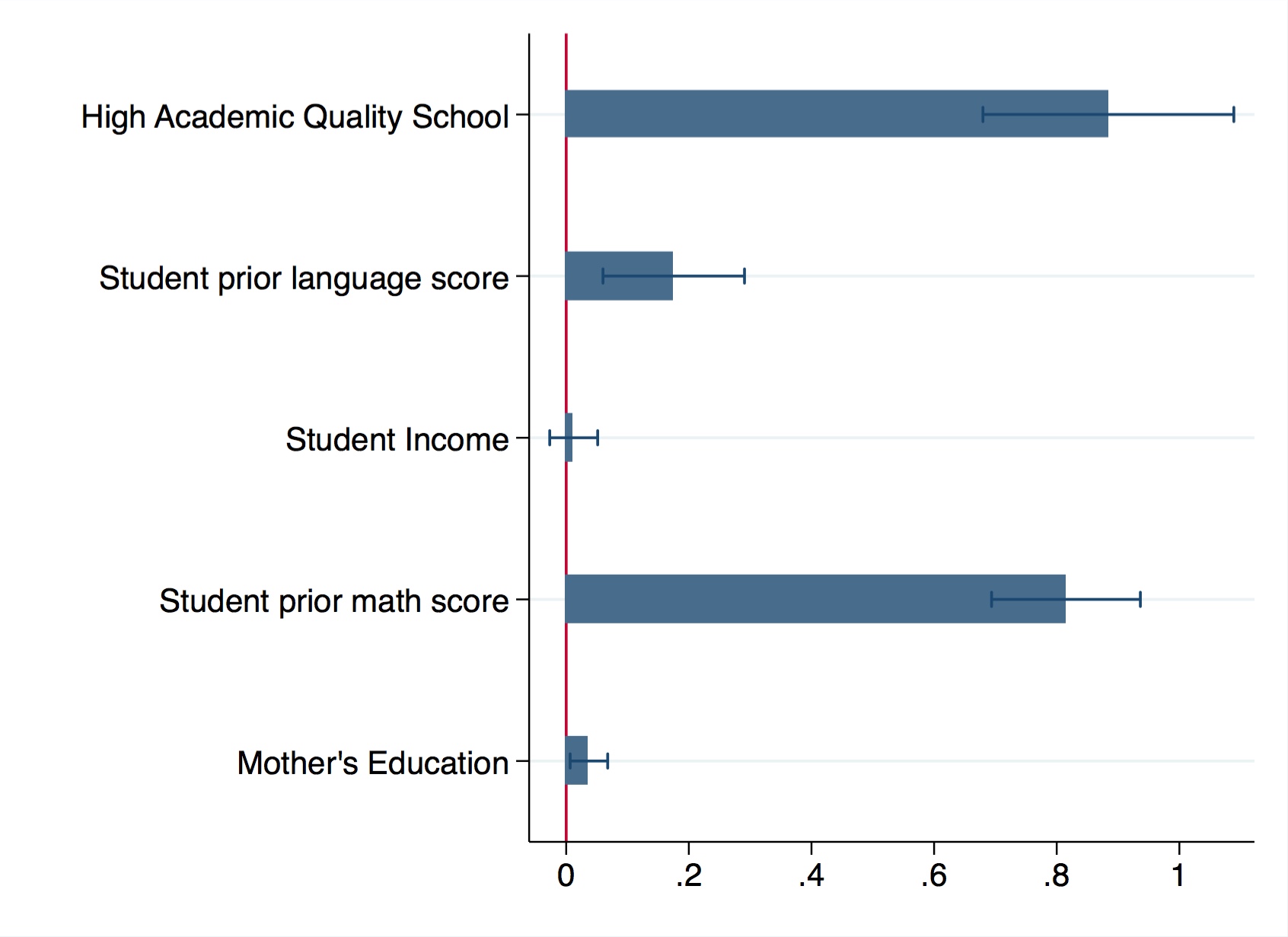} 
       \caption*{C. Attendance and school type} 
              \end{minipage} 
              \begin{minipage}{14.5 cm}{\footnotesize{Notes: The graph in panel (A) illustrates the kernel density plots for annual attendance rates by student income. Students with low income have a monthly income of less than CLP 300,000. Panel (B) plots the coefficients of a regression of attendance on school dummies and student covariates measuring academic ability, income and socio-economic status. The graph in panel (B) displays the differences in attendance rate by school quality. I control for observed student characteristics in this regression as well.}}
\end{minipage} 
  \end{figure}

\newpage

\begin{figure}[H] 
  \caption{School assignment of students that participated in DA, Magallanes in 2016}
  \centering
    \label{dataspatialall} 
    \begin{minipage}[b]{0.5\linewidth}
    \includegraphics[width=\linewidth]{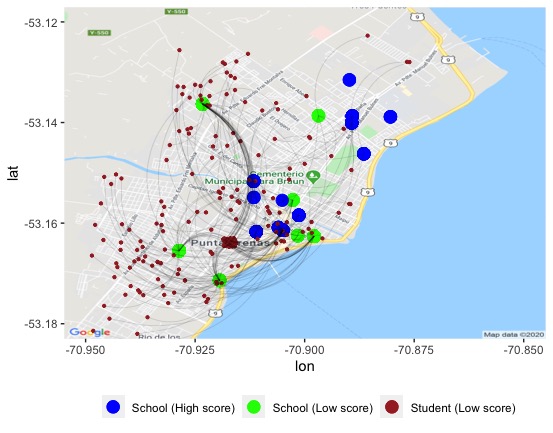}
    \caption*{A. Low ability } 
  \end{minipage}%%
    \begin{minipage}[b]{0.5\linewidth}
    \includegraphics[width=\linewidth]{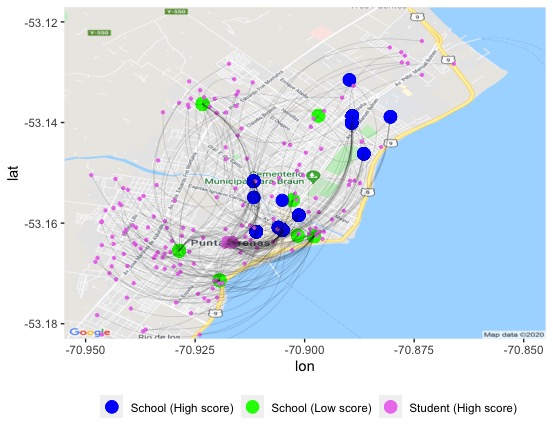}
    \caption*{B. High ability } 
  \end{minipage}
  \begin{minipage}{17.0cm}
\footnotesize{
    {Notes: The sample consists of students who participated in DA for ninth grade admissions in 2016 in Magallanes and therefore started ninth grade in the allocated school in 2017. The actual allocation data is taken from enrollment files for 2017.}}
    \end{minipage}
  \end{figure}
  
  \newpage
  \begin{figure}[H] 
  \caption{Segregation in Magallanes}
  \centering
    \label{segregation17} 
    \includegraphics[width=0.5\linewidth]{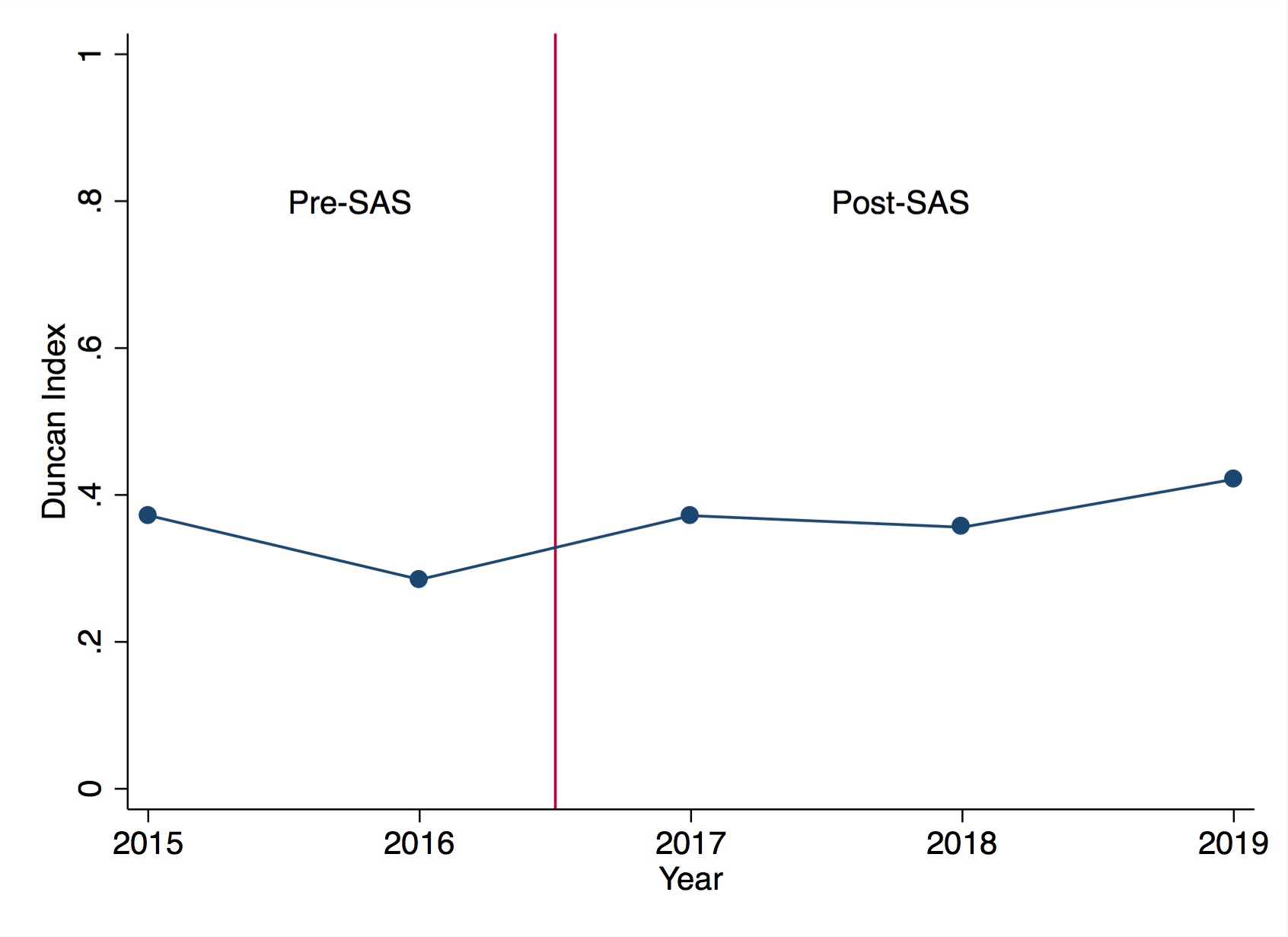}
\begin{minipage}{17.0cm}
\footnotesize{
    {Notes: The sample consists of ninth-grade students and schools in Magallanes. The enrollment under DA will get reflected in 2017 enrollment files.  I consider public and voucher schools for this analysis as these two types of schools participated in DA.}}
    \end{minipage}
  \end{figure}
  
  \newpage
  \begin{figure}[H] 
  \caption{Overlap between predicted ROL and the best schools in Magallanes}
    \label{abilitylatent} 
    \centering
    \includegraphics[width=0.60\linewidth]{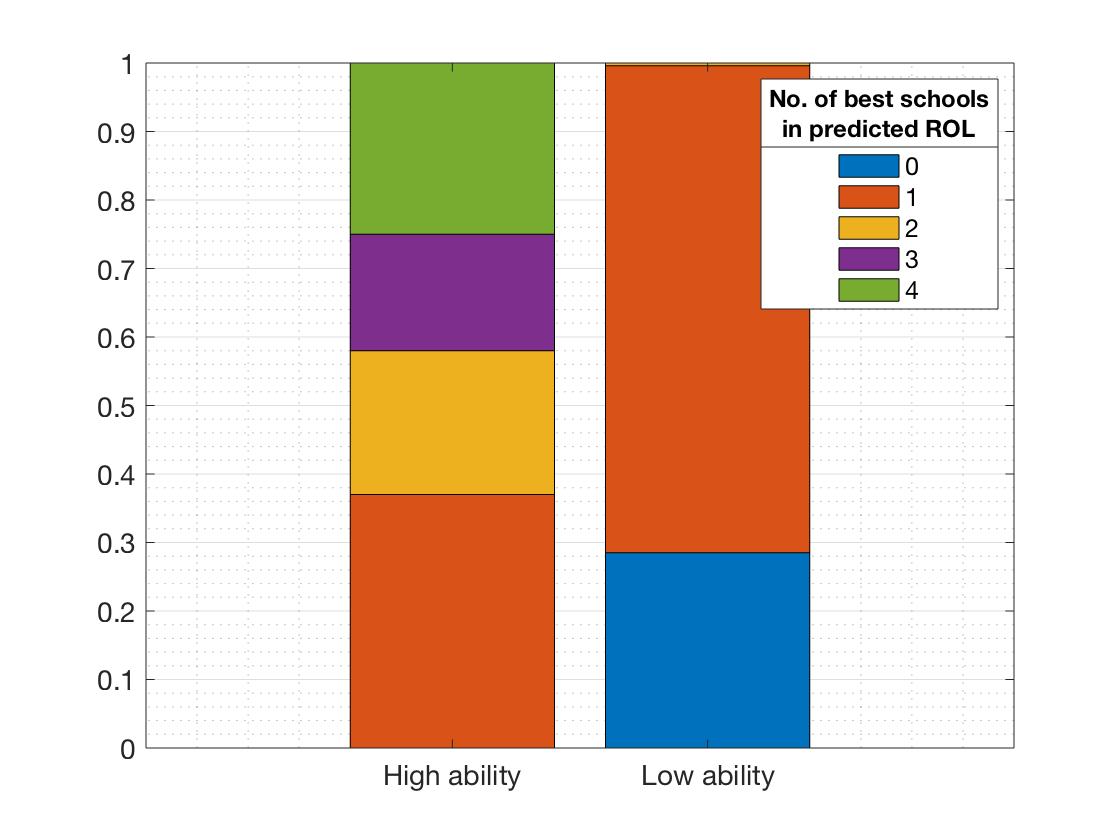}
    \begin{minipage}{14.0cm}
\footnotesize{
    {Notes: The predicted ROL has been generated using the estimates of model (2) in Table \ref{resultstable2}.}}
    \end{minipage}
\end{figure}
  
  \newpage
   \begin{figure}[H] 
  \centering
  \caption{Predicted probability: Outside option}
    \label{outsideoption} 
   \begin{minipage}[b]{0.40\linewidth}
    %\centering
    \includegraphics[width=\linewidth]{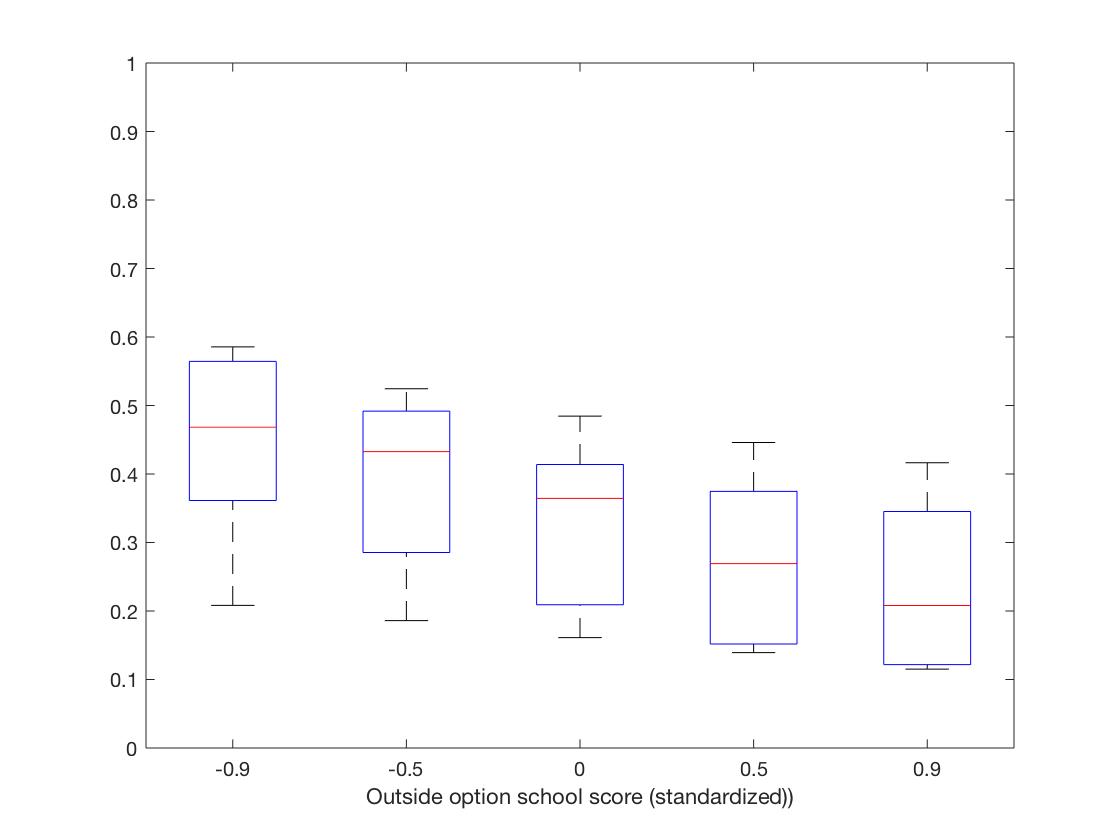} 
       \caption*{A. Mean school} 
  \end{minipage} 
\end{figure}
  
  %\newpage
\begin{figure}[H] 
  \caption{Predictive margins for high score school}
    \centering
    \label{allregions_dist1} 
  \begin{minipage}[b]{0.30\linewidth}
    %\centering
    \includegraphics[width=\linewidth]{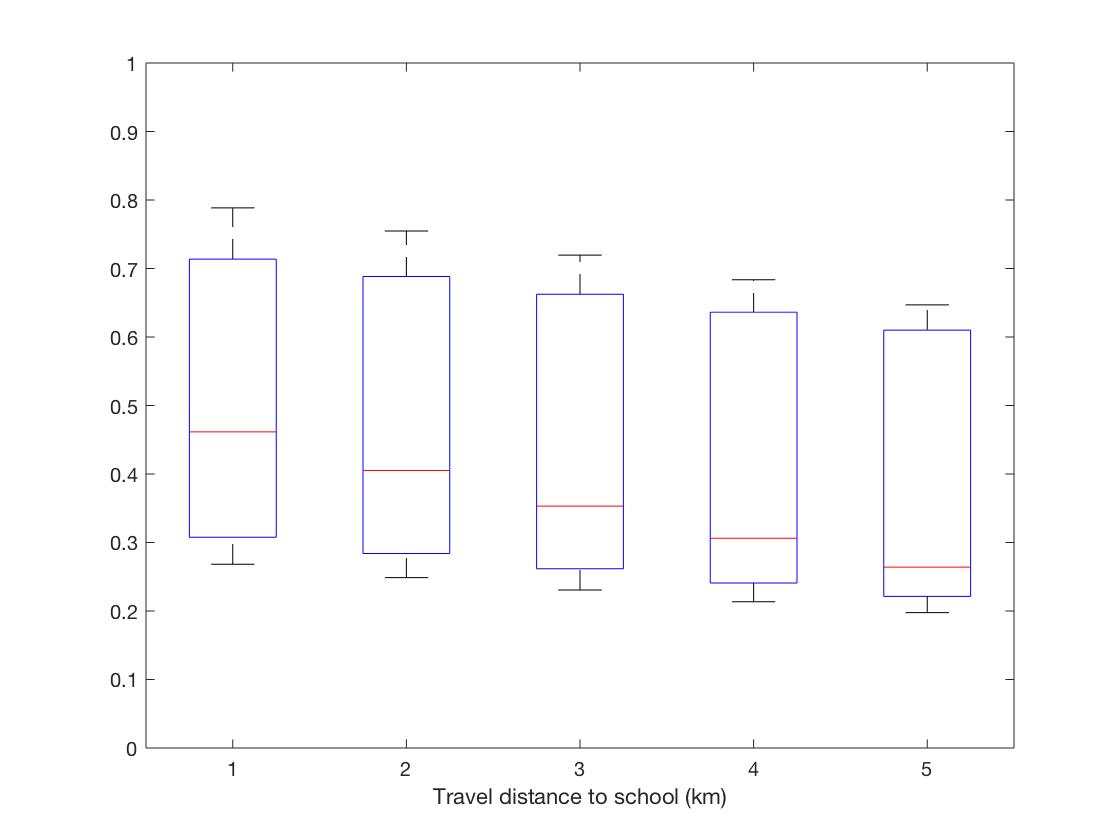} 
       \caption*{B. Low income high ability} 
           \vspace{0ex}
  \end{minipage} 
  \begin{minipage}[b]{0.30\linewidth}
    %\centering
      \includegraphics[width=\linewidth]{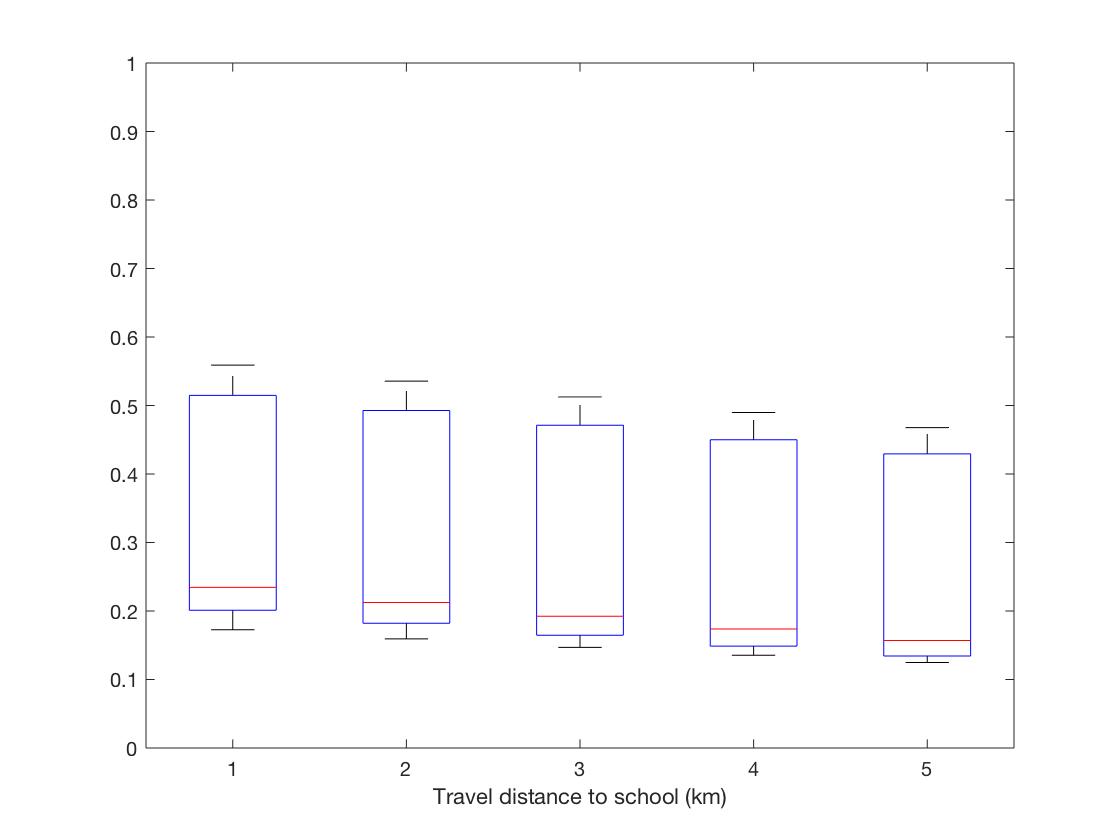}
       \caption*{C. Low income low ability} 
           \vspace{0ex}
  \end{minipage} 
    \begin{minipage}[b]{0.30\linewidth}
    %\centering
      \includegraphics[width=\linewidth]{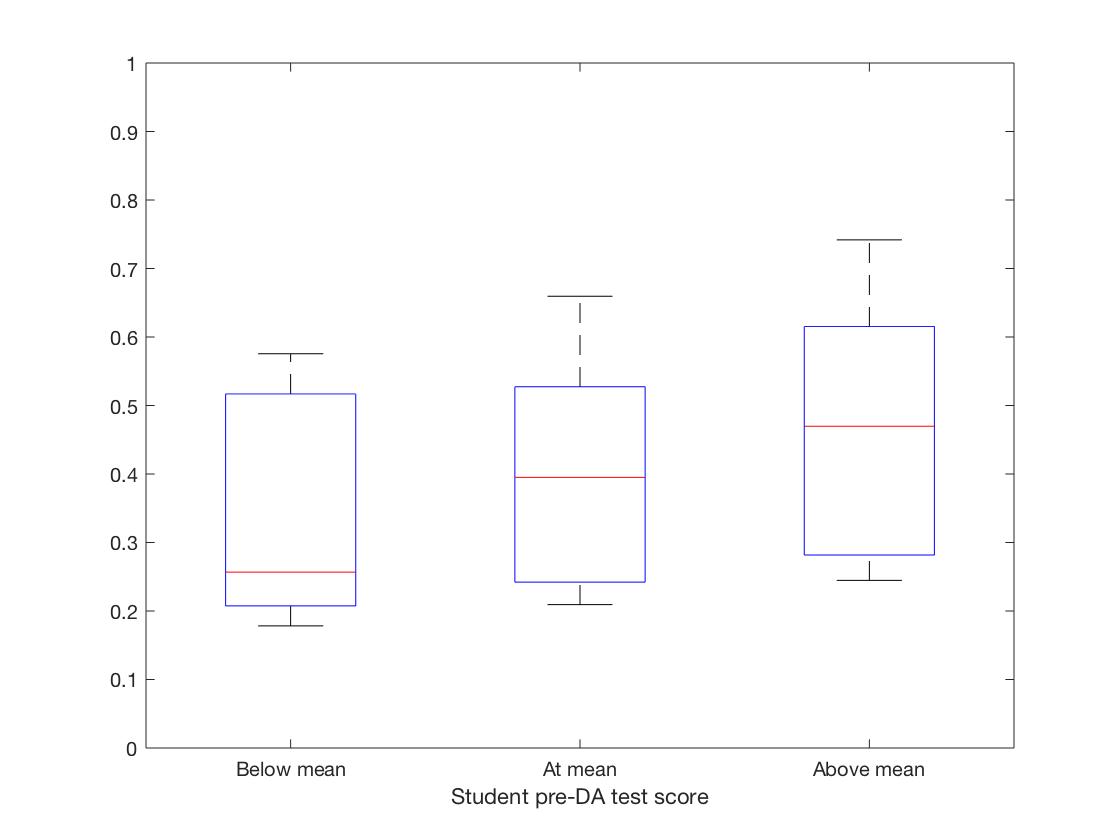}
       \caption*{C. Low income ability} 
           \vspace{0ex}
  \end{minipage} 
  \begin{minipage}{17.5 cm}{\footnotesize{}}
\end{minipage} 
  \end{figure}
\newpage
\begin{figure}[H] 
  \caption{Predictive probability of ranking high score school over low score school}
    \centering
    \label{allregions_ability2} 
  \begin{minipage}[b]{0.45\linewidth}
    %\centering
    \includegraphics[width=\linewidth]{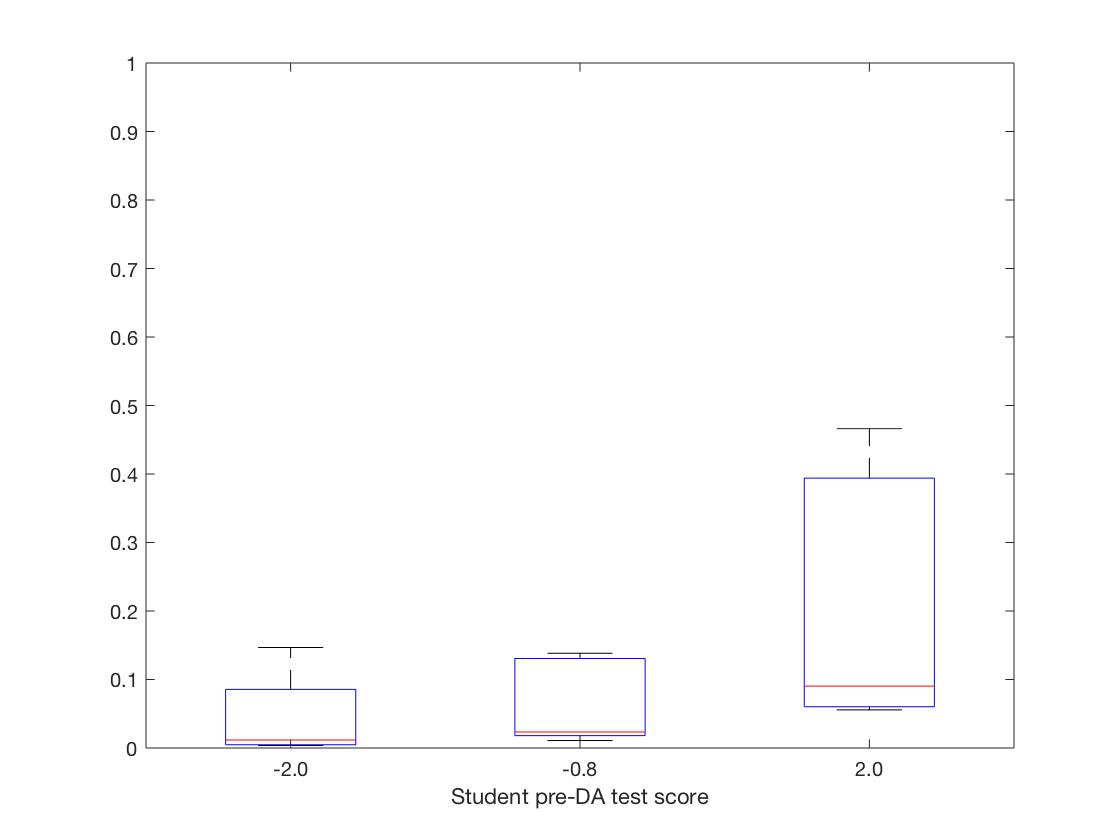} 
       \caption*{A. High score vs. low score} 
           \vspace{0ex}
  \end{minipage} 
  \begin{minipage}{17.5 cm}{\footnotesize{Notes: The predicted probability for this analysis is computed as $P(H>L>0)=\frac{e^{V_{H}}}{e^{V_{H}}+e^{V_{L}}}(1-F(-V_{L})-\frac{e^{V_{H}}}{e^{V_{H}}+e^{V_{L}}}F(-V_{L})(1-F(-V_{H}))$, where $H$ corresponds to the high score school and $L$ corresponds to the low score school. }}
\end{minipage} 
  \end{figure}

\newpage
\begin{figure}[H] 
  \caption{Simulated probability: Rank high score school over low score school}
    \centering
    \label{allregions_simu} 
  \begin{minipage}[b]{0.45\linewidth}
    %\centering
    \includegraphics[width=\linewidth]{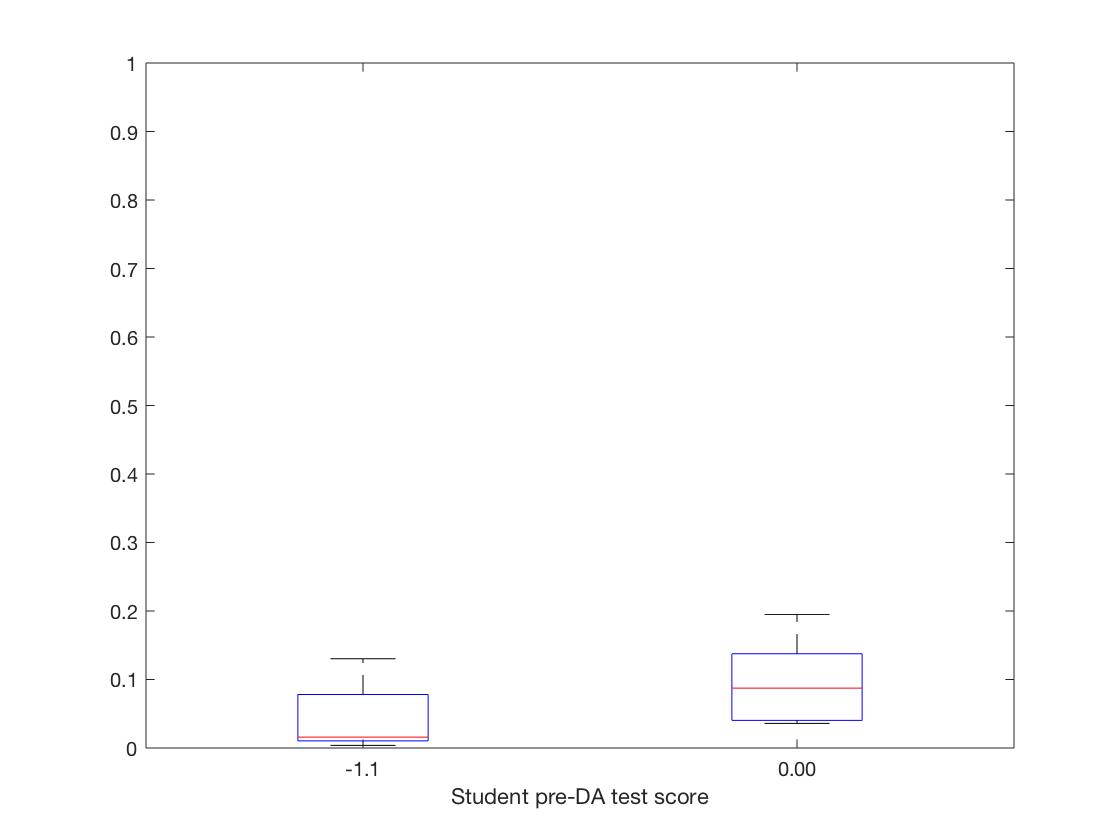} 
       \caption*{A. Actual predicted probability} 
           \vspace{0ex}
  \end{minipage} 
  \begin{minipage}[b]{0.45\linewidth}
    %\centering
      \includegraphics[width=\linewidth]{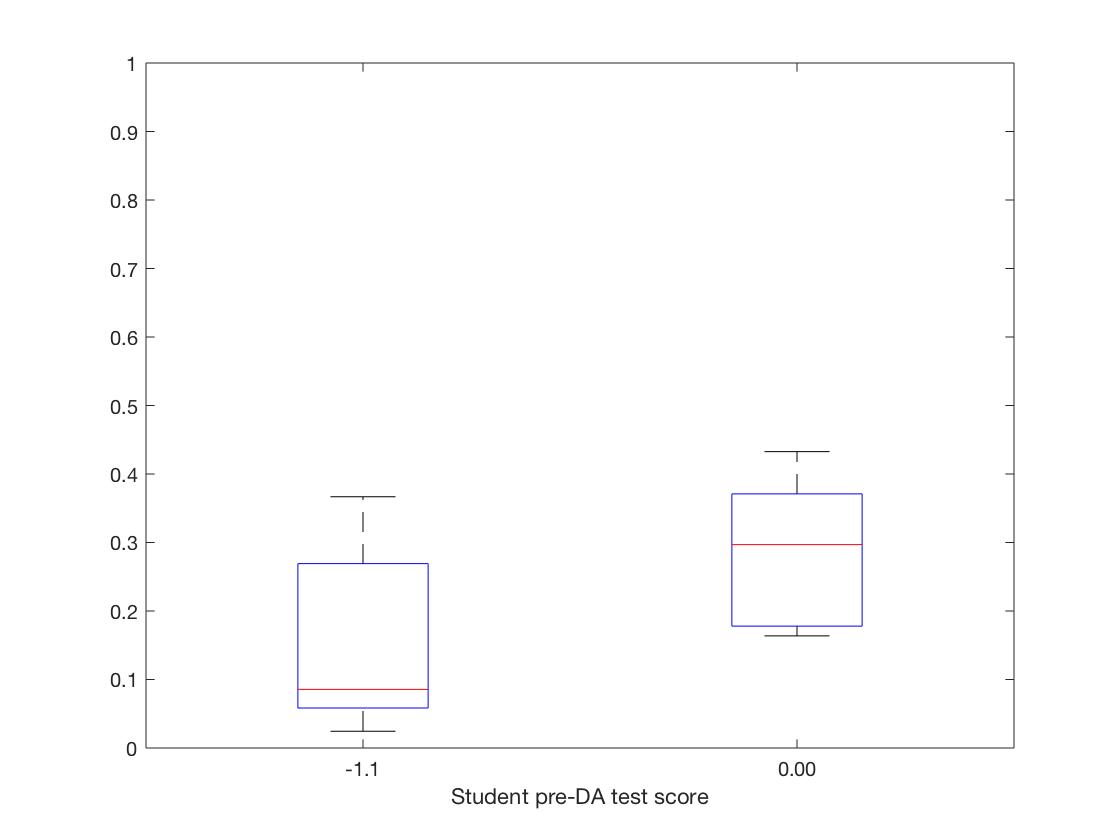}
       \caption*{B. Simulated predicted probability} 
           \vspace{0ex}
  \end{minipage} 
  \begin{minipage}{17.5 cm}{\footnotesize{}}
\end{minipage} 
  \end{figure}

\newpage
\bibliography{typ}

\end{document}